\documentclass[%
 reprint,
 superscriptaddress,
 amsmath,
 aps,
 prb
]{revtex4-2}

\usepackage[pdftex]{graphicx, color}
\usepackage[pdftex,
            colorlinks=true,
            citecolor=blue,
            linkcolor=blue]{hyperref}
\usepackage{newtxtext, newtxmath}
\usepackage{siunitx}
\usepackage{bm}
\usepackage{braket, mathtools}
\usepackage{bbm}
\usepackage{ulem}
\usepackage{tabularx}
\usepackage{comment}

\graphicspath{{./images/}}

\begin{document}

\title{Nonlinear optical responses in superconductors under magnetic fields:\\ quantum geometry and topological superconductivity}

\author{Hiroto Tanaka}
\email[]{tanaka.hiroto.54z@st.kyoto-u.ac.jp}
\affiliation{%
Department of Physics, Graduate School of Science, Kyoto University, Kyoto 606-8502, Japan
}%

\author{Hikaru Watanabe}
\affiliation{%
Research Center for Advanced Science and Technology, University of Tokyo, Komaba Meguro-ku, Tokyo 153-8904, Japan
}%

\author{Youichi Yanase}
\affiliation{%
Department of Physics, Graduate School of Science, Kyoto University, Kyoto 606-8502, Japan
}%

\date{\today}

\begin{abstract}
Noncentrosymmetric 
superconductors offer fascinating phenomena of quantum transport and optics 
such as nonreciprocal and nonlinear responses. 
Time-reversal symmetry breaking often plays an essential role in the emergence and enhancement of nonreciprocal transport. In this paper, we show the nonreciprocal optical responses in noncentrosymmetric superconductors arising from time-reversal symmetry breaking 
by demonstrating them in $s$-wave superconductors with a Rashba spin-orbit coupling and a magnetic field. Numerical results reveal the superconductivity-induced bulk photocurrent and second harmonic generation, which are forbidden at the zero magnetic field. 
We discuss the properties and mechanisms of the superconducting nonlinear responses emerging under the magnetic field.
In particular, we investigate the magnetic-field dependence of the photocurrent conductivity and clarify 
the essential ingredients 
which give a contribution unique to superconductors under the magnetic field. This contribution is dominant in the low carrier density regime although the corresponding joint density of states is tiny. We attribute the enhancement to the quantum geometry. 
Moreover, the nonlinear conductivity shows peculiar sign reversal at the transition to the topological superconducting state. We propose a bulk probe of topological transition and quantum geometry in superconductors. 
\end{abstract}

\maketitle

\section{Introduction}
Inversion symmetry breaking offers a variety of physical phenomena, which have been attracting attention from fundamental science to technology. Especially, nonreciprocal transport in noncentrosymmetric quantum materials is one of the central topics in condensed matter physics~\cite{Ideue2021,Tokura2018}. Rectification of electric current is an example, that occurs when the resistances for rightward and leftward electric currents are different~\cite{Rikken2001,Krstic2002,Pop2014,Rikken2005,Ideue2017,Wakatsuki2018,Hoshino2018,Wakatsuki2017,Qin2017,Yasuda2019,Ideue2020,Itahashi2020}. It is an effective tool for estimating the microscopic characteristics of noncentrosymmetric systems such as Rashba spin splitting in a polar semiconductor~\cite{Ideue2017}. 

The nonlinear optical response is also strongly correlated to the symmetry of quantum phases~\cite{Orenstein2021}. In particular, the second-order optical response is a useful probe for the microscopic parity violation in complex ordered states because the second-order response requires broken space inversion ($\mathcal{P}$) symmetry~\cite{Zhao2017,Torre2021}. 
The second-order optical response and nonreciprocal transport are described in a unified manner by the formula of the 
nonlinear conductivity, which is given by
\begin{align}\left<\mathcal{J}^{\alpha}(\omega)\right>_{(2)} =\int \frac{d\Omega}{2\pi}\sigma^{\alpha;\beta\gamma}(\omega;\Omega,\omega -\Omega)E^{\beta}(\Omega)E^{\lambda}(\omega -\Omega).
\end{align}
For example, the second-order optical responses for monochromatic light are constituted by the second harmonic generation and photogalvanic effect, which are denoted by the nonlinear conductivity $\sigma(2\omega,\omega,\omega)$ and $\sigma(0;\omega, -\omega)$, respectively. The second harmonic generation, which is a frequency doubling of the light through interaction with media, is particularly sensitive to the microscopic space inversion symmetry breaking in materials~\cite{Zhao2017,Torre2021,Fiebig2005}. The photogalvanic effect, which is a photo-induced direct current (photocurrent), is also a topic of current interest. Since the bulk photocurrent originating from the microscopic parity violation is influenced by the symmetry and geometric properties of the system, topological materials~\cite{Orenstein2021,Liu2020} and parity-violating magnets~\cite{Ogawa2016,Burger2020} are potential candidates of the novel type of photo-electric converter.

In this paper, we focus on superconductors, which host remarkable electromagnetic properties such as the zero resistivity phenomenon and the Meissner effect. It has been shown that Cooper pairs' condensation induces various nonlinear and nonreciprocal responses. For example, the amplitude mode of the fluctuating order parameter, which is called Higgs mode, gives rise to the reciprocal third-order optical responses~\cite{Matsunaga2014,Cea2016,Shimano-Tsuji}. 
The purpose of this study is to explore nonreciprocal optical responses, which are prohibited in centrosymmetric superconductors but arise from the interplay of space inversion symmetry breaking and superconductivity. 
In particular, we study the second-order optical responses of noncentrosymmetric superconductors. 

The noncentrosymmetric superconductor is an attractive material platform for exotic quantum phenomena such as the mixed singlet-triplet pairing~\cite{Gorkov2001, Frigeri2004,Bauer2012,Smidman2017} and topological superconductivity~\cite{Alicea2012, Sato2016}. The boosted upper critical field beyond the Pauli-Clogston-Chandrasekhar limit, which arises from the spin-momentum locking, is another characteristic property of noncentrosymmetric superconductors. Ising superconductivity has been reported in transition metal dichalcogenides, such as gated $\mathrm{MoS_{2}}$~\cite{Saito2016,Lu2015}, $\mathrm{NbSe_{2}}$~\cite{Xi2016}, and $\mathrm{TaS_{2}}$~\cite{delaBarrera2018}, in which the $\mathcal{P}$ symmetry breaking leads to a large spin-orbit coupling and huge upper critical field. Moreover, the $\mathcal{P}$ symmetry breaking causes unique nonreciprocal transport phenomena. For example, the superconducting fluctuation ~\cite{Wakatsuki2018, Wakatsuki2017,Qin2017} and the dynamics of vortices ~\cite{Hoshino2018, Ideue2020} enhance the magnetochiral anisotropy. Recently, the superconducting diode effect, which means zero resistivity in the forward current direction while a finite resistivity in the backward direction, has been 
intensively studied in various platforms~\cite{Ando2020,Nagaosa-Yanase,Nadeem2023}. These transport phenomena are useful for investigating unconventional superconducting states such as the helical superconducting state with finite-momentum Cooper pairs~\cite{Daido2022, He2022, Yuan2022, daido2023rectification}.

Although the nonlinear optics also potentially yield rich information on exotic superconducting states, the properties and mechanisms of the nonlinear optics in superconductors are yet-to-be uncovered. Here, we refer to a few papers that studied the second-order nonlinear responses in superconductors.
Experimentally, the second harmonic generation was observed in a $s$-wave superconductor thin film, where the $\mathcal{P}$ symmetry is broken not by the crystalline structure but by the supercurrent~\cite{Nakamura2020, Vaswani2020-ij}.
In theoretical studies, the second-order optical conductivity was formulated based on the Bogoliubov-de Gennes (BdG) Hamiltonian~\cite{Xu2019,Watanabe2022}, and numerical calculations have shown characteristic properties of the superconducting nonlinear responses~\cite{Tanaka2023}. For example, the resonant component, which originates from the transition between the electron and hole bands, shows a sharp peak with a resonant frequency around the superconducting gap. 
It was also verified that the coexistence of intraband and interband pairing is necessary for the second-order superconducting optical responses in time-reversal ($\mathcal{T}$) symmetric single-band superconductors~\cite{Tanaka2023, Huang2023}. It indicates that the superconducting nonlinear responses need spin-triplet or multiband superconductivity when the $\mathcal{T}$ symmetry is conserved.
Because of this constraint, it is possible to detect spin-triplet Cooper pairs by optical measurements. 

As we reviewed above, the recent progress unveiled the basic mechanism of the second-order nonlinear responses in $\mathcal{T}$-symmetric superconductors. 
Naturally, the next topic of interest is the effect of $\mathcal{T}$ symmetry breaking. 
The effects of $\mathcal{T}$ symmetry breaking are expected to be significant because of the following reasons. 
First, it was shown that the photogalvanic effect in the normal state is enhanced by the $\mathcal{T}$ symmetry breaking~\cite{Watanabe2021, Ahn2020} (Magneto-photogalvanic effect). 
Second, in the two-dimensional superconductors, which are suitable for optical measurements, the second-order responses are prohibited by $\mathcal{T}$ symmetry in most noncentrosymmetric point groups~\cite{Watanabe2021,Tanaka2023}. 
Third, the $s$-wave superconductors do not show the second-order optical responses characteristic of superconductors, when the $\mathcal{T}$ symmetry is conserved~\cite{Tanaka2023}.
Therefore, it is desirable to clarify how the $\mathcal{T}$ symmetry breaking influences the nonlinear responses in superconductors.

\begin{figure*}[htbp]
 \includegraphics[width=\linewidth]{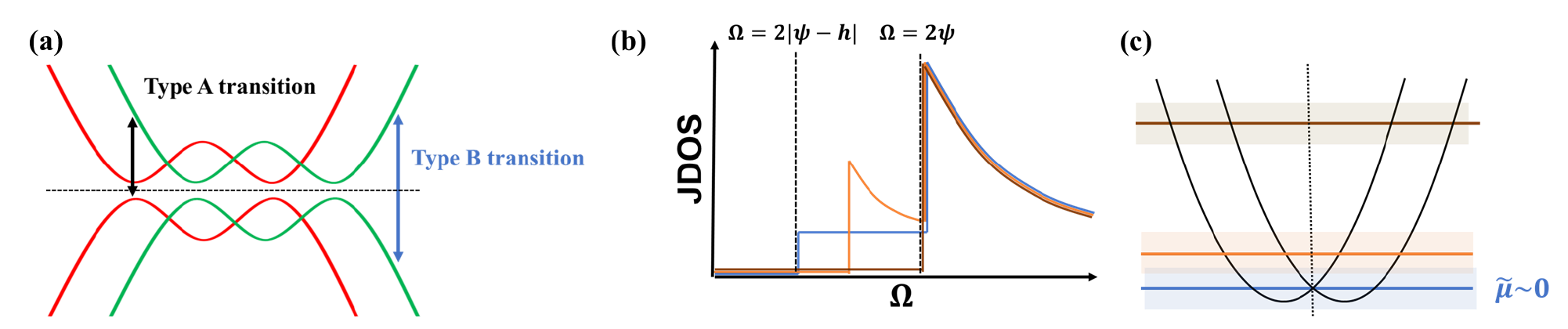}
 \caption{(a) Schematic illustration of the band structure and the optical transitions in the superconducting state. 
 The type A and type B transitions give resonant contributions to the second-order optical responses. The type A transition is an analog of optical transitions in the normal state. The type B transition gives rise to the superconducting nonlinear responses~\cite{Tanaka2023}. (b) Schematic illustration of the JDOS, which is derived from the type B transition. 
 The color of the lines corresponds to the Fermi level illustrated in Fig.~\ref{fig:schematic_figure}(c). A peak appears in the JDOS at $\Omega=2\psi$. In addition, a plateau (blue line) or a peak (orange line) may appear between $\Omega =2|\psi -h|$ and $\Omega =2\psi$, when the Fermi level is close to the Dirac point at the time-reversal invariant momentum ($\Gamma$ point in this paper). 
 (c) Approximate parabolic bands around the $\Gamma$ point in the normal state. The nonlinear optical responses show characteristic behaviors under the magnetic field when the chemical potential measured from the Dirac point $\tilde{\mu}$ is small.
}
 \label{fig:schematic_figure}
\end{figure*}

In the following part, we theoretically investigate the second-order optical responses of $s$-wave superconductors under the magnetic field. The photocurrent and second harmonic generation are demonstrated in a model with a two-dimensional $\mathrm{C_{4v}}$ crystal structure. We show that the $\mathcal{T}$ symmetry breaking due to the magnetic field enables the superconducting nonlinear responses to be finite although they disappear at zero magnetic fields. 
The nonlinear conductivities show the peak structure and low-frequency divergence which correspond to the resonant and nonresonant components of superconducting nonlinear responses, respectively. In addition, the dependence of the photocurrent conductivity on the magnetic field and the chemical potential is explained based on the approximately defined joint density of states (JDOS) [$\tilde{J}_{\mathrm{B(i)}}$ and $\tilde{J}_{\mathrm{B(ii)}}$ introduced later]. 
In particular, the contribution from $\tilde{J}_{\mathrm{B(ii)}}$ is sensitive to the magnetic field in the low-carrier-density regime as illustrated in Figs.~\ref{fig:schematic_figure}(b) and \ref{fig:schematic_figure}(c).
The chemical potential from the Dirac point at ${\bm k}=\bm{0}$ is the controlling parameter as is the case for topological $s$-wave superconductivity~\cite{Alicea2012,Sato2016}. We will see that quantum geometry~\cite{Resta2011} strongly enhances this field-sensitive component of superconducting nonlinear responses. Moreover, we show that the photocurrent conductivity drastically changes at the topological transition. Therefore, the second-order optical responses can be a probe of quantum geometry and topological superconductivity.

The outline of the paper is given below. In Sec.~\ref{sec:formulation}, we briefly explain the formulation and review the second-order nonlinear optical responses in superconductors. 
In Sec.~\ref{sec:joint_density_of_state}, we analyze the JDOS before showing the nonlinear optical conductivity. The JDOS is an essential quantity to discuss the characteristic behavior of photocurrent conductivity. In Sec.~\ref{sec:Numerical_results}, we demonstrate the photocurrent and second harmonic generation by numerical calculations. We explain the dependence of the photocurrent conductivity on the magnetic field and the chemical potential by the JDOS in Sec.~\ref{sec:Unique behaviors due to Dirac point}.
In particular, we reveal the characteristic behaviors due to the Dirac point. In Sec.~\ref{sec:enhancement by quantum geometry}, we show the enhancement of photocurrent conductivity due to the peculiar quantum geometry. In Sec.~\ref{sec:photocurrent conductivity around topological transition}, we show the specific behavior of photocurrent conductivity around the transition to the topological superconducting state. In Sec.~\ref{sec:discussion}, we discuss the multiple roles of the magnetic field and propose candidate materials for observing superconducting nonlinear responses. 
A brief summary is given in Sec.~\ref{sec:summary}.
Throughout this paper, we present formulas with $\hbar =1$ (Dirac constant) and $q=1$ (electron charge). 

\section{formulation}
\label{sec:formulation}
\subsection{Model Hamiltonian}
We consider s-wave superconductors with an antisymmetric spin-orbit coupling under a magnetic field.
The Hamiltonian is given by 
\begin{align}
    \mathcal{H} =& \mathcal{H}_{\mathrm{kin}}+\mathcal{H}_{\mathrm{ASOC}}+\mathcal{H}_{\mathrm{Zeeman}}+\mathcal{H}_{\mathrm{s-wave}}, \label{eq:normal_H} \\
    \mathcal{H}_{\mathrm{kin}} =& \sum_{\bm{k},s} \xi(\bm{k})c_{\bm{k},s}^{\dagger}c_{\bm{k},s}, \\
    \mathcal{H}_{\mathrm{ASOC}} =& \sum_{\bm{k},s,s^{\prime}}\bm{g}(\bm{k})\cdot\bm{\sigma}_{ss^{\prime}}c_{\bm{k},s}^{\dagger}c_{\bm{k},s^{\prime}}, \\
    \mathcal{H}_{\mathrm{Zeeman}} =& \sum_{\bm{k},s,s^{\prime}}\bm{h}\cdot\bm{\sigma}_{ss^{\prime}}c_{\bm{k},s}^{\dagger}c_{\bm{k},s^{\prime}}, \\
    \mathcal{H}_{\mathrm{s-wave}} =& \sum_{\bm{k}} \left(\psi c^{\dagger}_{\bm{k}+\bm{q},\uparrow}c^{\dagger}_{-\bm{k}+\bm{q},\downarrow} + \mathrm{H.c.}\right),
    \label{eq:H_swave}
\end{align}
where $\bm{\sigma}=(\sigma_{\mathrm{x}},\sigma_{\mathrm{y}},\sigma_{\mathrm{z}})$ is the vector of Pauli matrices, and $c_{\bm{k},s}$ ($c_{\bm{k},s}^{\dagger}$) is the annihilation (creation) operator with momentum $\bm{k}$ and spin $s$. In the following, we assume the two-dimensional $\mathrm{C_{4v}}$ crystal structure. $\mathcal{H}_{\mathrm{kin}}$ is a kinetic energy in the tight-binding approximation measured from a chemical potential $\mu$, and $\mathcal{H}_{\mathrm{ASOC}}$ represents a Rashba-type spin-orbit coupling. The kinetic energy and the g-vector of spin-orbit coupling are assumed as
\begin{align}
    \xi(\bm{k}) &= -2t_{1}(\cos k_{x} + \cos k_{y}) + 4t_{2}\cos k_{x}\cos k_{y} - \mu, \\
    \bm{g}(\bm{k}) &= \alpha (\sin k_{y}, -\sin k_{x}, 0).
\end{align}
Later, the parameters are set as $t_{1}=1$, $t_{2}=0.2$, and $\alpha =0.4$ for the numerical calculations.
$\mathcal{H}_{\mathrm{Zeeman}}$ is a Zeeman field, and $\mathcal{H}_{\mathrm{s-wave}}$ represents an $s$-wave superconducting order parameter introduced phenomenologically. The magnitude of the $s$-wave pair potential is set as $\psi = 0.09$. The Zeeman field is assumed as
\begin{align}
    \bm{h} = (0, h\cos\theta, h\sin \theta), \label{eq:magnetic_field}
\end{align}
with $\theta$ being the angle between the magnetic field and the two-dimensional plane. In the Rashba superconductor with an in-plane component of the magnetic field, the total momentum of Cooper pairs is finite without injecting an electric current~\cite{Smidman2017,Bauer2012}. This is called the helical superconducting state, and we introduce the Cooper pairs' momentum $2\bm{q}$ in Eq.~\eqref{eq:H_swave}. We can rewrite the Hamiltonian in the matrix form by using the Nambu spinor $\bm{c_{k,\,q}}=(c_{\bm{k}+\bm{q},\uparrow},c_{\bm{k}+\bm{q},\downarrow},c_{-\bm{k}+\bm{q},\uparrow}^{\dagger},c_{-\bm{k}+\bm{q},\downarrow}^{\dagger})^{\top}$,
\begin{align}
    \mathcal{H} &= \frac{1}{2}\sum_{\bm{k}}\bm{c_{k,\,q}}^{\dagger}H(\bm{k},\bm{q})\bm{c_{k,\,q}} \\
    &= \frac{1}{2}\sum_{\bm{k}}\bm{c_{k,\,q}}^{\dagger}
    \begin{pmatrix}
    H_{\mathrm{N}}(\bm{k}+\bm{q}) & \psi(i\sigma_{y}) \\
    \psi(i\sigma_{y})^{\top} & -H_{\mathrm{N}}(-\bm{k}+\bm{q})^{\top}
    \end{pmatrix}
    \bm{c_{k,\,q}}.
\end{align}
Here, $H_{\mathrm{N}}(\bm{k})$ is the normal-state Hamiltonian given by
\begin{align}
    H_{\mathrm{N}}(\bm{k}) = \xi(\bm{k}) + \left(\bm{g}(\bm{k})+\bm{h}\right)\cdot\bm{\sigma}.
\end{align}

The total momentum of Cooper pairs $2\bm{q}$ is determined so as to minimize the free energy, which is calculated by
\begin{align}
    F &= \left<\mathcal{H}\right>_{\mathrm{eq}}-TS, \\
    &=-\frac{T}{V}\sum_{\bm{k},\alpha}\log\left(1+e^{-E_{\alpha}(\bm{k},\,\bm{q})/T}\right).
    \label{eq:free_energy}
\end{align}
See Appendix~\ref{app:finite_momentum} for details.

\subsection{General formula for nonlinear conductivity}
Next, this subsection briefly explains the formulation of the nonlinear optical response in superconductors. With the velocity gauge $\bm{E}=-\partial_{t}\bm{A}(t)$, the coupling to external electric fields is introduced by the minimal coupling prescription $H_{\mathrm{N}}(\bm{k})\rightarrow H_{\mathrm{N}}(\bm{k}-e\bm{A})$. Thus, the vector potential $\bm{A}$ dependence of the Hamiltonian is expressed as
\begin{align}
    H(\bm{k},\bm{q},\bm{A}) = 
    \begin{pmatrix}
    H_{\mathrm{N}}(\bm{k}+\bm{q}-e\bm{A}) & \psi(i\sigma_{y}) \\
    \psi(i\sigma_{y})^{\dagger} & -H_{\mathrm{N}}(-\bm{k}+\bm{q}-e\bm{A})^{\top}
    \end{pmatrix}.
\end{align}
The perturbative Hamiltonian $\Delta H(\bm{k},\bm{q}, t)$ is given by
\begin{align}
    \Delta H(\bm{k},\bm{q}, t) &\equiv H(\bm{k},\bm{q}, \bm{A}(t)) - H(\bm{k},\bm{q}, \bm{0}) \\
    &= \sum_{n=1} \frac{1}{n!}A^{\alpha_{1}}(t)\cdots A^{\alpha_{n}}(t)J^{\alpha_{1}\cdots\alpha_{n}}(\bm{k},\bm{q}),
\end{align}
where we introduce the generalized velocity operator
\begin{align}
    J^{\alpha_{1}\cdots\alpha_{n}}(\bm{k},\bm{q})=(-1)^{n}\left.\frac{\partial^{n}H(\bm{k},\bm{q},\bm{A})}{\partial A^{\alpha_{1}}\cdots \partial A^{\alpha_{n}}}\right|_{\bm{A}=\bm{0}}.
\end{align}
We define the electric current density operator $\mathcal{J}^{\alpha}(t)$ as
\begin{align}
    \mathcal{J^{\alpha}}(t) = \frac{1}{2}\sum_{\bm{k}}\bm{c_{k,\,q}}^{\dagger}J_{\bm{A}}^{\alpha}(\bm{k},\bm{q}, t)\bm{c_{k,\,q}}, 
\end{align}
where we define $J_{\bm{A}}^{\alpha}(\bm{k},\bm{q}, t)$ by
\begin{align}
    J_{\bm{A}}^{\alpha}(\bm{k},\bm{q}, t) = \sum_{m=0}\frac{1}{m!}(-1)^{m}A^{\beta_{1}}(t)\cdots A^{\beta_{m}}(t)J^{\alpha\beta_{1}\cdots\beta_{m}}(\bm{k},\bm{q}).
\end{align}
Following the standard perturbative treatment, we evaluate the expectation value of the electric current density 
\begin{align}
\left<\mathcal{J}^{\alpha}(\omega)\right>=\sum_{n=1}\left<\mathcal{J}^{\alpha}(\omega)\right>_{(n)},
\end{align}
where $\left<\mathcal{J}^{\alpha}(\omega)\right>_{(n)}$ is the electric current of the $n$-th order in the electromagnetic field $\bm{A}$. The formula for the second-order nonlinear conductivity is obtained as
\begin{align}
\label{eq:tot_Res}
&\sigma^{\alpha;\beta\gamma}(\omega;\omega_{1},\omega_{2}) = \notag \\
& \frac{1}{2(i\omega_{1} - \eta)(i\omega_{2} - \eta)}\sum_{\bm{k}}\Biggl[\sum_{a}\frac{1}{2}J^{\alpha\beta\gamma}_{aa}f_{a} \notag \\
&+ \sum_{a,b}\frac{1}{2}\left(\frac{J^{\alpha\beta}_{ab}J^{\gamma}_{ba}f_{ab}}{\omega_{2}+i\eta-E_{ba}} + \frac{J^{\alpha\gamma}_{ab}J^{\beta}_{ba}f_{ab}}{\omega_{1}+i\eta-E_{ba}}\right) \notag \\
&+ \sum_{a,b}\frac{1}{2}\frac{J^{\alpha}_{ab}J^{\beta\gamma}_{ba}f_{ab}}{\omega+2i\eta-E_{ba}} \notag \\
&+ \sum_{a,b,c} \frac{1}{2}\frac{J^{\alpha}_{ab}}{\omega+2i\eta-E_{ba}}\left(\frac{J^{\beta}_{bc}J^{\gamma}_{ca}f_{ac}}{\omega_2+i\eta-E_{ca}}-\frac{J^{\beta}_{ca}J^{\gamma}_{bc}f_{cb}}{\omega_2+i\eta-E_{bc}}\right) \notag \\
&+ \sum_{a,b,c} \frac{1}{2}\frac{J^{\alpha}_{ab}}{\omega+2i\eta-E_{ba}}\left(\frac{J^{\gamma}_{bc}J^{\beta}_{ca}f_{ac}}{\omega_1+i\eta-E_{ca}}-\frac{J^{\gamma}_{ca}J^{\beta}_{bc}f_{cb}}{\omega_1+i\eta-E_{bc}}\right)\Biggr],
\end{align}
where indices $a,b,c$ are spanned by the energy eigenvalues $E_{a}$ of the unperturbed Hamiltonian $H(\bm{k},\bm{q})$ and the energy difference is $E_{ab} \equiv E_{a}-E_{b}$. We omit the description $(\bm{k},\bm{q})$ from $J^{\alpha_{1}\cdots\alpha_{n}}(\bm{k},\bm{q})$ and others. We introduce the Fermi-Dirac distribution function $f_{a}=(e^{\beta E_{a}} + 1)^{-1}$ and defined $f_{ab} \equiv f_{a}-f_{b}$. The infinitesimal positive parameter $\eta$ appears due to the adiabatic application of the external fields.

\subsection{Normal and anomalous photocurrent responses}
In this subsection, we discuss the photocurrent response given by the second-order nonlinear conductivity $\sigma^{\alpha;\beta\gamma}(0;\Omega,-\Omega)$. In the gapful superconductors at low temperatures, the total photocurrent conductivity is decomposed into the two components~\cite{Watanabe2022,Nagaosa-Yanase}
\begin{align}
    \sigma = \sigma_{\mathrm{n}} + \sigma_{\mathrm{a}}.
\end{align}
The first term $\sigma_{\mathrm{n}}$ is a normal photocurrent that can be finite even in the normal state, while the second term $\sigma_{\mathrm{a}}$ is an anomalous photocurrent unique to the superconducting state~\cite{Watanabe2022}. Note that the superconducting nonlinear responses appear even in the normal photocurrent due to the optical transition unique to the superconducting state~\cite{Tanaka2023}. As we show below, the normal photocurrent corresponds to the resonant components characterized by the optical transition, whereas the anomalous photocurrent arises from the non-resonant components that show divergent behaviors in the low-frequency regime.

The normal photocurrent consists of four contributions
\begin{align}
    \sigma_{\mathrm{n}} = \sigma_{\mathrm{Einj}}+\sigma_{\mathrm{Minj}}+\sigma_{\mathrm{shift}}+\sigma_{\mathrm{gyro}},
\end{align}
which are termed electric injection current~\cite{Sipe2000}, magnetic injection current~\cite{Zhang2019}, shift current~\cite{Von_Balz1981, Sipe2000}, and gyration current~\cite{Watanabe2021,Ahn2020}, respectively. The formulas have been given by~\cite{Watanabe2022}
\begin{subequations}
\begin{align}
\label{eq:Einj_term}
\sigma^{\alpha;\beta\gamma}_{\mathrm{Einj}} &= -\frac{i\pi}{8\eta}\sum_{a\neq b}(J^{\alpha}_{aa}-J^{\alpha}_{bb})\Omega^{\lambda_{\beta}\lambda_{\gamma}}_{ba}F_{ab}, \\
\label{eq:Minj_term}
\sigma^{\alpha;\beta\gamma}_{\mathrm{Minj}} &= -\frac{\pi}{4\eta}\sum_{a\neq b}(J^{\alpha}_{aa}-J^{\alpha}_{bb})g^{\lambda_{\beta}\lambda_{\gamma}}_{ba}F_{ab}, \\
\label{eq:shift_term}
\sigma^{\alpha;\beta\gamma}_{\mathrm{shift}} &= -\frac{\pi}{4}\sum_{a\neq b}\Im\left[\left[D_{\lambda_{\alpha}}\xi^{\lambda_{\beta}}\right]_{ab}\xi^{\lambda_{\gamma}}_{ba} + \left[D_{\lambda_{\alpha}}\xi^{\lambda_{\gamma}}\right]_{ab}\xi^{\lambda_{\beta}}_{ba}\right]F_{ab}, \\
\sigma^{\alpha;\beta\gamma}_{\mathrm{gyro}} &= \frac{i\pi}{4}\sum_{a\neq b}\Re\left[\left[D_{\lambda_{\alpha}}\xi^{\lambda_{\beta}}\right]_{ab}\xi^{\lambda_{\gamma}}_{ba} - \left[D_{\lambda_{\alpha}}\xi^{\lambda_{\gamma}}\right]_{ab}\xi^{\lambda_{\beta}}_{ba}\right]F_{ab}. \label{eq:gyro_term}
\end{align}    
\end{subequations}
These formulas contain geometric quantities such as the Berry curvature ($\Omega^{\lambda_{\alpha}\lambda_{\beta}}_{ab}$), quantum metric ($g^{\lambda_{\alpha}\lambda_{\beta}}_{ab})$, and the covariant derivative $D_{\lambda_{\alpha}}$. These geometric quantities are based on the connection $\xi^{\lambda_{\alpha}}_{ab}$ which is different from the Berry connection in the normal state. In the superconducting state, we introduce the connection $\xi^{\lambda_{\alpha}}_{ab}$ by
\begin{align}
    \xi^{\lambda_{\alpha}}_{ab} = i\left<a_{\bm{\lambda}}\middle| \frac{\partial b_{\lambda}}{\partial\lambda_{\alpha}}\right>,
\end{align}
where $\bm{\lambda}$ is the variational parameter introduced by
\begin{align}
    H(\bm{k},\bm{q}, \bm{A}=\bm{0})\rightarrow H_{\bm{\lambda}}(\bm{k},\bm{q})=H(\bm{k},\bm{q}, \bm{\lambda}).
\end{align}
The normal photocurrent includes a resonant component characterized by $F_{ab} \equiv f_{ab}\delta(\Omega - E_{ba})$, which means the Pauli exclusion principle at the optical transition. 

The anomalous photocurrent consists of two contributions
\begin{align}
    \sigma_{\mathrm{a}} = \sigma_{\mathrm{NRSF}} + \sigma_{\mathrm{CD}},
\end{align}
which are termed the nonreciprocal superfluid density term and the conductivity derivative term, respectively. The formulas are obtained as~\cite{Watanabe2022}
\begin{align}
\label{NRSF}
\sigma^{\alpha;\beta\gamma}_{\mathrm{NRSF}} &= \lim_{\bm{\lambda}\rightarrow0}-\frac{1}{2\Omega^2}\partial_{\lambda_{\alpha}}\partial_{\lambda_{\beta}}\partial_{\lambda_{\gamma}}F_{\bm{\lambda}}, \\
\sigma^{\alpha;\beta\gamma}_{\mathrm{CD}} &= \lim_{\bm{\lambda}\rightarrow0}\frac{1}{4\Omega^{2}}\partial_{\lambda_{\alpha}}\left[\sum_{a\neq b}J^{\beta}_{ab}J^{\gamma}_{ba}f_{ab}\left(\frac{1}{\Omega-E_{ab}}+\frac{1}{E_{ab}}\right)\right], \label{eq:CD_term}
\end{align}
where $F_{\bm{\lambda}}$ is the free energy of the BdG Hamiltonian $H_{\bm{\lambda}}(\bm{k},\bm{q})$. A unique property of the anomalous photocurrent is the low-frequency divergence, $\sigma_{\mathrm{NRSF}} \propto \Omega^{-2}$ and $\sigma_{\mathrm{CD}} \propto \Omega^{-1}$. Therefore, the low-frequency photocurrent is dominated by the anomalous photocurrent, and it can realize a giant photogalvanic effect in superconductors.

\section{Joint density of states}
\label{sec:joint_density_of_state}

\begin{figure*}[htbp]
 \includegraphics[width=\linewidth]{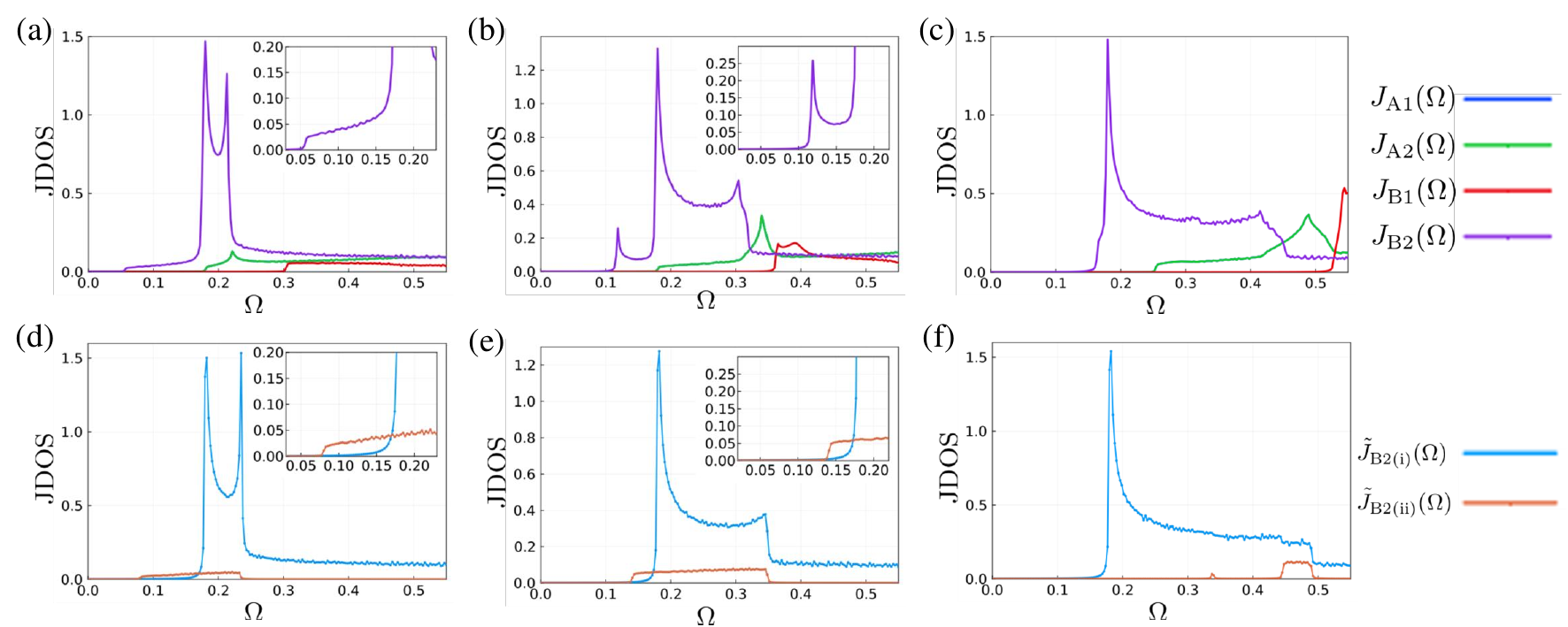}
 \caption{
 (Upper panels) The JDOS for the chemical potential (a) $\mu =-3.2$, (b) $-3.12$, and (c) $-3.0$. 
 The JDOS $J_{\mathrm{A1}}(\Omega)$, $J_{\mathrm{A2}}(\Omega)$, $J_{\mathrm{B1}}(\Omega)$, and $J_{\mathrm{B2}}(\Omega)$ correspond to the optical transitions $E_{3\bm{k}}\leftrightarrow E_{1\bm{k}}$, $E_{4\bm{k}}\leftrightarrow E_{2\bm{k}}$, $E_{4\bm{k}}\leftrightarrow E_{1\bm{k}}$, and $E_{3\bm{k}}\leftrightarrow E_{2\bm{k}}$, respectively. 
 (Lower panels) Blue and orange lines plot the approximate JDOS, $\tilde{J}_{\mathrm{B2(i)}}(\Omega)$ and $\tilde{J}_{\mathrm{B2(ii)}}(\Omega)$, respectively. The effective chemical potential in (d) $\tilde{\mu}=0$, (e) $0.08$, and (f) $0.2$ correspond to the panels (a), (b), and (c), respectively.
 We adopt $t =0.6$ consistent with the tight-binding parameters $t_{1}=1$ and $t_{2}=0.2$.
}
 \label{fig:JDOS2}
\end{figure*}
\begin{figure*}
\includegraphics[width=0.9\linewidth]{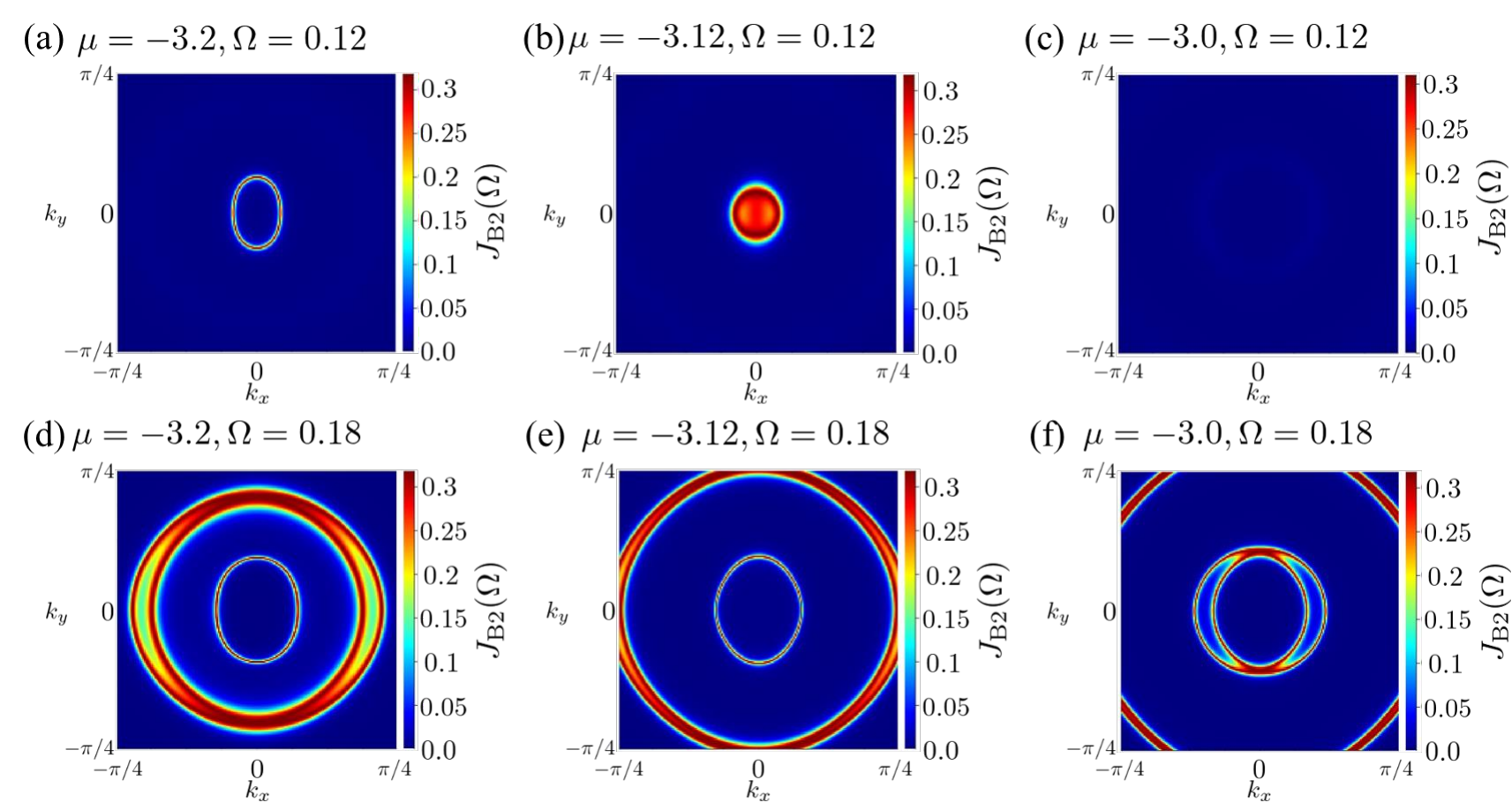}
\caption{Distribution of the JDOS $J_{\mathrm{B2}}(\Omega)$ in the $\bm{k}$-space $[-\pi/4,\pi/4]\times [-\pi/4,\pi/4]$.
Upper panels (a,b,c) for the case of $\Omega = 0.12 <2\psi$ and lower panels (d,e,f) for $\Omega = 0.18 = 2\psi$.
We set the chemical potentials (a)(d) $\mu =-3.2$, (b)(e) $\mu =-3.12$, and (c)(f) $\mu =-3.0$.
We fix $h=0.05$, $\theta =\ang{0}$, and $\psi =0.09$.
}
\label{fig:JDOS_k_profile}
\end{figure*}

Before showing the nonlinear optical responses in detail, we discuss the JDOS in the model and analytically calculate it. The JDOS is an essential quantity for the nonlinear optical responses, because the formulas for the normal photocurrent conductivity, Eqs.~\eqref{eq:Einj_term}-\eqref{eq:gyro_term}, contain the factor $\delta(\Omega-E_{ba})$, which corresponds to the optical transition between the $a$ and $b$ bands. Below we show the characteristic frequency dependence of the JDOS and effects of the finite total momentum of Cooper pairs $2\bm{q}$. 
In addition, we identify the essential optical transition and approximately decompose the corresponding JDOS to $\tilde{J}_{\mathrm{B2(i)}}$ and $\tilde{J}_{\mathrm{B2(ii)}}$. 
These approximated JDOS show characteristic dependence on the magnetic field and chemical potential.
In the later sections, the photocurrent conductivity is discussed based on the analysis of the JDOS.

In the following part, we mainly discuss the electron systems with the Fermi level near the Dirac point at the time-reversal invariant momentum [see Fig.~\ref{fig:schematic_figure}(c)]. This is because of the following reasons. First, it has been shown that several nonreciprocal and nonlinear responses show peculiar chemical potential dependence around the Dirac point~\cite{Tokura2018, Ideue2017, Hoshino2018, Okada2016, Morimoto2016, Hosur2011}. For example, the magnetochiral anisotropy in semiconductors is significantly enhanced around the Dirac point~\cite{Tokura2018,Ideue2017}. The band geometry around the Dirac point also gives a giant circular photogalvanic effect~\cite{Hosur2011}. We will see that the superconducting nonlinear optical responses are sensitive to the 
presence of the Dirac point. 
Second, the $s$-wave superconductors with a Fermi level near the Dirac point are candidates for topological superconductors~\cite{Alicea2012,Sato2016}. We show that the nonlinear optical responses show peculiar behaviors at the topological transition. Therefore, the following studies will be helpful in the search and identification of topological superconductors.

\subsection{Numerical results of JDOS}\label{sec:numerical_JDOS}
We define the JDOS, $J_{\mathrm{A1}}$, $J_{\mathrm{A2}}$, $J_{\mathrm{B1}}$, and $J_{\mathrm{B2}}$, as
\begin{align}
    J_{\mathrm{A1}}(\Omega)=\sum_{\bm{k}}\delta(\Omega - E_{3\bm{k}} + E_{1\bm{k}}), \\
    J_{\mathrm{A2}}(\Omega)=\sum_{\bm{k}}\delta(\Omega - E_{4\bm{k}} + E_{2\bm{k}}), \\
    J_{\mathrm{B1}}(\Omega)=\sum_{\bm{k}}\delta(\Omega - E_{4\bm{k}} + E_{1\bm{k}}), \\
    J_{\mathrm{B2}}(\Omega)=\sum_{\bm{k}}\delta(\Omega - E_{3\bm{k}} + E_{2\bm{k}}),
\end{align}
where 
$E_{i\bm{k}}$ are eigenenergies of Bogoliubov quasiparticles $(E_{1\bm{k}}\leq E_{2\bm{k}} \leq E_{3\bm{k}} \leq E_{4\bm{k}})$ in the superconducting state. Because $E_{1\bm{k}}$ and  $E_{2\bm{k}}$ are hole bands and  $E_{3\bm{k}}$ and $E_{4\bm{k}}$ are electron bands, the above four JDOS are relevant for the optical responses.
In particular, the JDOS with subscript $\mathrm{B}$, namely, $J_{\mathrm{B1}}$ and $J_{\mathrm{B2}}$, correspond to the type B transition specified in Fig.~\ref{fig:schematic_figure}(a), which was shown to be essential for the nonlinear optical responses unique to superconductors~\cite{Tanaka2023}. 
Figures~\ref{fig:JDOS2}(a), \ref{fig:JDOS2}(b), and \ref{fig:JDOS2}(c) show the JDOS for $\mu =-3.2$, $-3.12$, and $-3.0$, respectively.
We see that the JDOS $J_{\mathrm{B2}}(\Omega)$ is dominant at low energies. Thus, it is expected that the type B optical transition between the $E_{2\bm{k}}$ and $E_{3\bm{k}}$ bands dominates the superconducting nonlinear responses to the irradiation of low-frequency light. 
%

\begin{figure*}
\includegraphics[width=\linewidth]{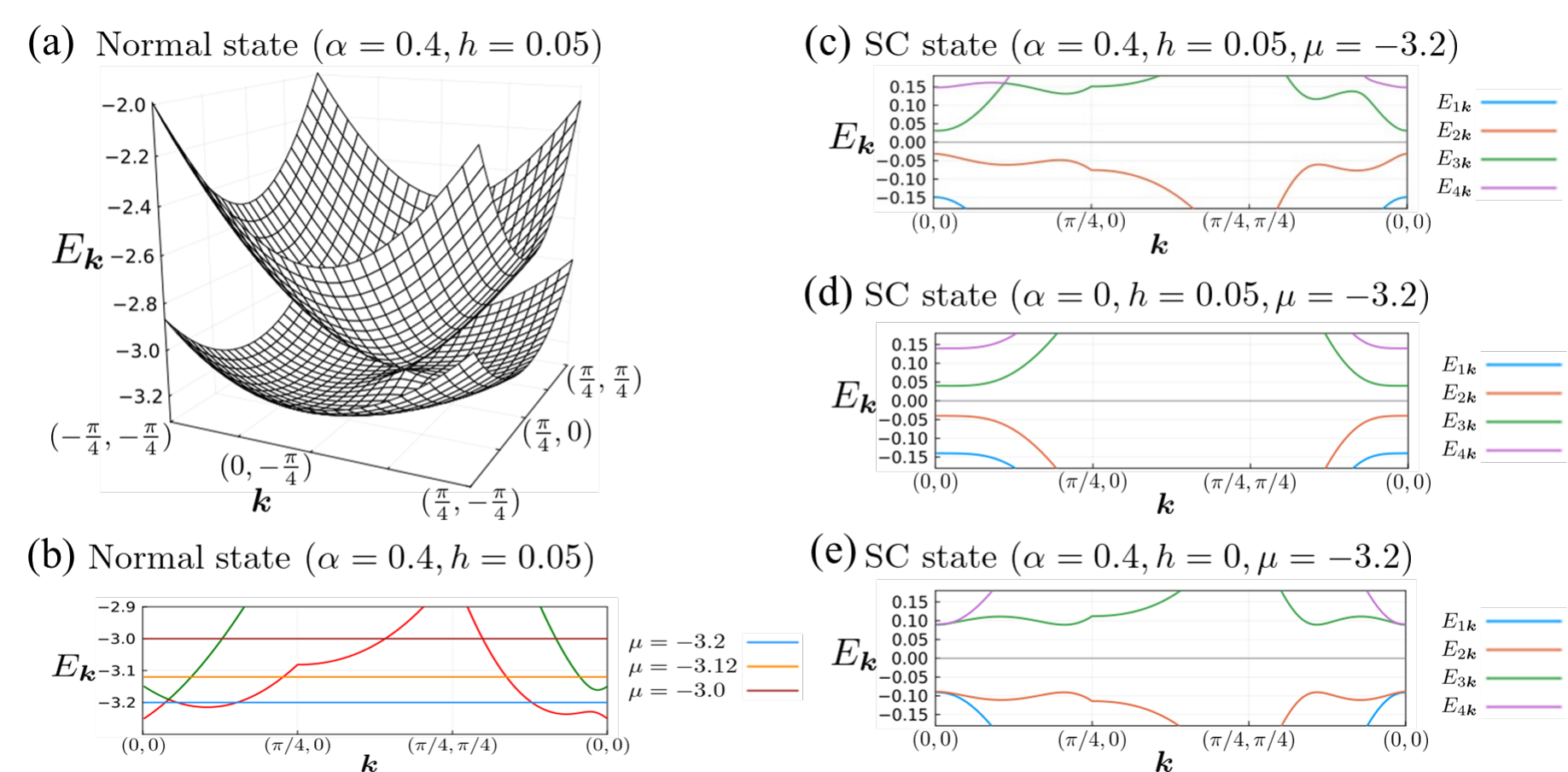}
\caption{The energy bands of quasiparticles around the $\Gamma$ point [$\bm{k}=(0,0)$]. 
(a)(b) The energy dispersion of electrons in the normal state. We set $\alpha =0.4$, $h=0.05$, and $\mu =0$. The horizontal lines in Fig.~\ref{fig:band_figure}(b) indicate the Fermi levels for $\mu=-3.2$, $-3.12$, and $-3.0$. (c) The energy dispersion of Bogoliubov quasiparticles in the superconducting state for $\psi = 0.09$, $\alpha =0.4$, $h =0.05$, and $\mu =-3.2$. (d) Same as (c), but without the spin-orbit coupling ($\alpha =0$). (e) Same as (c), but without the magnetic field ($h =0$). The angle of the magnetic field is fixed as $\theta =\ang{0}$.}  
\label{fig:band_figure}
\end{figure*}

In our model, the Dirac point exists at $\bm{k} \simeq \bm{0}$ [see Fig.~\ref{fig:band_figure}(a)]. When the magnetic field is absent, the chemical potential $\mu =-3.2$ lies on the Dirac point [blue lines in Fig.~\ref{fig:schematic_figure}(c) and Fig.~\ref{fig:band_figure}(b)], while $\mu =-3.12$ and $-3.0$ lie above the Dirac point [orange and brown lines in Fig.~\ref{fig:schematic_figure}(c), respectively]. The chemical potential dependence of the JDOS $J_{\mathrm{B2}}(\Omega)$ was illustrated in Fig.~\ref{fig:schematic_figure}(b) based on the results in Figs.~\ref{fig:JDOS2}(a)-\ref{fig:JDOS2}(c).
Specifically, the JDOS 
has characteristic structures. First, a sharp peak exists around $\Omega \simeq 2\psi =0.18$ independent of the chemical potential. Second, Fig.~\ref{fig:JDOS2}(a) ($\mu =-3.2$) and Fig.~\ref{fig:JDOS2}(b) ($\mu =-3.12$) show finite $J_{\mathrm{B2}}(\Omega)$ below the frequency $\Omega=2\psi$, while Fig.~\ref{fig:JDOS2}(c) does not show it. Especially, the JDOS in Fig.~\ref{fig:JDOS2}(b) has a sharp peak at a low frequency $\Omega \simeq 0.12$.
The origin of the similarities and differences between JDOS spectra for different chemical potentials is explained below.

Here, we focus on the JDOS at low frequencies, $\Omega < 2\psi=0.18$, namely, in the in-gap region for $h=0$. For the analysis, we show Fig.~\ref{fig:JDOS_k_profile} for the momentum-resolved contribution to the JDOS. 
Figures~\ref{fig:JDOS_k_profile}(a) and \ref{fig:JDOS_k_profile}(b) show that the low-frequency JDOS originates from the momentum space around the Dirac point $\bm{k} \simeq \bm{0}$. The spin-orbit coupling is negligible in this regime due to its momentum dependence, and therefore the band dispersion in the superconducting state can be approximated by taking $\alpha=0$
[Compare Fig.~\ref{fig:band_figure}(c) for $\alpha = 0.4$ with Fig.~\ref{fig:band_figure}(d) for $\alpha = 0$]. 
Then, the energy difference $\Delta E_{32} \equiv E_{3}-E_{2}$ is approximated as $\Delta E_{32} \simeq 2\left|\sqrt{\xi^{2} + \psi^{2}}- h\right|$. This approximation is exact at $\bm{k}=\bm{0}$ when the total momentum of Cooper pairs $\bm{q}$ is zero. The energy gap $\Delta E_{32(\bm{k}=\bm{0})}$ is obtained as $\Delta E_{32(\bm{k}=\bm{0})}=0.08$ and $\Delta E_{32(\bm{k}=\bm{0})} = 0.141$ when we adopt the parameters in Figs.~\ref{fig:JDOS2}(a) and \ref{fig:JDOS2}(b) but set $\bm{q}=\bm{0}$. 
In reality, the gap edge in the JDOS appears at lower energies because of the finite momentum of Cooper pairs.

When the chemical potential is far away from the Dirac point, the JDOS does not show the low-frequency component. 
This is because the energy gap is not determined at $\bm{k}=\bm{0}$.
For $\mu=-3.0$ in the case of Fig.~\ref{fig:JDOS2}(c), the approximation in the previous paragraph gives an estimate of $\Delta E_{32(\bm{k}=\bm{0})} = 0.339$. However, we do not see any characteristic behavior around $\Omega \simeq 0.339$ because this energy is far above the superconducting gap. 

\begin{table}[htbp]
\centering
\renewcommand{\arraystretch}{1.5}
\caption{The values of $q_{x}$ and $(\Delta \bm{h}_{\mathrm{eff}})_{y}$ for the parameters in Figs.~\ref{fig:JDOS2}(a) and \ref{fig:JDOS2}(b). The total momentum of Cooper pairs $2\bm{q}$ is determined so as to minimize the free energy (see Appendix \ref{app:finite_momentum}).}
\begin{tabular}{c|c|c|c}
Figure of JDOS & $\mu$ & $q_{x}$ & $(\Delta \bm{h}_{\mathrm{eff}})_{y}$ \\
\hline
\hline
Figure~\ref{fig:JDOS2}(a) & -3.2 & $-3.0\times 10^{-2}$ & $1.2\times 10^{-2}$ \\
\hline
Figure~\ref{fig:JDOS2}(b) & -3.12 & $-2.8\times 10^{-2}$ & $1.1\times 10^{-2}$ \\
\end{tabular}
\label{tab:q_x_and_h_eff}
\end{table}

Although we have neglected the momentum of Cooper pairs $\bm{q}$ in the above discussion, it plays a quantitatively essential role. 
Figures~\ref{fig:JDOS2}(a) and \ref{fig:JDOS2}(b) show that the JDOS becomes finite above a frequency lower than the above estimate. 
For example, the finite $J_{\mathrm{B2}}$ appears above $\Omega \simeq 0.05$ in Fig.~\ref{fig:JDOS2}(a) is spite of the previous estimate $\Delta E_{32(\bm{k}=\bm{0})}=0.08$.
By taking into account a small but finite total momentum of Cooper pairs $2\bm{q} = (2q_{x}, 0, 0)$, 
the normal Hamiltonian can be approximated around $\bm{k}\simeq \bm{0}$ as
\begin{align}
    \label{eq:effective_magnetic_field}
    H_{\mathrm{N}}(\bm{k}+\bm{q}) &\simeq H_{\mathrm{N}}(\bm{k}) +\Delta\bm{h}_{\mathrm{eff}}\cdot\bm{\sigma},\\
    \Delta\bm{h}_{\mathrm{eff}} &\equiv (0,-\alpha q_{x},0)^{\top}.
\end{align}
Thus, it is estimated that the JDOS is finite above the frequency $\Omega = 2\left|\sqrt{{\xi_{\bm{k}=\bm{0}}}^{2} + \psi^{2}}- h_{\mathrm{eff}}\right|$, where we define $h_{\mathrm{eff}}$ by  $\bm{h}_{\mathrm{eff}} \equiv \bm{h}+\Delta\bm{h}_{\mathrm{eff}}$. In the cases of Figs.~\ref{fig:JDOS2}(a) and \ref{fig:JDOS2}(b), 
$q_x$ and $h_{\mathrm{eff}}$ are obtained as summarized in Table \ref{tab:q_x_and_h_eff}, 
and the gap edge is estimated as $2\left|\sqrt{{\xi_{\bm{k}=\bm{0}}}^{2} + \psi^{2}}- h_{\mathrm{eff}}\right|=0.056$ and $0.119$, respectively. These values are more precisely consistent with the numerical results than the previous estimate where $h$ is adopted instead of $h_{\mathrm{eff}}$. 

While the JDOS in the low-frequency region $\Omega < 2\psi=0.18$ originates from the momentum space around the Dirac point $|\bm{k}| \simeq 0$, the sharp peak at $\Omega =2\psi$ is dominantly contributed from the region $|\bm{k}| \simeq \pi/4$, as shown in  Figs.~\ref{fig:JDOS_k_profile}(d), \ref{fig:JDOS_k_profile}(e), and \ref{fig:JDOS_k_profile}(f). For the momentum away from the Dirac point, the energy dispersion of Bogoliubov quasiparticles is not significantly affected by the Zeeman magnetic field and is well approximated by setting $h=0$. [Compare Fig.~\ref{fig:band_figure}(c) with  Fig.~\ref{fig:band_figure}(e) where $h=0$.] A notable effect of the magnetic field is the asymmetry in the band dispersion. 
Although the energy bands satisfy the relation $E_{2\bm{k}}=-E_{3\bm{k}}$ if either the spin-orbit coupling or Zeeman field is absent [Figs.~\ref{fig:band_figure}(d) and \ref{fig:band_figure}(e)], the relation breaks down in Fig.~\ref{fig:band_figure}(c) because the system is neither $\mathcal{T}$- nor $\mathcal{P}$-symmetric because of the finite Rashba spin-orbit coupling and the Zeeman magnetic field. 
The asymmetry in energy is roughly estimated by the energy nonreciprocity in the normal state, $\epsilon_{a\bm{k}}-\epsilon_{a -\bm{k}}\propto \bm{g_{k}}\cdot \bm{h}$, where $\epsilon_{a\bm{k}}$ is the energy band of normal Hamiltonian [see Appendix~\ref{app:energy_appro_asym} for detail].
However, the effect of the energy-spectrum asymmetry is ignorable in the energy difference $\Delta E_{32}$ because the energy shift cancels out between $E_{2}$ and $E_{3}$. 
Therefore, the effect of the magnetic field on the JDOS is negligible when the JDOS arises from the momentum away from the Dirac point. This is the reason why the sharp peak in the JDOS at $\Omega \simeq 2 \psi$ is robust.

\subsection{Approximate JDOS}\label{sec:approximate_JDOS}

To analyze the system with a Fermi level near the Dirac point it is essential to understand the JDOS illustrated in Fig.~\ref{fig:schematic_figure}(b) and obtained in Fig.~\ref{fig:JDOS2}. For this purpose, we introduce the approximate JDOS, 
$\tilde{J}_{\mathrm{B2(i)}}(\Omega)$ and $\tilde{J}_{\mathrm{B2(ii)}}(\Omega)$, into which $J_{\mathrm{B2}}(\Omega)$ is approximately decomposed.
In this subsection, we approximate the normal state Hamiltonian with a focus on the $\Gamma$ point (${\bm k}=\bm{0}$) of the Brillouin zone as 
\begin{align}
    \xi_{\bm{k}} &= t\left({k_{x}}^{2} + {k_{y}}^{2}\right) - \tilde{\mu}, \label{eq:simple_xi_2}\\
    \bm{g}_{\bm{k}} &= \alpha (k_{y}, - k_{x}, 0),\label{eq:simple_g_2}\\
    \tilde{\mu} &\equiv \mu + 4(t_{1} + t_{2}), 
\end{align}
where $\tilde{\mu}$ is the effective chemical potential indicating the difference of the Fermi level from the Dirac point.
For our choice of parameters, $\mu=-3.2$ corresponds to $\tilde{\mu}=0$ while $\mu=-3.12$ and $\mu=-3.0$ to $\tilde{\mu}=0.08$ and $\tilde{\mu}=0.2$, respectively.
For simplicity, we ignore the total momentum of Cooper pairs and set $2\bm{q}=0$ in this subsection. 
The detailed calculation is presented in Appendix~\ref{app:appro_JDOS}.

First, we derive the approximate JDOS $\tilde{J}_{\mathrm{B2(i)}}(\Omega)$, which gives a sharp peak of $J_{\mathrm{B2}}(\Omega)$ around $\Omega = 2\psi$. In the derivation, we ignore $\bm{g_{k}}\cdot \bm{h}$ because its effect is almost canceled between $E_{2}$ and $E_{3}$ in the energy difference $\Delta E_{32}$. 
Then, we obtain the energies of Bogoliubov quasiparticles as 
\begin{align}
E_{\bm{k}} = \pm\sqrt{{\xi_{\bm{k}}}^{2} + {g_{\bm{k}}}^{2} + h^{2} + \psi^{2} \pm^{\prime} 2\sqrt{{\xi_{\bm{k}}}^{2}({g_{\bm{k}}}^{2} + h^{2}) + \psi^{2}h^{2}}}.
\label{eq:approximate_energy}
\end{align}
The approximate energy satisfies the symmetry $E_{\bm{k}} = E_{-\bm{k}}$ because we ignore $\bm{g_{k}}\cdot \bm{h}$. 
When $|\xi_{\bm{k}}||\bm{g_{k}}|\gg\psi h$, the energies of Bogoliubov quasiparticles are furthermore approximated by
%
\begin{align}
E_{\bm{k}} = & \pm\sqrt{\left(|\xi_{\bm{k}}|\pm^{\prime}\sqrt{{g_{\bm{k}}}^{2} + h^{2}}\right)^{2}+\psi^{2}}.
\end{align}
To estimate the JDOS, we consider an electron band of $E_{3\bm{k}}$,
\begin{align}
E_{3\bm{k}} = \sqrt{\left(|\xi_{\bm{k}}|-\sqrt{{g_{\bm{k}}}^{2} + h^{2}}\right)^{2}+\psi^{2}}.
\label{eq:appro_energy}
\end{align}
Since $E_{2\bm{k}}$ and $E_{3\bm{k}}$ are related to each other under the charge-conjugation transformation $E_{2\bm{k}} = -E_{3\bm{-k}}$, the energy difference of the transition between the electron and hole bands is given by
\begin{align}
\label{eq:appro_delta_E}
\Delta E_{\bm{k}}= E_{3\bm{k}} - E_{2\bm{k}} = 2\sqrt{\left(|\xi_{\bm{k}}|-\sqrt{{g_{\bm{k}}}^{2} + h^{2}}\right)^{2}+\psi^{2}},
\end{align}
which is equivalent to the energy difference between $E_{3\bm{k}}$ and $-E_{3\bm{-k}}$.
The approximate JDOS $\tilde{J}_{\mathrm{B2(i)}}(\Omega)$ is introduced corresponding to Eq.~\eqref{eq:appro_delta_E},
\begin{align}
    \tilde{J}_{\mathrm{B2(i)}}(\Omega) = \sum^
{\prime}_{\bm{k}} \left(\Omega - \Delta E_{\bm{k}} \right),
    \label{eq:definition_JBi}
\end{align}
where integration is carried out on the region $|\xi_{\bm{k}}||\bm{g_{k}}|>\psi h$.
When we assume $g_{\bm{k}} \gg h$, which is satisfied for large $k$, $\tilde{J}_{\mathrm{B2(i)}}(\Omega)$ shows the divergence at $\Omega =2\psi$.
In fact, $\tilde{J}_{\mathrm{B2(i)}}(\Omega)$ reproduces the sharp peak of the JDOS around $\Omega= 2\psi$, as shown in Figs.~\ref{fig:JDOS2}(d), \ref{fig:JDOS2}(e), and \ref{fig:JDOS2}(f).

Next, we introduce the approximate JDOS $\tilde{J}_{\mathrm{B2(ii)}}(\Omega)$ based on another assumption $\psi h\gg|\xi_{\bm{k}}||\bm{g_{k}}|$. In this case, we obtain the electron band of $E_{3\bm{k}}$,
\begin{align}
E_{3\bm{k}} =  \sqrt{{\xi_{\bm{k}}}^{2} + {g_{\bm{k}}}^{2} +h^{2} + \psi^{2} -2h\sqrt{{\xi_{\bm{k}}}^{2} + \psi^{2}} }.
\end{align}
Then, the energy difference for the optical transition between the electron and hole bands is given by
\begin{align}
\Delta E_{\bm{k}} = 2\sqrt{{\xi_{\bm{k}}}^{2} + {g_{\bm{k}}}^{2} +h^{2} + \psi^{2} -2h\sqrt{{\xi_{\bm{k}}}^{2} + \psi^{2}} }.
\label{eq:appro_delta_E_2}
\end{align}
Thus, the approximate JDOS $\tilde{J}_{\mathrm{B2(ii)}}(\Omega)$ is introduced as
\begin{align}
    \tilde{J}_{\mathrm{B2(ii)}}(\Omega) = \sum^{\prime\prime}_{\bm{k}} \delta \left(\Omega -2\sqrt{{\xi_{\bm{k}}}^{2} + {g_{\bm{k}}}^{2} +h^{2} + \psi^{2} -2h\sqrt{{\xi_{\bm{k}}}^{2} + \psi^{2}} }\right).
    \label{eq:definition_JBii}
\end{align}
where integration is carried out on the region $\psi h>|\xi_{\bm{k}}| |{\bm g}_{\bm{k}}|$. 
Integrating the momentum around $k=0$,
$\tilde{J}_{\mathrm{B2(ii)}}(\Omega)$ is approximated by
\begin{align}
    \tilde{J}_{\mathrm{B2(ii)}}(\Omega) =
    \begin{cases}
        0 & \left(0\leq \Omega <2\left|\sqrt{{\tilde{\mu}}^{2} + \psi^{2}}-h\right| \right) \\
    \frac{\Omega}{8\pi \tilde{\alpha}^{2}} & \left(2\left|\sqrt{{\tilde{\mu}}^{2} + \psi^{2}}-h\right| \leq \Omega\right) 
    \end{cases},
    \label{eq:appro_B_ii}
\end{align}
with
\begin{align}
    \tilde{\alpha} \equiv \sqrt{\alpha^{2} -2t\tilde{\mu} + \frac{2t\tilde{\mu}h}{\sqrt{{\tilde{\mu}}^{2} + \psi^{2}}}}.
\end{align}


We see that the JDOS $\tilde{J}_{\mathrm{B2(ii)}}(\Omega)$ is finite above $\Omega = 2\left|\sqrt{{\tilde{\mu}}^{2} + \psi^{2}}-h\right|$. 
We expect that the contribution of $\tilde{J}_{\mathrm{B2(ii)}}(\Omega)$ is important in the low-frequency region because it can be finite for the frequency $\Omega$ smaller than $2\psi$ 
contrary to $\tilde{J}_{\mathrm{B2(i)}}(\Omega)$ when $|\tilde{\mu}|$ is sufficiently small. Corresponding to this fact, we will see the nonlinear optical responses in the region $2|\psi -h| \leq \Omega \leq 2|\psi|$ when the Fermi level lies on the Dirac point. 
As shown in Fig.~\ref{fig:JDOS2}(d),  $\tilde{J}_{\mathrm{B2(ii)}}(\Omega)$ for $\tilde{\mu} = 0$ shows a plateau starting from $\Omega=2|\psi -h|$. A similar structure is seen in the numerical results of the JDOS [Fig.~\ref{fig:JDOS2}(a)]. 
Indeed, the JDOS $J_{\mathrm{B2}}(\Omega)$ can be approximated by summation of $\tilde{J}_{\mathrm{B2(i)}}(\Omega)$ and $\tilde{J}_{\mathrm{B2(ii)}}(\Omega)$, as $J_{\mathrm{B2}}(\Omega) \simeq \tilde{J}_{\mathrm{B2(i)}}(\Omega)+\tilde{J}_{\mathrm{B2(ii)}}(\Omega)$.
This feature was illustrated in Fig.~\ref{fig:schematic_figure}(b). 

\subsection{Comparison of JDOS and approximate JDOS}

Here, we compare the JDOS calculated in Sec.~\ref{sec:numerical_JDOS} with the approximate JDOS introduced in Sec.~\ref{sec:approximate_JDOS}. First, let us compare Fig.~\ref{fig:JDOS2}(a) 
with Fig.~\ref{fig:JDOS2}(d), 
where the chemical potential lies on the Dirac point.
The plateau of $J_{\mathrm{B2}}(\Omega)$ at $\Omega <2\psi$ is attributed to $\tilde{J}_{\mathrm{B2(ii)}}(\Omega)$, and the sharp peak of $J_{\mathrm{B2}}(\Omega)$ at $\Omega =2\psi$ is attributed to $\tilde{J}_{\mathrm{B2(i)}}(\Omega)$. Thus, we see a qualitative agreement between the numerically calculated JDOS and the approximate JDOS. Finite $J_{\mathrm{B2}}(\Omega)$ is observed above $\Omega\simeq 0.05$, while $\tilde{J}_{\mathrm{B2(ii)}}(\Omega)$ becomes finite at $\Omega = 2|\psi -h|=0.08$. As previously discussed, this difference is due to the effect of the finite total momentum of Cooper pairs, which was neglected in Sec.~\ref{sec:approximate_JDOS}. 

We discuss the ${\bm k}$-space where $\tilde{J}_{\mathrm{B2(i)}}(\Omega)$ and $\tilde{J}_{\mathrm{B2(ii)}}(\Omega)$ originate from. Figure~\ref{fig:JDOS_k_profile}(a) shows that the low-frequency JDOS $J_{\mathrm{B2}}(\Omega=0.12<2\psi)$ arises from the region around the Dirac point ($\bm{k} =0$). This is consistent with the 
assumption for
$\tilde{J}_{\mathrm{B2(ii)}}(\Omega)$. On the other hand, the JDOS $J_{\mathrm{B2}}(\Omega=0.18=2\psi)$ is dominantly contributed from the region $k\simeq \alpha/t \simeq 0.21\pi$ [Fig.~\ref{fig:JDOS_k_profile}(d)], which gives $\tilde{J}_{\mathrm{B2(i)}}(\Omega)$ [Fig.~\ref{fig:JDOS2}(d)]. It indicates that the contribution corresponds to the $\tilde{J}_{\mathrm{B2(i)}}$. 
Note that a small contribution to 
$\tilde{J}_{\mathrm{B2(ii)}}$ also comes from the ${\bm k}$-space $k \simeq 6.42 \times 10^{-2} \pi$. The detailed discussion is presented in Appendix \ref{app:appro_JDOS}. 

\begin{figure}[htbp]
 \includegraphics[width=0.9\linewidth]{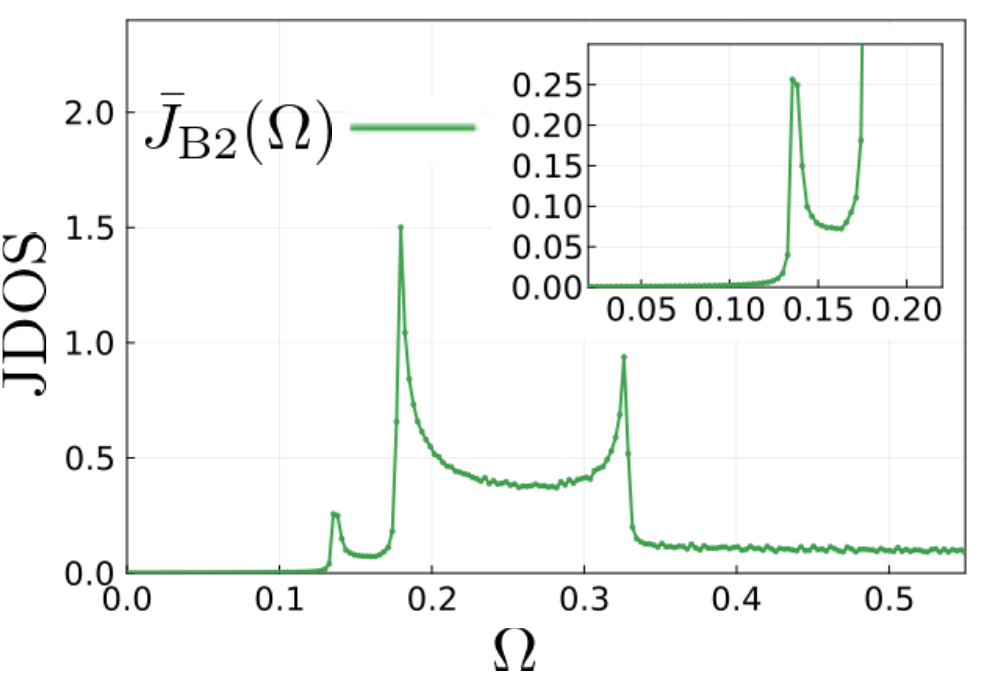}
 \caption{The approximate JDOS $\bar{J}_{\mathrm{B2}}(\Omega)$ defined by Eq.~\eqref{eq:approximate_energy}. We set the effective chemical potential $\tilde{\mu}=0.08$. Other parameters are also the same as Fig.~\ref{fig:JDOS2}(e).}
 \label{fig:bar_JDOS}
\end{figure}

Next, Fig.~\ref{fig:JDOS2}(b) 
is compared with Fig.~\ref{fig:JDOS2}(e), 
where the Fermi level slightly deviates from the Dirac point. We again see qualitative agreement between the JDOS and the approximate JDOS. The plateau of the JDOS $J_{\mathrm{B2}}(\Omega)$ at $0.12 < \Omega < 0.18$ is attributed to $\tilde{J}_{\mathrm{B2(ii)}}(\Omega)$, and the sharp peak of $J_{\mathrm{B2}}(\Omega)$ at $\Omega =0.18$ is attributed to $\tilde{J}_{\mathrm{B2(i)}}(\Omega)$. However, $\tilde{J}_{\mathrm{B2(i)}}(\Omega)$ and $\tilde{J}_{\mathrm{B2(ii)}}(\Omega)$ do not reproduce the sharp peak of $J_{\mathrm{B2}}(\Omega)$ at $\Omega \simeq 0.12$. This discrepancy can be resolved when we approximate the JDOS by the energy band in Eq.~\eqref{eq:approximate_energy} as 
\begin{align}
&\bar{J}_{\mathrm{B2}}(\Omega) = \sum_{\bm{k}} \left(\Omega - \Delta E_{\bm{k}} \right), 
\end{align}
with
\begin{align}
&\Delta E_{\bm{k}} = 2\sqrt{{\xi_{\bm{k}}}^{2} + {g_{\bm{k}}}^{2} + h^{2} + \psi^{2} - 2\sqrt{{\xi_{\bm{k}}}^{2}({g_{\bm{k}}}^{2} + h^{2}) + \psi^{2}h^{2}}}.
\end{align}
As shown in Fig.~\ref{fig:bar_JDOS}, $\bar{J}_{\mathrm{B2}}(\Omega)$ reproduces the sharp peak of $J_{\mathrm{B2}}(\Omega)$ at $\Omega \simeq 0.12$. This result implies that the peak originates from the ${\bm k}$-space where $|\xi_{\bm{k}}||\bm{g_{k}}|\simeq \psi h$ is satisfied. 
Because this condition is satisfied near the Dirac point, the position of this peak is roughly estimated as $\Omega \simeq 2|\sqrt{{\tilde{\mu}}^{2} + \psi^{2}}-h|$. 

Contrary to the above cases, we do not see the contribution of $\tilde{J}_{\mathrm{B2(ii)}}(\Omega)$ in the JDOS for $\mu =-3.0$ [Fig.~\ref{fig:JDOS2}(c)] corresponding to an effective chemical potential $\tilde{\mu}=0.2$ deviating from the Dirac point. The sharp peak of the JDOS still appears around $\Omega =2\psi$ and is attributed to $\tilde{J}_{\mathrm{B2(i)}}(\Omega)$. On the other hand, $\tilde{J}_{\mathrm{B2(ii)}}(\Omega)$ vanishes in the low-frequency region $\Omega \leq 2\psi$ because a finite contribution is obtained only above $\Omega=2|\sqrt{{\tilde{\mu}}^{2} + \psi^{2}}-h|\simeq 0.34>2\psi$.

Based on the above results, we expect characteristic behaviors in the nonlinear optical responses, such as the peculiar magnetic field dependence, when the Fermi level is close to the Dirac point. As indicated by Eq.~\eqref{eq:appro_B_ii}, finite photocurrent conductivity appears in the in-gap region $\Omega < 2\psi$ under the magnetic fields. On the other hand, a sharp peak of the JDOS around $\Omega =2\psi$ is not sensitive to the magnetic field except for the change of the pair potential $\psi$, and therefore, we expect that the nonlinear optical responses in this frequency range are robust under the magnetic field. These observations are verified in Sec.~\ref{sec:Unique behaviors due to Dirac point}.

\section{Nonlinear optical responses}
\label{sec:Numerical_results}

In this section, we numerically demonstrate the nonlinear optical conductivity in noncentrosymmetric superconductors. 
In the two-dimensional superconductors with the $\mathrm{C_{4v}}$ crystal structure, typical for Rashba superconductors on interfaces and surfaces, 
all the second-order optical responses are prohibited at the zero magnetic field~\cite{Watanabe2021}.
However, the second-order responses may occur in the magnetic field, which breaks not only the $\mathcal{T}$ symmetry but also the crystalline symmetry. 
Actually, we show the field-induced nonlinear optical responses in the following.

For a quantitative estimation, we set $t_{1}=1\mathrm{eV}$ and calculate the response coefficients in the SI unit. Thus, the results of the nonlinear conductivity are given in the unit $\mathrm{AV^{-2}}$. Numerical calculations are performed on the $N^{2}$-discretized Brillouin zone ($N=3000$).
For numerical convergence, we replace the parameter $\eta$ with a phenomenological scattering rate $\gamma=2.0\times 10^{-4}$ and introduce a finite temperature $T=10^{-4}$ for the Fermi-Dirac distribution function.

\begin{figure*}[htbp]
 \includegraphics[width=\linewidth]{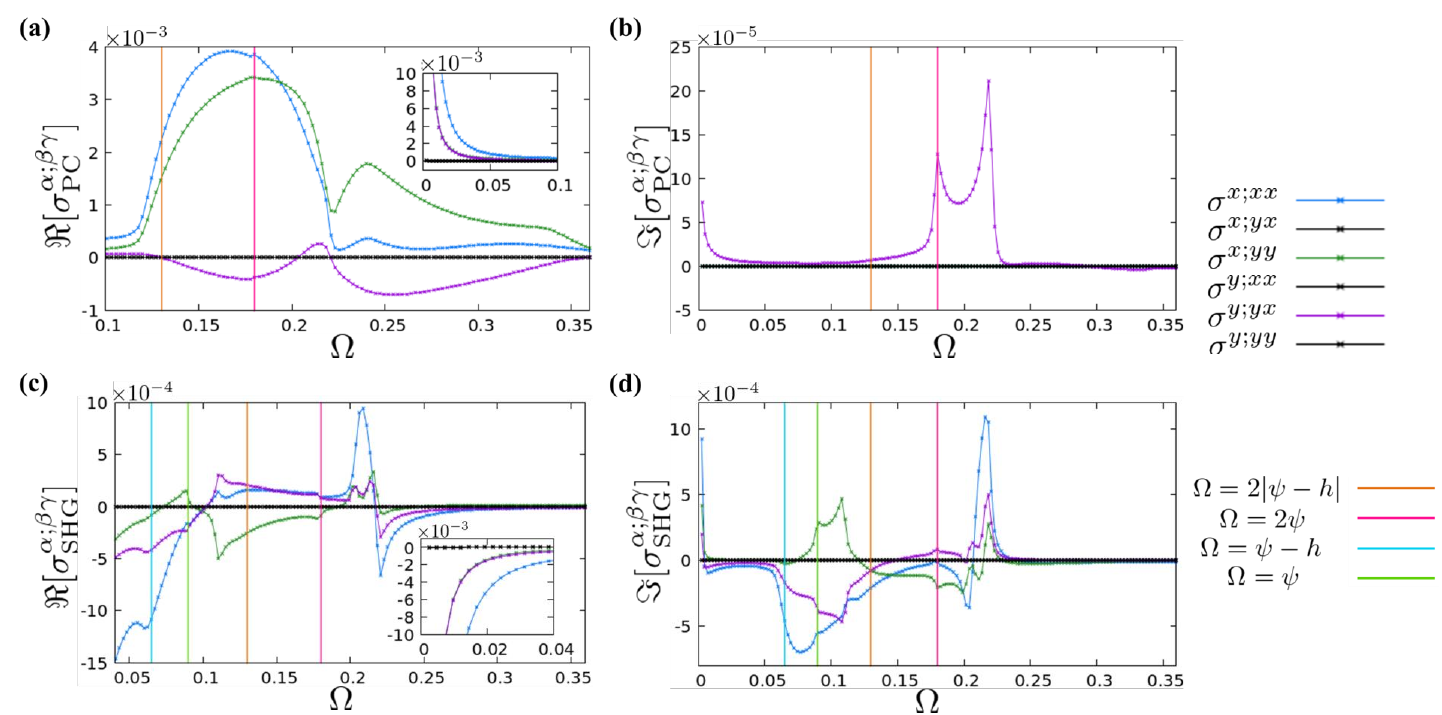}
 \caption{Frequency dependence of the photocurrent generation and the second harmonic generation under the in-plane magnetic field ($\theta =\ang{0}$). (a) Real part $\Re[\sigma^{\alpha;\beta\gamma}_{\mathrm{PC}}]$ and (b) imaginary part $\Im[\sigma^{\alpha;\beta\gamma}_{\mathrm{PC}}]$ of the photocurrent generation. (c) Real part $\Re[\sigma^{\alpha;\beta\gamma}_{\mathrm{SHG}}]$ and (d) imaginary part $\Im[\sigma^{\alpha;\beta\gamma}_{\mathrm{SHG}}]$ of the second harmonic generation. We set the parameters $h=0.025$, $\psi=0.09$, and $\mu =-3.2$. Note that $\sigma^{x;yx}=\sigma^{y;xx}=\sigma^{y;yy}=0$ since these components are prohibited by the $\mathrm{m^{\prime}m2^{\prime}}$ magnetic point group symmetry. The vertical lines illustrate the frequencies $\Omega=2\psi, 2|\psi -h|$ for the photocurrent generation and $\Omega=\psi, 2\psi, \psi -h, 2|\psi -h|$ for the second harmonic generation. }
 \label{fig:general2.pdf}
\end{figure*}

\begin{table*}[htbp]
\centering
\renewcommand{\arraystretch}{1.5}
\caption{Constrains of second-order optical responses and the symmetry class with $\theta$ being the angle between the magnetic field and the two-dimensional plane [$\bm{h}=(0, h\cos \theta, h\sin \theta)$]. The subscript "$\mathcal{T}\mathrm{-even}$" means the contribution which is even under the $\mathcal{T}$ operation (see also Table~\ref{tab:mechanism_class}).}
\begin{tabular}{c|c|c}
Angle $\theta$ & Symmetry & Constraints  \\
\hline \hline
$\bm{h}\parallel \hat{\bm{y}}$ ($\theta =\ang{0}$) & $\mathrm{m^{\prime}}m2^{\prime}$ & $\sigma^{x;yx}=\sigma^{y;xx}=\sigma^{y;yy}=0$, \,\, $\sigma^{\alpha;\beta\gamma}_{\mathcal{T}\mathrm{-even}}=0$ for all $\alpha, \beta, \gamma$\\
\hline
$\ang{0} < \theta < \ang{90}$ & $\mathrm{m^{\prime}}$ & $\sigma^{x;xx}_{\mathcal{T}\mathrm{-even}}=\sigma^{x;yy}_{\mathcal{T}\mathrm{-even}}=\sigma^{y;yx}_{\mathcal{T}\mathrm{-even}}=0$ \\
\hline
$\bm{h}\parallel \hat{\bm{z}}$ ($\theta =\ang{90}$) & $\mathrm{4m^{\prime}m^{\prime}}$ & $\sigma^{\alpha;\beta\gamma}=0$ for all $\alpha, \beta, \gamma$\\
\end{tabular}
\label{tab:constraints}
\end{table*}

\subsection{Photocurrent and second harmonic generation}

We show the photocurrent generation and second harmonic generation under the in-plane magnetic field ($\theta =0$) in Fig.~\ref{fig:general2.pdf}.
It should be noted that some components vanish due to the constraints discussed below. 
First, the constraint
\begin{align} \sigma_{\mathrm{PC}}^{\alpha;\beta\gamma}=\left(\sigma_{\mathrm{PC}}^{\alpha;\gamma\beta}\right)^{*}, \,\,\,\,\, \sigma_{\mathrm{SHG}}^{\alpha;\beta\gamma}=\sigma_{\mathrm{SHG}}^{\alpha;\gamma\beta},
\end{align}
has to be satisfied by definition. This constraint requires $\Im[\sigma^{\alpha;\beta\beta}_{\mathrm{PC}}]=0$.
Second, the symmetry of the system prohibits some components of the nonlinear optical responses. In this respect, the magnetic field plays a key role. 
In our model, the $\mathrm{C_{4v}}$ symmetry of the crystal structure lowers under the magnetic field, and thus the magnetic field is indispensable because the $\mathrm{C_{4v}}$ point group symmetry prohibits the second-order nonlinear responses in two-dimensional systems.  The relation between the symmetry constraints and the magnetic fields is summarized in Table~\ref{tab:constraints} and discussed in the following. First, when the magnetic field $\bm{h}=(0, h\cos \theta, h\sin \theta)$ is parallel to the $y$-axis ($\theta=\ang{0}$), the model is characterized by the $\mathrm{m^{\prime}m2^{\prime}}$ magnetic point group symmetry. The symmetry prohibits the nonlinear conductivity $\sigma^{x;yx}$, $\sigma^{y;xx}$, and $\sigma^{y;yy}$. Figure~\ref{fig:general2.pdf} is consistent with these constraints. Moreover, all the $\mathcal{T}$-even contributions symmetric under the $\mathcal{T}$ operation (see Table~\ref{tab:mechanism_class} in Appendix~\ref{app:general_property} for classification of the photocurrent conductivity based on the mechanism and symmetry) disappear in the two-dimensional system due to the $\mathcal{T}\mathrm{C_{2z}}$ symmetry. Second, when $\theta$ satisfies $\ang{0}<\theta<\ang{90}$, the model has the $\mathcal{T}\mathrm{m_{x}}$ symmetry, which is classified into the magnetic point group $\mathrm{m}^{\prime}$. The $\mathcal{T}$-even contributions of $\sigma^{x;xx}$, $\sigma^{x;yy}$, and $\sigma^{y;yx}$ disappear due to the symmetry. Finally, when the magnetic field is perpendicular to the $xy$-plane ($\theta=\ang{90}$), the symmetry of the system is $4\mathrm{m}^{\prime}\mathrm{m}^{\prime}$, and all the components of nonlinear conductivity vanish in the two-dimensional system due to the $\mathrm{C_{2z}}$ rotation symmetry. 
In Sec.~\ref{sec:Unique behaviors due to Dirac point}, we focus on the photocurrent conductivity $\sigma^{x;xx}_{\mathrm{PC}}$ and $\sigma^{y;yx}_{\mathrm{PC}}$. These components have no $\mathcal{T}$-even contribution for any angle $\theta$ due to the  $\mathcal{T}m_{x}$ symmetry. In Appendix~\ref{app:general_property}, we show that the magnetic injection current and gyration current [Eqs.~\eqref{eq:Minj_term} and \eqref{eq:gyro_term}], which are $\mathcal{T}$-odd, are dominant in the real and imaginary parts of these photocurrent conductivity components, respectively.

In Fig.~\ref{fig:general2.pdf}, we see a characteristic frequency dependence of superconducting nonlinear responses. 
First, in the low-frequency region, the divergent nonlinear responses are observed in the real part of $\sigma^{\alpha;\beta\gamma}_{\mathrm{PC}}$ and $\sigma^{\alpha;\beta\gamma}_{\mathrm{SHG}}$. The nonreciprocal superfluid density term $\sigma^{\alpha;\beta\gamma}_{\mathrm{NRSF}}$, which is odd under the $\mathcal{T}$ symmetry, causes the anomalous divergent behaviors. The static conductivity derivative term $\sigma^{\alpha;\beta\gamma}_{\mathrm{sCD}}$ disappears because this term is even for the $\mathcal{T}$ symmetry. Although the imaginary part also shows a weakly diverging behavior at low frequencies, this behavior is artificial and comes from the phenomenological treatment of the scattering rate~\cite{Watanabe2022}.

Second, the type B transition illustrated in Fig.~\ref{fig:schematic_figure}(a) gives the characteristic resonant contribution to the photocurrent conductivity and second harmonic generation. This contribution is often characterized by the superconducting gap. The peak around $\Omega=2\psi$ is a typical behavior and observed in Figs.~\ref{fig:general2.pdf}(a) and \ref{fig:general2.pdf}(b). Contrary to the photocurrent, not only the transition $\Delta E \sim \Omega$ but also $\Delta E \sim 2\Omega$ may give rise to essential contributions to the second harmonic generation, where $\Delta E$ is an energy difference between transition bands. Figures~\ref{fig:general2.pdf}(c) and \ref{fig:general2.pdf}(d) actually show the peak structure and sign reversal at $\Omega = \psi$. The frequency dependence of the second harmonic generation is more complex than that of the photocurrent generation.
In the remaining part, we discuss the photocurrent conductivity in detail.

\section{Unique behaviors due to Dirac point}
\label{sec:Unique behaviors due to Dirac point}
In this section, we show the characteristic behaviors of the nonlinear optical responses in the systems with a chemical potential close to the Dirac point. For this purpose, we focus on the resonant components of the superconducting photocurrent, which are characterized by the JDOS. We investigate in detail the dependence on the chemical potential and magnetic field. Consistent with our conjecture in Sec.~\ref{sec:joint_density_of_state}, the presence of the Dirac point near the Fermi level gives rise to unique behaviors in the superconducting nonlinear responses. In the following, we verify the conjecture with numerical calculations.



\begin{figure*}[htbp]
 \includegraphics[width=\linewidth]{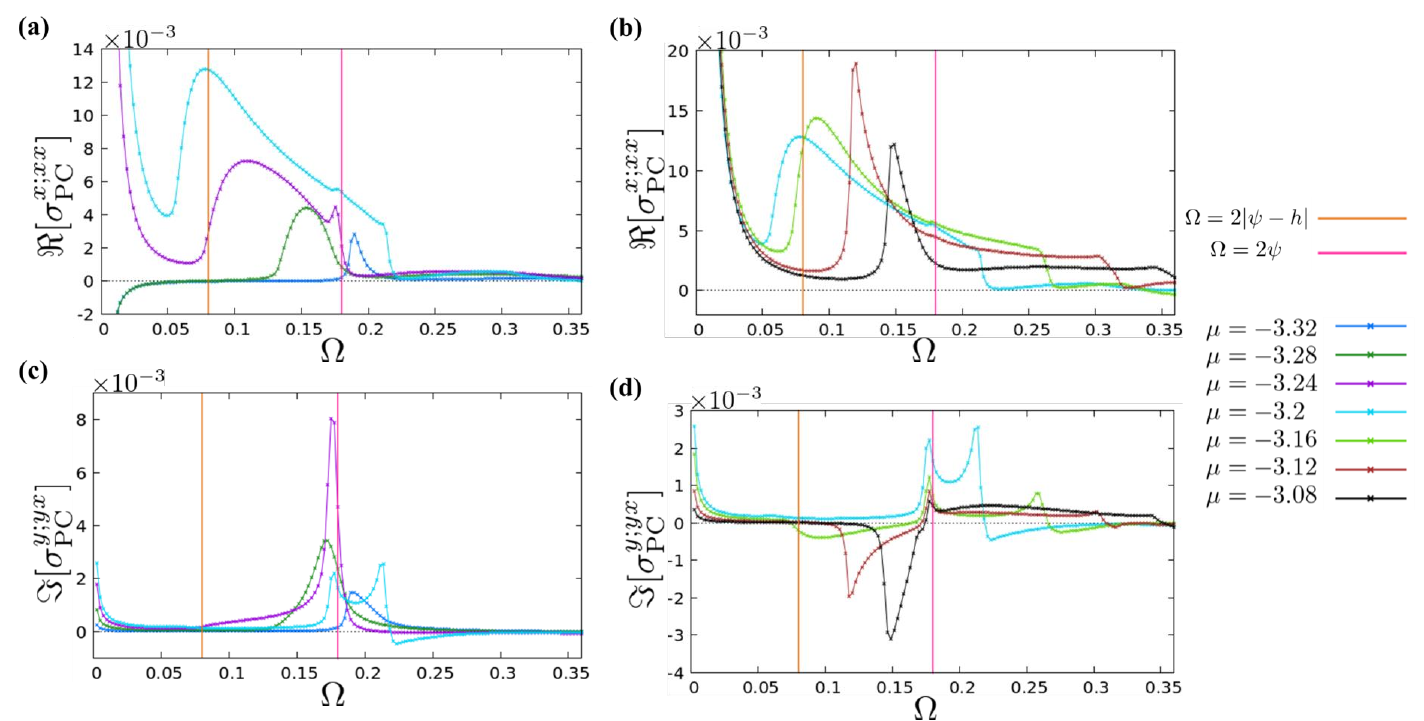} \caption{Photocurrent conductivity $\Re[\sigma^{x;xx}_{\mathrm{PC}}]$ and $\Im[\sigma^{y;yx}_{\mathrm{PC}}]$ for various chemical potentials $\mu$. (Upper panels) $\Re[\sigma^{x;xx}_{\mathrm{PC}}]$ for (a) $-3.32\leq\mu\leq -3.2$ and (b) $-3.2\leq\mu\leq -3.08$. (Lower panels) $\Im[\sigma^{y;yx}_{\mathrm{PC}}]$ for (c) $-3.32\leq\mu\leq -3.2$ and (d) $-3.2\leq\mu\leq -3.08$. We set $\psi =0.09$, $h=0.05$, and $\theta =\ang{30}$.}
 \label{fig:mu_depend}
\end{figure*}

\subsection{Chemical potential dependence}
\label{subsec:dependence_chemi}

First, we discuss the dependence of photocurrent conductivity on the chemical potential $\mu$. Figures~\ref{fig:mu_depend}(a) and \ref{fig:mu_depend}(b) plot the real part of the photocurrent conductivity $\sigma^{x;xx}_{\mathrm{PC}}$ with varying the chemical potential below and above the Dirac point $\mu = -3.2$, respectively. At $\mu=-3.32$, the photocurrent conductivity shows a sharp peak around $\Omega =2\psi$. The peak broadens when the chemical potential approaches the Dirac point. For $\mu=-3.2$ on the Dirac point, the photocurrent conductivity is enhanced in the region between $\Omega \simeq 2|\psi -h|$ and $\Omega\simeq 2\psi$. 
Figure~\ref{fig:mu_depend}(b) shows that a sharp peak appears at a frequency between $\Omega=2|\psi-h|$ and $\Omega =2\psi$ for $\mu=-3.12$ and $\mu=-3.08$. 
We also find a little peak around $\Omega =2\psi$ independent of the chemical potential, which is hard to see in the figures for some parameters. 

Based on the discussions in Sec.~\ref{sec:joint_density_of_state}, we compare the photocurrent conductivity with the JDOS. 
Indeed, the behaviors of the photocurrent conductivity are related to the approximate JDOS studied in Sec.~\ref{sec:joint_density_of_state}. 
The photocurrent conductivity enhanced at $2|\psi -h| < \Omega <2\psi$ is similar to the approximate JDOS $\tilde{J}_{\mathrm{B2(ii)}}(\Omega)$ in the dependence on the frequency and chemical potential. At $\mu=-3.2$, the photocurrent conductivity is enhanced in the whole region $2|\psi -h| < \Omega <2\psi$. 
When the chemical potential goes away from the Dirac point, the enhancement starts at a higher frequency $\Omega\simeq 2|\sqrt{{\tilde{\mu}}^{2} + \psi^{2}}-h|$. Thus, the photocurrent conductivity in this frequency region is attributed to the contribution from the approximate JDOS $\tilde{J}_{\mathrm{B2(ii)}}(\Omega)$. More precisely, we see the enhancement starting at $\Omega\simeq 0.06 <2|\psi-h|$ because the finite total momentum of Cooper pairs induces the effective magnetic field $\Delta \bm{h}_{\mathrm{eff}}$ and influence the JDOS and the photocurrent conductivity. 
In other words, the estimation based on $\tilde{J}_{\mathrm{B2(ii)}}(\Omega)$ becomes more precise when we use $\bm{h}_{\mathrm{eff}}$ instead of $\bm{h}$ as in Eq.~\eqref{eq:effective_magnetic_field}. On the other hand, a tiny anomaly around $\Omega = 2 \psi$ is robust against the change of the chemical potential, and thus it is attributed to the contribution of the approximate JDOS $\tilde{J}_{\mathrm{B2(i)}}(\Omega)$.


Figures~\ref{fig:mu_depend}(c) and \ref{fig:mu_depend}(d) show the imaginary part of the photocurrent conductivity $\sigma^{y;yx}_{\mathrm{PC}}$ with the same parameters as Figs.~\ref{fig:mu_depend}(a) and \ref{fig:mu_depend}(b). Unlike $\Re[\sigma^{x;xx}_{\mathrm{PC}}]$ in Figs.~\ref{fig:mu_depend}(a) and \ref{fig:mu_depend}(b), the existence of a large peak around $\Omega =2\psi$ is pronounced in $\Im[\sigma^{y;yx}_{\mathrm{PC}}]$. 
At $\mu=-3.2$, the photocurrent conductivity is broadly distributed in the region $2|\psi-h| < \Omega < 2\psi$. These behaviors are also consistent with the above discussions based on the JDOS. The differences from $\Re[\sigma^{x;xx}_{\mathrm{PC}}]$ are attributed to the relative contributions of $\tilde{J}_{\mathrm{B2(i)}}(\Omega)$ and $\tilde{J}_{\mathrm{B2(ii)}}(\Omega)$.
For $\Re[\sigma^{x;xx}_{\mathrm{PC}}]$, although $\tilde{J}_{\mathrm{B2(ii)}}(\Omega)$ is tiny, the contribution of $\tilde{J}_{\mathrm{B2(ii)}}(\Omega)$ is dominant because it is remarkably enhanced by quantum geometry. Later this enhancement is discussed in detail [Sec.~\ref{sec:enhancement by quantum geometry}]. For $\Im[\sigma^{y;yx}_{\mathrm{PC}}]$, the contributions of $\tilde{J}_{\mathrm{B2(i)}}(\Omega)$ and $\tilde{J}_{\mathrm{B2(ii)}}(\Omega)$ are comparable and may have an opposite sign. 

Note that the large peaks in the photocurrent conductivity at $2|\psi -h| < \Omega < 2\psi$ for $\mu=-3.12$ and $\mu=-3.08$ are not expected from $\tilde{J}_{\mathrm{B2(ii)}}$ but from $\bar{J}_{\mathrm{B}}(\Omega)$. This means a substantial contribution from the k-space where $|\xi_{\bm{k}}||\bm{g_{k}}|\simeq \psi h$.

\begin{figure*}[htbp]
\includegraphics[width=\linewidth]{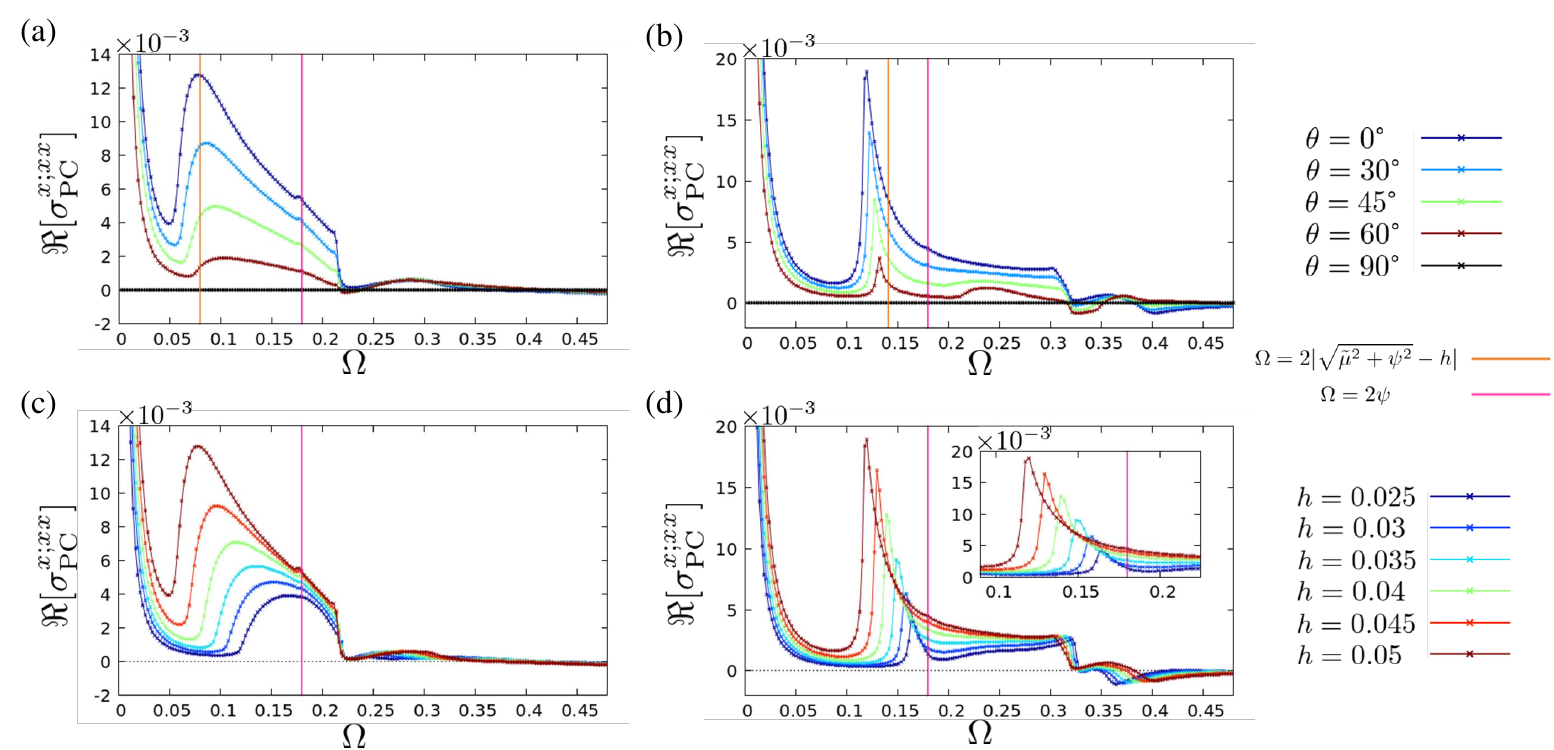}
\caption{Photocurrent conductivity $\Re[\sigma^{x;xx}_{\mathrm{PC}}]$ with varying the magnetic field $\bm{h}$. (a) and (b) 
The angle of the magnetic field is varied by fixing the magnitude $h=0.05$. (c) and (d) 
The magnitude of the magnetic field is varied by fixing the angle $\theta =\ang{0}$.  The chemical potential is close to the Dirac point as (a) (c) $\mu=-3.2$ while (b) (d) $\mu=-3.12$.}
\label{fig:general5}
\end{figure*}

\subsection{Magnetic field dependence}

As shown in the previous subsection, the photocurrent conductivity is related to the JDOS. Therefore, we also expect a unique magnetic field dependence of the photocurrent conductivity. That is verified below.

First, we discuss the angle $\theta$ of the magnetic field $\bm{h}=(0, h\cos\theta, h\sin\theta)$. Figures~\ref{fig:general5}(a) and \ref{fig:general5}(b) plot the real part of the photocurrent conductivity $\sigma^{x;xx}_{\mathrm{PC}}$ for various angles. 
Here, we see a consequence of the symmetry constraint. 
The photocurrent conductivity disappears under the perpendicular magnetic field with $\theta = \ang{90}$ because the $\mathrm{C_{2z}}$ rotation symmetry prohibits the second-order optical responses. 
For angles different from $\theta = \ang{90}$, we see qualitatively the same frequency dependence. 
In Fig.~\ref{fig:general5}(a) with $\mu=-3.2$, the photocurrent conductivity is enhanced between $\Omega =2|\psi -h|$ ($=2|\sqrt{{\tilde{\mu}}^{2} + \psi^{2}}-h|$) and $\Omega =2\psi$. Figure~\ref{fig:general5}(b) for $\mu=-3.12$ shows sharp peaks around $\Omega\simeq 2|\sqrt{{\tilde{\mu}}^{2} + \psi^{2}}-h|$ for various angles.  
Given the total momentum of Cooper pairs $2\bm{q}$, these estimations are corrected so that the peaks are predicted to be at lower frequencies due to the correction to the effective magnetic field $\bm{h}_{\mathrm{eff}}$. The momentum $q_{y}$ is roughly proportional to $h_{y}$ when we fix $\psi$. Therefore, the frequency shift of the photocurrent conductivity is expected to be larger in the result with $h_{y}=\cos \ang{0}$ than that with $h_{y}=\cos \ang{60}$. In agreement with this expectation, we see a larger peak shift to the low-frequency side with a smaller angle $\theta$ in Fig.~\ref{fig:general5}(b). 
The qualitative behaviors of the photocurrent conductivity remain unchanged with varying the angle $\theta$ of the magnetic fields unless we set $\theta = \ang{90}$. 
Therefore, the effect of the total momentum of Cooper pairs, namely, the helical superconductivity, on the photocurrent conductivity can be ignored when we neglect the small shift of frequency dependence.
Naturally, the magnitude of the photocurrent conductivity becomes smaller when the angle approaches $\theta=\ang{90}$.

Next, we show the change in the photocurrent conductivity when varying the magnitude of the magnetic field. Figures~\ref{fig:general5}(c) and \ref{fig:general5}(d) plot the real part of the photocurrent conductivity $\sigma^{x;xx}_{\mathrm{PC}}$ with varying the magnetic field $h$. 
For $\mu=-3.2$ on the Dirac point [Fig.~\ref{fig:general5}(c)], the photocurrent conductivity begins rising at a lower frequency with a larger magnetic field. This is consistent with the JDOS which is finite above $\Omega \simeq 2|\psi -h|$ and attributed to the contribution of $\tilde{J}_{\mathrm{B2(ii)}}(\Omega)$. 
For $\mu=-3.12$ slightly above the Dirac point [Fig.~\ref{fig:general5}(d)], the peak of the photocurrent conductivity shifts to a lower frequency with enlarging the magnetic field. 
These peaks have been identified as a contribution from the ${\bm k}$-space region where $|\xi_{\bm{k}}||\bm{g_{k}}|\simeq \psi h$ is satisfied, and the magnetic field dependence is consistent with this interpretation. 


\begin{figure*}
\includegraphics[width=\linewidth]{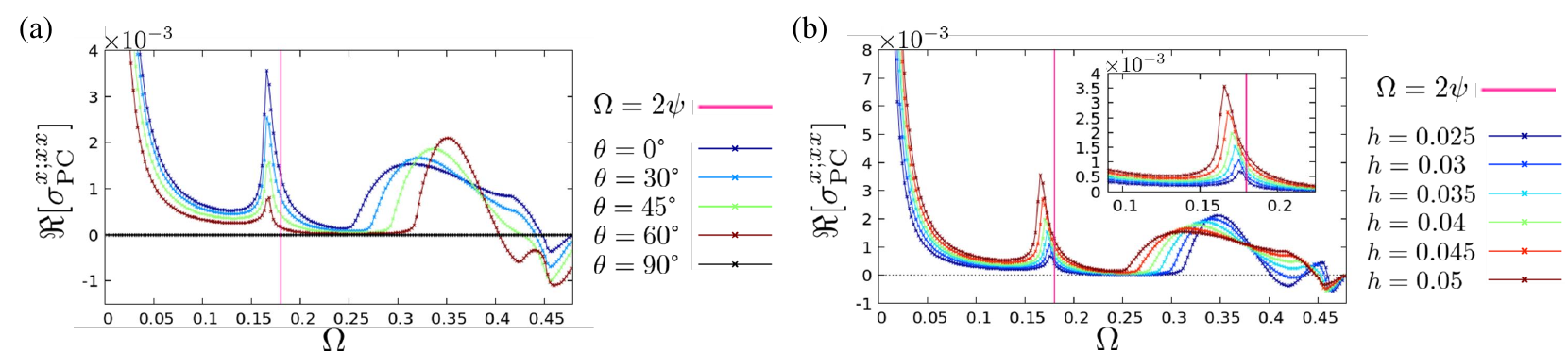}
\caption{Photocurrent conductivity $\Re[\sigma^{x;xx}_{\mathrm{PC}}]$ 
for $\mu=-3.0$ away from the Dirac point. (a) 
We fix $h =0.05$ and change the angle of the magnetic field $\theta$. (b) 
We fix the angle $\theta =\ang{0}$
and varies the magnitude of the magnetic field $h$.}
\label{fig:general6}
\end{figure*}

However, the above peculiar behaviors of the photocurrent conductivity disappear when the chemical potential is far away from the Dirac point. 
This is expected from the analysis of the JDOS. Because the unique contribution in the in-gap region $\Omega <2\psi$ disappears at $\mu =-3.0$, we expect that the contribution of $\tilde{J}_{\mathrm{B2(i)}}$ is dominant. In Fig.~\ref{fig:general6}, where we plot $\Re[\sigma^{x;xx}_{\mathrm{PC}}]$ for $\mu =-3.0$, 
the photocurrent conductivity shows a peak around $\Omega =2\psi$, and the position of the peak does not significantly change with varying the angle and magnitude of the magnetic field. These results are consistent with the above expectation that 
the photocurrent conductivity is mainly related to the JDOS
$\tilde{J}_{\mathrm{B2(i)}}(\Omega)$ when the chemical potential is not close to the Dirac point.

We summarize the results in this subsection. 
For chemical potentials near the Dirac point, a component of 
the JDOS from the region $\bm{k}\simeq \bm{0}$, namely $\tilde{J}_{\mathrm{B2(ii)}}(\Omega)$, gives significant contributions to the photocurrent conductivity in the superconducting state. Since these contributions are sensitive to magnetic fields, the photocurrent conductivity shows characteristic parameter dependence in the low-frequency regime $\Omega<2\psi$. 
On the other hand, the contribution of $\tilde{J}_{\mathrm{B2(i)}}(\Omega)$ is dominant when the chemical potential is not in the vicinity of the Dirac point. This contribution gives a sharp peak around $\Omega =2\psi$, which is robust against the change in the magnetic field. 
Thus, the discussions in Sec.~\ref{sec:approximate_JDOS}, such as the decomposition of the JDOS to the approximate JDOS $\tilde{J}_{\mathrm{B2(i)}}(\Omega)$ and $\tilde{J}_{\mathrm{B2(ii)}}(\Omega)$, are useful to elucidate the photocurrent conductivity in noncentrosymmetric superconductors in the in-gap frequency range $\Omega \lesssim 2\psi$.

\subsection{Comparison between normal and superconducting states}
\label{subsec:comparison_normal_SC}

\begin{figure*}[htbp]
 \includegraphics[width=\linewidth]{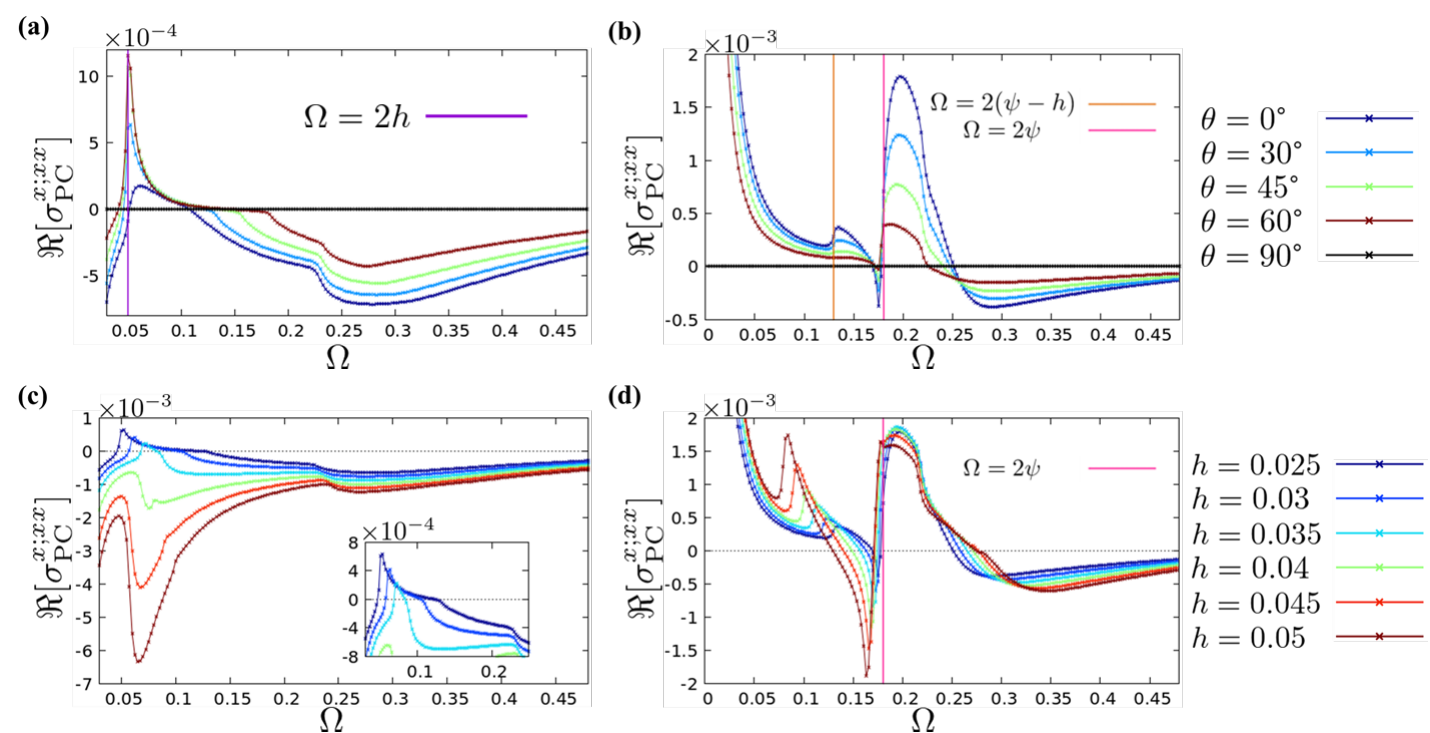}
 \caption{
 The left panels (a) and (c) show the photocurrent conductivity $\Re[\sigma^{x;xx}_{\mathrm{PC}}]$ in the normal state ($\psi=0$), while the right panels (b) and (d) show the superconducting state ($\psi=0.09$). 
 In the upper panels (a) and (b), the angle $\theta$ of the magnetic field is varied with fixing $h=0.025$, while in the lower panels (c) and (d), the magnitude of the magnetic field $h$ is changed with fixing $\theta =\ang{30}$.  
 We set the chemical potential $\mu=-0.8$. 
 }
 \label{fig:SC_normal}
\end{figure*}

At the end of this section, we compare the photocurrent conductivity between the normal and superconducting states. We again focus on the resonant components. When we choose the parameter $\mu=-3.2$, the resonant component is negligible in the normal state under the in-plane magnetic field. 
Therefore, we conclude that the resonant component of the photocurrent conductivity originates from superconductivity.

To study a qualitatively different case, we calculate the photocurrent conductivity of the model at the chemical potential $\mu=-0.8$, which is 
equivalent to the energy of Dirac points at $\bm{k} = (\pi,0)$ and $(0,\pi)$.
Therefore, the low-frequency photocurrent conductivity mainly comes from the region $\bm{k}\simeq (\pi,0)$ and $\simeq(0,\pi)$ in the Brillouin zone.
First, we discuss the difference between $\mu=-0.8$ and $\mu=-3.2$.
Figures~\ref{fig:SC_normal}(b) and \ref{fig:SC_normal}(d) show the real part of the photocurrent conductivity $\sigma^{x;xx}_{\mathrm{PC}}$ in the superconducting state for $\mu=-0.8$. 
The photocurrent conductivity is enhanced and changes the sign around $\Omega \simeq 2\psi$. This behavior is different from the results for $\mu=-3.2$ [Fig.~\ref{fig:general5}] and owing to the cancellation of contributions from $\bm{k}\simeq (\pi,0)$ and $\simeq (0,\pi)$. 
As we show the ${\bm k}$-resolved contribution in Fig.~\ref{fig:k_profile} of Appendix~\ref{app:model_with_mu}, the regions $\bm{k}\simeq (\pi,0)$ and $\simeq (0,\pi)$ give opposite contributions to the photocurrent conductivity. Thus, the sign of the photocurrent conductivity depends on the frequency, and indeed the sign reversal occurs. A detailed discussion on the sign reversal is given in Appendix~\ref{app:model_with_mu}. 

Next, we compare the normal state and superconducting state for $\mu=-0.8$. 
As shown in Fig.~\ref{fig:SC_normal}, the photocurrent conductivity shows qualitatively different behaviors between the normal and superconducting states. 
Figure~\ref{fig:SC_normal}(a) shows that the photocurrent conductivity in the normal state has a sharp peak around $\Omega=2h$ for angles of the magnetic field $\ang{30}\leq\theta\leq\ang{60}$. However, the peak is broadly rounded at $\theta=\ang{0}$. Contrary to the normal state, the qualitative behaviors of the photocurrent conductivity remain unchanged with varying the angle $\theta$ in the superconducting state, and we observe characteristic frequency dependence around $\Omega=2\psi$ [Fig.~\ref{fig:SC_normal}(b)]. 
Note that the photocurrent conductivity generally decreases with increasing the angle of the magnetic field and disappears at $\theta=\ang{90}$.
The normal state photocurrent also shows non-monotonic magnetic field dependence. 
Figure~\ref{fig:SC_normal}(c) with the inset shows that the peak position shifts to a higher frequency with a larger magnetic field from $h=0.025$ to $h=0.035$, consistent with $\Omega=2h$. 
However, in the range $0.04\leq h \leq 0.05$, we do not find the peak around $\Omega =2h$ probably because the peak is smeared out by another contribution that grows with a magnetic field. 
Contrary to the normal state, the photocurrent conductivity in the superconducting state shows the peaks with sign reversal around $\Omega=2\psi$ without the change of qualitative behaviors as increasing $h$. 

\section{Enhancement of Photocurrent generation by quantum geometry}
\label{sec:enhancement by quantum geometry}

Here, we highlight the effect of quantum geometry on the photocurrent conductivity. In Sec.~\ref{sec:Unique behaviors due to Dirac point}, we have shown that a component of the JDOS $\tilde{J}_{\mathrm{B2(ii)}}$ gives dominant contribution to the photocurrent conductivity when the Fermi level lies near the Dirac point [for example, see Figs.~\ref{fig:general5}(a) and \ref{fig:general5}(c)]. However, the value of $\tilde{J}_{\mathrm{B2(ii)}}$ is tiny and much smaller than $\tilde{J}_{\mathrm{B2(i)}}$ [Fig.~\ref{fig:JDOS2}], and the discussion of the JDOS is not sufficient to explain the large resonant photocurrent in the low-frequency regime. 
In this section, we focus on the quantum geometric properties of Bogoliubov bands because it has been shown that quantum geometry gives colossal nonlinear responses in some normal states~\cite{Orenstein2021}. Indeed, we show that quantum geometry plays a key role in enhancing the superconducting nonlinear responses in the low-frequency in-gap regime.

When the Fermi level is close to the Dirac point, the magnetic injection current mechanism gives a dominant contribution to the real part of the photocurrent conductivity $\Re[\sigma^{x;xx}_{\mathrm{PC}}]$ [Appendix~\ref{app:general_property}]. The formula of the magnetic injection current [Eq.~\eqref{eq:Minj_term}] contains the quantum metric $(g^{\alpha\beta}_{ab})$, where $\alpha$ means $\lambda_{\alpha}$ for short hand notation. The JDOS $J_{\mathrm{B2}}(\Omega)$, which originates from the transition between the hole $E_{2}$ and particle $E_{3}$ bands, is dominant in the low-frequency regime $\Omega \leq 2\psi$, and thus we focus on the magnetic injection current originating from the B2 transition $\sigma^{x;xx}_{\mathrm{Minj(B2)}}$ defined as
\begin{align}
    \sigma^{x;xx}_{\mathrm{Minj(B2)}} = -\frac{\pi}{4\eta}\sum_{\bm{k}}(J^{x}_{22}-J^{x}_{33})g_{32}^{xx}(f_{2} - f_{3})\delta(\Omega -E_{32}).
\end{align}
This formula implies that the band-resolved quantum metric $g_{32}^{xx}$ affects the photocurrent conductivity in the low-frequency regime. 

\begin{figure*}
\includegraphics[width=0.9\linewidth]{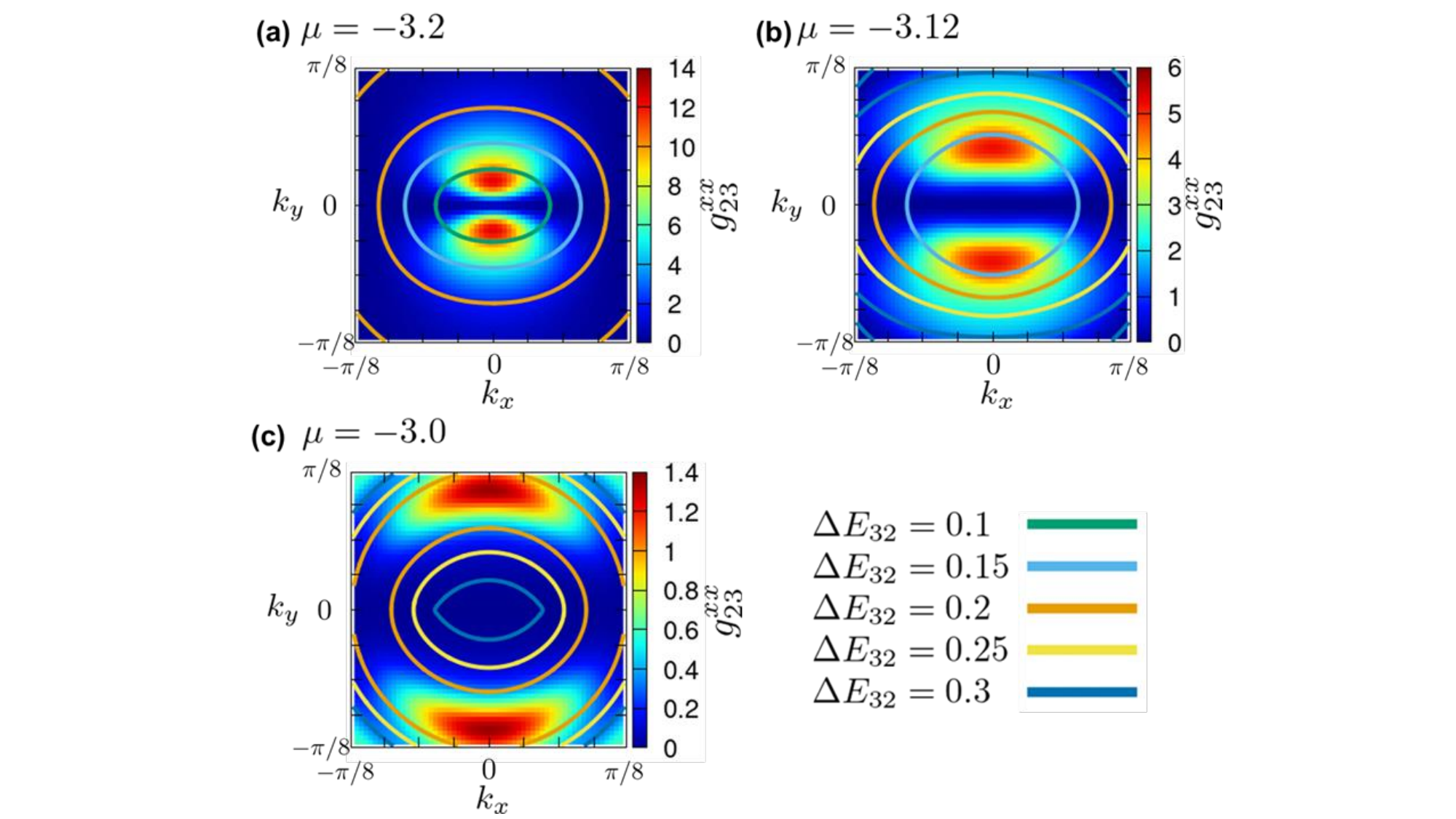}
\caption{Distribution of the quantum metric $g^{xx}_{23}$ in the $\bm{k}$-space. We fix $h=0.05$, $\theta=\ang{0}$, and $\psi=0.09$ and set (a) $\mu=-3.2$, (b) $\mu=-3.12$, and (c) $\mu =-3.0$. Colored lines plot contour lines of $\Delta E_{23}$. Note that the mapping range of the color plot is different between the figures because the typical values of the quantum metric are vastly different among the figures (a), (b), and (c).}
\label{fig:metric_plot}
\end{figure*}

We show the distribution of the quantum metric $g_{32}^{xx}$ in the $\bm{k}$-space in Fig.~\ref{fig:metric_plot}. The parameters in Fig.~\ref{fig:metric_plot}(a) are the same as Fig.~\ref{fig:general5}(a) with $\theta =\ang{0}$. The color map shows the huge quantum metric $g_{32}^{xx}$ around $\bm{k}=\bm{0}$, as is generally expected for Dirac electrons. As shown by the contour lines of $E_{32}=E_{3}-E_{2}$, the quantum metric enhances the photocurrent conductivity in the low-frequency regime $\Omega<2\psi =0.18$. These results indicate that the quantum metric selectively enhances the photocurrent conductivity originating from the JDOS $\tilde{J}_{\mathrm{B2(ii)}}$. 

Next, we discuss the effect of the quantum metric when the Fermi level slightly deviates from the Dirac point. For $\mu=-3.12$, the enhancement by $g_{32}^{xx}$ is prominent on the peaks of the photocurrent conductivity in Fig.~\ref{fig:general5}(b) around $\Omega\simeq 0.12<2\psi$. As shown in Fig.~\ref{fig:JDOS_k_profile}(b), the JDOS $J_{\mathrm{B2}}(\Omega=0.12)$ originates from the region $k<\frac{\pi}{16}$, where $g_{32}^{xx}$ is large as shown in Fig.~\ref{fig:metric_plot}(b). 
Therefore, the photocurrent conductivity in the in-gap region is cooperatively enhanced by the JDOS and quantum metric and shows a large peak around $\Omega = 0.12$. 

When the chemical potential is far away from the Dirac point, the contribution of $\tilde{J}_{\mathrm{B2(ii)}}$ vanishes in the low-frequency regime $\Omega <2\psi$. The quantum metric for $\mu=-3.0$ is shown in Fig.~\ref{fig:metric_plot}(c). In contrast to the cases of $\mu=-3.2$ and $-3.12$ [Figs.~\ref{fig:metric_plot}(a) and \ref{fig:metric_plot}(b)], the quantum metric  $g_{32}^{xx}$ enhances the contribution of $\tilde{J}_{\mathrm{B2(i)}}$ around $\Omega =2\psi$. However, the value of the photocurrent conductivity 
for $\mu=-3.0$ 
is smaller 
than that for chemical potentials near the Dirac point
[compare Fig.~\ref{fig:general6}(a)
with Figs.~\ref{fig:general5}(a) and \ref{fig:general5}(b)]. 
Optical responses tend to decrease as the Fermi level moves away from the Dirac point due to the suppression of quantum geometric quantities.
The photocurrent conductivity $\Re[\sigma^{x;xx}_{\mathrm{PC}}]$ is also reduced because of the decrease in the size of the quantum metric $g_{32}^{xx}$. 

From the above results, we conclude that the quantum metric enhances the photocurrent conductivity in the low-frequency in-gap region. This is the reason why we have observed the pronounced photocurrent conductivity below $\Omega \leq 2 \psi$ in the previous sections. Here, we note that both magnetic field and spin-orbit coupling are necessary for a finite quantum metric $g_{32}^{\alpha\beta}$.
The $g_{32}^{\alpha\beta}$ is given by the current operator as
\begin{align}
g_{32}^{\alpha\beta} = \frac{1}{{E_{23}}^{2}}\Re[J^{\alpha}_{32}J^{\beta}_{23}].
\end{align}
Thus, the matrix element of the paramagnetic current operator $J^{\alpha}_{32}$ needs to be finite for a nonzero quantum metric $g_{32}^{\alpha\beta}$. When the magnetic field is absent, $J^{\alpha}_{23}$ and $g_{32}^{\alpha\beta}$ vanish. Moreover, a finite g-vector of spin-orbit coupling $\bm{g_{k}}$ is also necessary even when the derivative $\partial_{\alpha}\bm{g_{k}}$ is not zero. This indicates that $g_{32}^{\alpha\beta}(\bm{k})$ vanishes at $\bm{k}=\bm{0}$ where the g-vector is zero. This is consistent with our numerical results in Fig.~\ref{fig:metric_plot}, where we see $g_{32}^{xx}=0$ at $\bm{k}=\bm{0}$.

We have shown the enhancement of the photocurrent conductivity 
in the superconducting state due to the giant quantum metric $g_{32}^{xx}$. In the normal state, the quantum metric diverges at the Dirac point due to the energy degeneracy. Although the superconducting gap $E_{23}\neq 0$ prevents the divergence of the quantum metric, the large quantum metric appears when the Fermi level lies near the Dirac point.
It is implied that the large quantum metric in the superconducting states inherits peculiar geometric properties of Dirac states. 
The magnetic field and spin-orbit coupling are essential not only for satisfying the symmetry constraint but also for realizing the quantum-geometrically enhanced optical responses in noncentrosymmetric superconductors.

\section{Photocurrent conductivity around Topological transition}
\label{sec:photocurrent conductivity around topological transition}

Next, we investigate the relation between superconducting nonlinear responses and topological superconductivity. We show that the photocurrent conductivity dramatically changes when the system becomes a topological superconducting state.

In this study, we have considered $s$-wave Rashba superconductors with the Zeeman field. This model is a candidate for the realization of topological superconductivity hosting Majorana zero modes. The $s$-wave superconductor becomes a topological superconductor with a nonzero Chern number when the Zeeman field $h$ is larger than the critical value $h^{c}$ ~\cite{Sato2009, Sato2010, Sau2010}. The critical value is given as $h^{c}=\sqrt{{\tilde{\mu}}^{2} + \psi^{2}}$ when we set the magnetic field perpendicular to the plane. Note that we assume Eqs.~\eqref{eq:simple_xi_2} and \eqref{eq:simple_g_2} when we estimate $h^{c}$. 
Much effort has been devoted to the search for topological superconductivity, as it attracts much attention from viewpoints ranging from fundamental science to engineering~\cite{Alicea2012}. However, no firm evidence for topological superconductivity has yet been obtained. This is partly because the physical phenomena manifesting the signature of topological superconductivity are limited.
The existence of the Majorana edge mode is a unique characteristic of topological superconductivity and some theoretical works propose the optical probe for the Majorana edge mode~\cite{He2021PRL, He2021PRB, Bi2023}. On the other hand, the exploration of bulk probes for topological superconductivity has also been awaited. Therefore, it is highly desirable to elucidate the optical responses that signal the transition to the topological superconducting phase.

\begin{figure*}
\includegraphics[width=\linewidth]{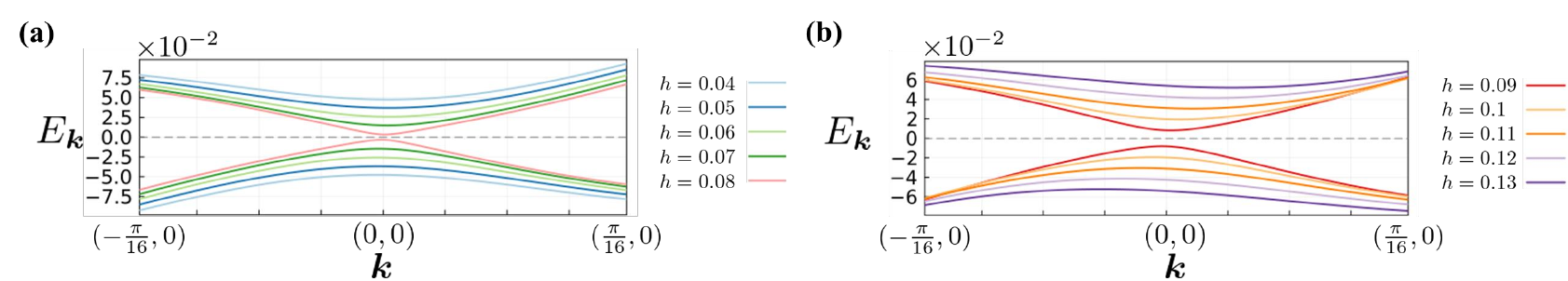}
\caption{The energy band of Bogoliubov quasiparticles around the $\Gamma$ point [$\bm{k}=(0,0)$] with varying the magnitude of the magnetic field. We set $\mu=-3.2$, $\psi =0.09$, and $\theta =\ang{60}$.}
\label{fig:top_trans_band}
\end{figure*}

In this section, we investigate the photocurrent conductivity around the topological transition between the topologically trivial and nontrivial superconducting phases. In the previous work~\cite{Sato2009,Sato2010, Sau2010}, 
it was shown that the topological transition occurs under the magnetic field perpendicular to the plane. In the following, we consider tilted magnetic fields to the plane because the photocurrent conductivity vanishes under the perpendicular magnetic field. We set the angle $\theta$ as $\theta=\ang{60}$. Because the topological property is robust against small changes in the Hamiltonian, 
the topological transition occurs even though the tilted magnetic fields. 
Gap closing is one of the characterizations of the topological transition. Figure~\ref{fig:top_trans_band} shows the energy band around $\bm{k}=\bm{0}$ when we set $\mu =-3.2$ and $\psi =0.09$. As we enlarge the magnitude of the magnetic field in $0.04\leq h \leq 0.08$, the energy gap becomes smaller [Fig.~\ref{fig:top_trans_band}(a)]. On the other hand, the energy gap opens as the magnetic field is further increased in  $0.09\leq h\leq 0.13$ [Fig.~\ref{fig:top_trans_band}(b)].  These results indicate that the gap closing and topological transition occur in the region $0.08< h < 0.09$. Note that the critical value $h_{c}$ is estimated to be 0.09 when we ignore the total momentum of Cooper pairs. In fact, the critical value is obtained as $h_{c}=0.09$ when we set the angle $\theta=\ang{90}$. When the total momentum of Cooper pairs is finite, the effective magnetic field $\Delta \bm{h}_{\mathrm{eff}}$ slightly shifts the critical value $h_{c}$.

\begin{figure*}
\includegraphics[width=\linewidth]{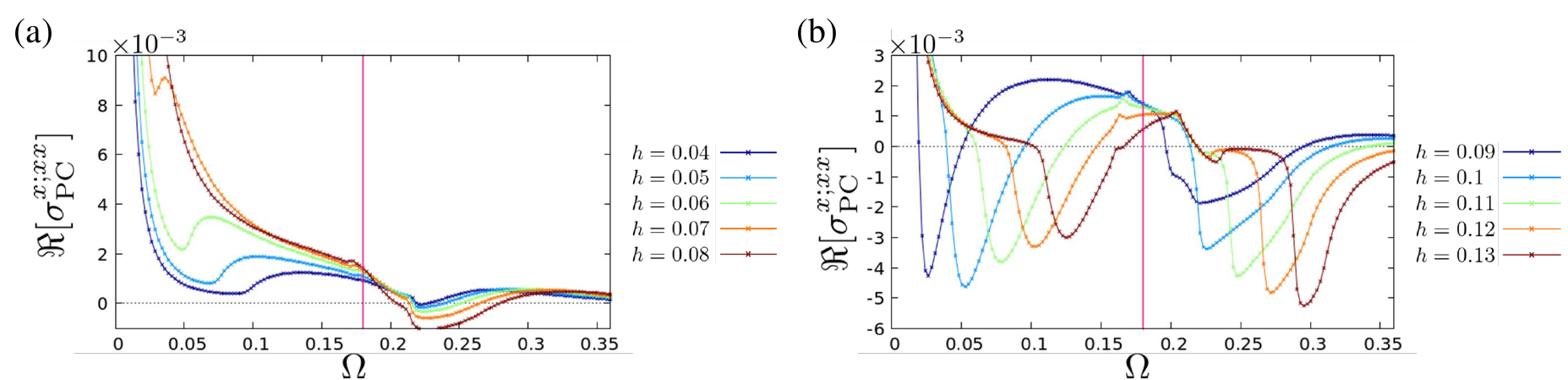}
\caption{The photocurrent conductivity for various magnitudes of the magnetic field.  We show $\Re[\sigma^{x;xx}_{\mathrm{PC}}]$ for (a) $0.04\leq h \leq 0.08$ and (b) $0.09\leq h \leq 0.13$.}
\label{fig:topo_trans}
\end{figure*}

Figure~\ref{fig:topo_trans} shows the photocurrent conductivity $\sigma^{x;xx}_{\mathrm{PC}}$ for the magnetic fields (a) $0.04\leq h\leq 0.08$ and (b) $0.09\leq h \leq 0.13$. We focus on the resonant components in the low-frequency region $\Omega <2\psi$. In the topologically trivial region $0.04\leq h \leq 0.08$, the photocurrent conductivity begins to increase at a lower frequency with a larger magnetic field [Fig.~\ref{fig:topo_trans}(a)]. When the magnetic field is further increased in the topological superconducting phase, the photocurrent conductivity is negatively enhanced and shows peaks at a higher frequency with a larger magnetic field [Fig.~\ref{fig:topo_trans}(b)].
The drastic changes in frequency dependence and sign are associated with the closing of the energy gap around $\bm{k}=\bm{0}$ [Fig.~\ref{fig:top_trans_band}]. It is naturally expected that these low-frequency contributions originate from the JDOS $\tilde{J}_{\mathrm{B2(ii)}}$. 
%
We would like to emphasize the sign change of the resonant photocurrent before and after the topological transition. Below we discuss the impact of the topological transition and attribute the origin of the sign change 
to the band inversion, which plays a key role in various topological transitions.

\begin{figure*}
\includegraphics[width=0.9\linewidth]{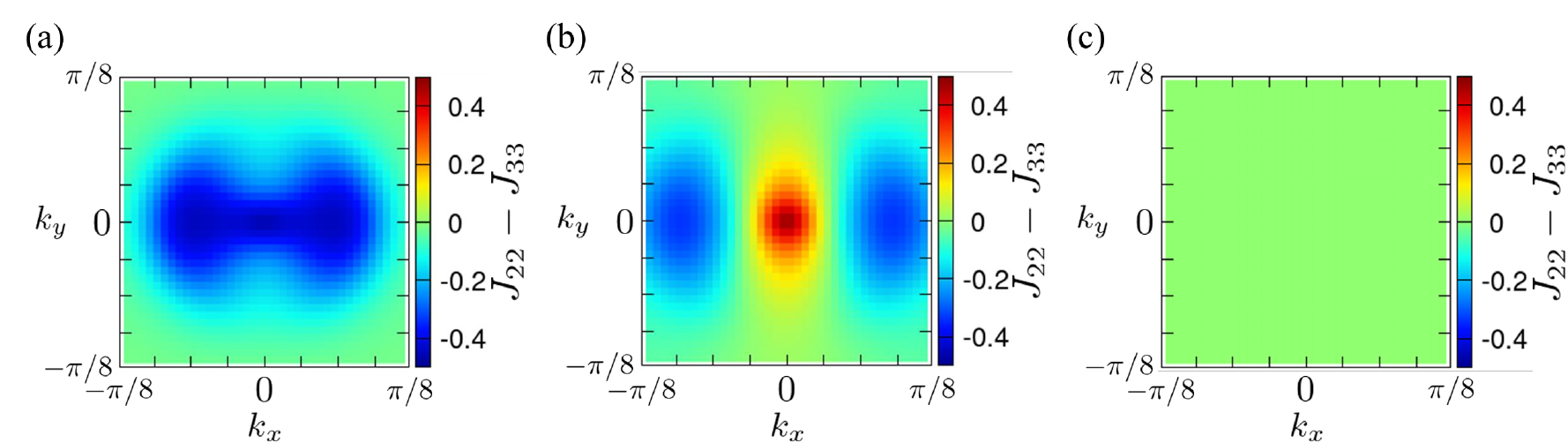}
\caption{Distribution of $J^{x}_{22}-J^{x}_{33}$ in the $\bm{k}$-space. (a) (b) We fix the angle of the magnetic field $\theta =\ang{60}$ and set the magnitude (a) $h=0.06$ and (b) $h=0.1$ below and above the topological transition, respectively. (c) We set $\theta =\ang{90}$ and $h =0.05$. We see that $J^{x}_{22}-J^{x}_{33}$ vanishes.}
\label{fig:group_velocity_k_pro}
\end{figure*}

In the low-frequency regime, the magnetic injection current originating from the B2 transition $\sigma^{x;xx}_{\mathrm{Minj(B2)}}$ gives dominant contributions to the photocurrent conductivity $\sigma^{x;xx}_{\mathrm{PC}}$. In the zero temperature limit, $\sigma^{x;xx}_{\mathrm{Minj(B2)}}$ is given by
\begin{align}
    \sigma^{x;xx}_{\mathrm{Minj(B2)}} = -\frac{\pi}{4\eta}\sum_{\bm{k}}(J^{x}_{22}-J^{x}_{33})g_{32}^{xx}\delta(\Omega -E_{32}).
\end{align}
The quantum metric $g_{32}^{xx}$ is finite and positive. Thus, the sign of $\sigma^{x;xx}_{\mathrm{Minj(B2)}}$ is determined by the difference of the paramagnetic current $J^{x}_{22}-J^{x}_{33}$. The sign of $J^{x}_{22}-J^{x}_{33}$ around $\bm{k}=\bm{0}$ is opposite between the trivial state [Fig.~\ref{fig:group_velocity_k_pro}(a)] and the topological state [Fig.~\ref{fig:group_velocity_k_pro}(b)]. This behavior causes the sign change of the photocurrent conductivity $\sigma^{x;xx}_{\mathrm{PC}}$.

The sign change of $J^{x}_{22}-J^{x}_{33}$ is caused by the band inversion. At the $\Gamma$ point ($\bm{k}=\bm{0}$), the eigenenergies of Bogoliubov quasiparticles $E_{2\bm{k}}$ and $E_{3\bm{k}}$ are obtained as
\begin{align}
    (E_{2\bm{k}}, E_{3\bm{k}}) =
    \begin{cases}
        \left(E^{(-)}, E^{(+)}\right) & \left(\sqrt{{\tilde{\mu}}^{2} +\psi^{2}} > h\right) \\
        \left(E^{(+)}, E^{(-)}\right) & \left(\sqrt{{\tilde{\mu}}^{2} +\psi^{2}} < h\right)
    \end{cases},
\end{align}
where we define $E^{(-)}$ and $E^{(+)}$ as
\begin{align}
    E^{(-)} =  -\sqrt{{\xi_{\bm{k}=\bm{0}}}^{2} + \psi^{2}} + h, \quad E^{(+)} =  \sqrt{{\xi_{\bm{k}=\bm{0}}}^{2} + \psi^{2}} - h.
\end{align}
This indicates the band inversion between the $E^{(-)}$ band and the $E^{(+)}$ band at the topological transition.
Note that the subscript $a$ of $E_{a\bm{k}}$ is defined as $E_{1\bm{k}}\leq E_{2\bm{k}} \leq 0  \leq E_{3\bm{k}} \leq E_{4\bm{k}}$. 
For the $E^{(-)}$ and $E^{(+)}$ bands, the $J^{\alpha}_{aa}$ is obtained as
\begin{align}
    J^{\alpha}_{aa} &= \partial_{\alpha}\xi_{\bm{k}} + \partial_{\alpha}\bm{g}_{\bm{k}}\cdot\hat{\bm{h}} \quad \left(E^{(-)} \text{ band}\right), \\
    J^{\alpha}_{aa} &= \partial_{\alpha}\xi_{\bm{k}} - \partial_{\alpha}\bm{g}_{\bm{k}}\cdot\hat{\bm{h}} \quad \left(E^{(+)} \text{ band}\right).
\end{align}
Thus, $J^{x}_{22} - J^{x}_{33}$ is given by
\begin{align}
    J^{x}_{22} - J^{x}_{33} = 
    \begin{cases}
        2(\partial_{x} \bm{g}_{\bm{k}}\cdot \hat{\bm{h}}) &  \left(\sqrt{{\tilde{\mu}}^{2} +\psi^{2}} > h\right) \\
        -2(\partial_{x} \bm{g}_{\bm{k}}\cdot \hat{\bm{h}}) &  \left(\sqrt{{\tilde{\mu}}^{2} +\psi^{2}} < h\right).
    \end{cases}
\end{align}
Because the $E^{(-)}$ and $E^{(+)}$ bands are inverted at the topological transition, the sign of $J^{x}_{22} - J^{x}_{33}$ changes resulting in the sign change of the photocurrent conductivity.

In the normal state, the diagonal element of the paramagnetic current operator $J^{\alpha}_{aa}$ is 
equivalent to the group velocity. With the Hellmann-Feynman theorem, the group velocity $v_{a}^{\alpha}$ is given by
$
    v^{\alpha}_{a} = \partial_{\alpha}E_{a} = J^{\alpha}_{aa}
$.
In the superconducting state, the Hellmann-Feynman relation fails in the Bogoliubov de-Gennes formalism. However, the $J^{\alpha}_{aa}$ is closely related to the group velocity of electron and hole bands in the normal state. 
Below, we discuss the origin of $J^{\alpha}_{22} - J^{\alpha}_{33}$ from the perspective of the normal state band structure.
Around $\bm{k}=\bm{0}$, the $J^{\alpha}_{aa}$ in the superconducting state is written as
\begin{align}
    J^{\alpha}_{aa} &= w_{-}v^{\alpha}_{\mathrm{e+}} - w_{+}v^{\alpha}_{\mathrm{h-}} \quad \left(E^{(-)} \text{ band}\right), \label{eq:J_alpha_aa_m} \\
    J^{\alpha}_{aa} &= w_{+}v^{\alpha}_{\mathrm{e-}} - w_{-}v^{\alpha}_{\mathrm{h+}} \quad \left(E^{(+)} \text{ band}\right) \label{eq:J_alpha_aa_p}, 
\end{align}
where we define $w_{\pm}$ as
\begin{align}
    w_{-}=\frac{-\xi_{\bm{k}} + u}{2u}, \quad w_{+}=\frac{\xi_{\bm{k}} + u}{2u},
\end{align}
with $u=\sqrt{{\xi_{\bm{k}}}^{2}+\psi^{2}}$ [Appendix \ref{app:diagonal_elements}].
We introduce $v^{\alpha}_{\mathrm{e\pm}}$ and $v^{\alpha}_{\mathrm{h\pm}}$ as the group velocity of the electron band $\epsilon^{\mathrm{e}\pm}_{\bm{k}} = \xi_{\bm{k}} \pm |\bm{g_{k}} + \bm{h}|$ and the hole band $\epsilon^{\mathrm{h}\pm}_{\bm{k}} = -\epsilon^{\mathrm{e}\pm}_{-\bm{k}}$ in the normal state, respectively. 
In the normal state ($\psi=0$), Eqs.~\eqref{eq:J_alpha_aa_m} and \eqref{eq:J_alpha_aa_p} reproduce the Hellmann-Feynman relation.
The difference  $|J^{\alpha}_{22} - J^{\alpha}_{33}|$ is rewritten with $v^{\alpha}_{\mathrm{e\pm}}$ and $v^{\alpha}_{\mathrm{h\pm}}$ as
\begin{align}
    |J^{\alpha}_{22}-J^{\alpha}_{33}|= \left|w_{-}(v^{\alpha}_{\mathrm{e+}}+v^{\alpha}_{\mathrm{h+}}) - w_{+}(v^{\alpha}_{\mathrm{e-}}+v^{\alpha}_{\mathrm{h-}})\right|.
    \label{eq:main_J22_J_33}
\end{align}
The difference between the effective currents of the electron and hole bands, which leads to $\mathrm{e}v^{\alpha}_{\mathrm{e\pm}}\neq -\mathrm{e}v^{\alpha}_{\mathrm{h\pm}}$, is essential for finite $|J^{\alpha}_{22}-J^{\alpha}_{33}|$. Note that we write the charge $\mathrm{e}=1$ explicitly here. 

From the above discussion, we notice that the nonreciprocity in the band structure, namely $\epsilon^{\mathrm{e}\pm}_{\bm{k}}\neq\epsilon^{\mathrm{e}\pm}_{-\bm{k}}$, is essential for the difference $|J^{\alpha}_{22}-J^{\alpha}_{33}|$. 
Indeed, when we assume $\epsilon^{\mathrm{e}\pm}_{\bm{k}}= \epsilon^{\mathrm{e}\pm}_{-\bm{k}}$, the group velocity of the hole bands is obtained as
\begin{align}
v^{\alpha}_{\mathrm{h}\pm}=\partial_{\alpha}(-\epsilon_{-\bm{k}}^{\mathrm{e\pm}})=-\partial_{\alpha}\epsilon_{\bm{k}}^{\mathrm{e\pm}}=-v^{\alpha}_{\mathrm{e}\pm},
\end{align}
and the difference $|J^{\alpha}_{22}-J^{\alpha}_{33}|$ vanishes as shown in Eq.~\eqref{eq:main_J22_J_33}.
The nonreciprocity $\Delta \epsilon_{\bm{k}}^{\pm}$ in the normal state is given by
\begin{align}
    \Delta \epsilon_{\bm{k}}^{\pm} \equiv \frac{1}{2}\left(\epsilon_{\bm{k}}^{\mathrm{e\pm}} - \epsilon_{-\bm{k}}^{\mathrm{e\pm}}\right) = \frac{\bm{g_{k}}\cdot\bm{h}}{\sqrt{{g_{\bm{k}}}^{2} + h^{2}}} + O\left((\bm{g_{k}}\cdot\bm{h})^{2}\right).
\end{align}
When we set the magnetic field perpendicular to the plane, the $\bm{g_{k}}$ is normal to the magnetic field in the Rashba system. Consistent with this fact, the $J^{\alpha}_{22}-J^{\alpha}_{33}$ vanishes when we set the angle $\theta =\ang{90}$ [Fig.~\ref{fig:group_velocity_k_pro}(c)].

In this section, we have shown the characteristic behavior of the photocurrent conductivity $\sigma^{x;xx}_{\mathrm{PC}}$ around the topological transition. Corresponding to the gap closing, the peak position shifts to a lower frequency with approaching the topological transition. More interestingly, we predicted the sign change of the resonant photocurrent conductivity before and after the topological transition. This sign change is due to the band inversion. Because the occupied band ($E_{2}$) and the unoccupied band ($E_{3}$) are inverted,  the sign of $J^{\alpha}_{22}-J^{\alpha}_{33}$ and that of the magnetic injection current $\sigma^{x;xx}_{\mathrm{Minj(B2)}}$ change. 
The close relationship between topological transition and photocurrent conductivity may allow the detection of topological superconductivity by optical measurements. 
The background of such a relationship is the fact that the velocity difference is an essential ingredient of the injection current. Therefore, the magnetic injection current, which is dominant in our setup, is sensitive to the band inversion, and in turn, useful for detecting the topological transition. 
A model of class DIII topological superconductor also shows a similar sign change of photocurrent conductivity between trivial and time-reversal invariant topological superconducting phases~\cite{Raj2023}. Thus, the sign change may be a universal characteristic of topological superconductors. However, the mechanism of the sign change in class DIII topological superconductors is left to be elucidated. On the other hand, the sign change in our results for class D topological superconductors is due to the magnetic injection current, which is sensitive to the band inversion at the topological transition. This mechanism of the sign-reversal phenomenon is different from the case of class DIII topological superconductors because the magnetic injection current vanishes in $\mathcal{T}$-symmetric systems. The elucidation of the mechanism of the sign change in the class DIII system is set for future study.


\section{Discussion}
\label{sec:discussion}
We have shown the superconducting nonlinear optical responses in the noncentrosymmetric $s$-wave superconductors under the magnetic field. Before summarizing the paper, in this section, we discuss the role of the magnetic field in 
the superconducting nonlinear responses from several points of view, such as symmetry, quantum geometry, and topological superconductivity. 
We also discuss candidate superconductors.
The presence of magnetic fields broadens the range of candidate materials for superconducting nonlinear optics. 

We would like to emphasize that the magnetic field or the Zeeman field induced by the ferromagnetic proximity effect plays an essential role in the existence of superconducting nonlinear responses, as discussed below. 
First, we refer to the effect of $\mathcal{T}$-symmetry breaking. In our previous work, the nonlinear responses in $\mathcal{T}$-symmetric superconductors were investigated, and it was shown that the nonlinear superconducting responses vanish in the single-band $s$-wave superconductors~\cite{Tanaka2023}. Consistent with this fact, the quantum metric $g^{\alpha\beta}_{32}$ and the matrix elements of the paramagnetic current operator $J^{\alpha}_{23}$, which correspond to the type B transition unique to superconductors, vanish in the absence of the magnetic field [Appendix~\ref{app:q_metric_without_MF}]. It is also known that the linear optical response through the transition between electron and hole bands is forbidden in the $\mathcal{T}$-symmetric single-band $s$-wave superconductors~\cite{Ahn2020}. Thus, $\mathcal{T}$-symmetry breaking is required for the emergence of linear and nonlinear superconducting responses unless unconventional Cooper pairing exists. 
Next, the crystal symmetry reduction due to the magnetic field is also essential for the nonlinear responses. In this paper, we assume the two-dimensional $\mathrm{C_{4v}}$ crystal structure. In such models with the $\mathrm{C_{2z}}$ two-fold rotation symmetry, the second-order optical responses are prohibited. In general, nonlinear responses are strictly forbidden under some crystal symmetries, regardless of the properties of Cooper pairing~\cite{Watanabe2021}. However, the magnetic field can break the crystal symmetries and induce finite nonlinear responses as we have demonstrated in this paper. 
Summarizing, the magnetic fields enable superconducting nonlinear responses to be finite even in the conventional $s$-wave superconductors by breaking $\mathcal{T}$ symmetry and crystal symmetry simultaneously. Therefore, the effect of magnetic fields significantly broadens classes of candidate materials for superconducting nonlinear responses because neither unconventional Cooper pairing such as spin-triplet pairing nor low-symmetry crystal structure is needed.

Here, we discuss platforms of the second-order optical responses in superconductors. 
Space inversion symmetry has to be broken for the second-order responses. 
There are various mechanisms of parity violation in superconductors. The most typical and ubiquitous class is the superconductors lacking the inversion symmetry in the crystal structures. Such noncentrosymmetric superconductors exist in a broad range from surfaces and heterostructures~\cite{Saito2016,Lu2015,Xi2016,delaBarrera2018,Ye2012,Reyren2007,Ueno2008,Ando2020,Sekihara2013} to bulk compounds~\cite{Bauer2012, Smidman2017}. Given the stability of superconductivity, the $s$-wave pairing is expected to be dominant in typical superconductors. Thus, our study analyzing such superconductors would be helpful for the investigation of nonlinear optics in a wide range of superconductors. 
In particular, atomically thin two-dimensional superconductors under in-plane magnetic fields are a promising setup for the observation of nonlinear responses originating from the quasiparticle dynamics, which we have discussed, because the vortex production is suppressed. 

Besides the symmetry constraints, the magnetic fields induce the characteristic contribution to the resonant photocurrent in the low-carrier-density superconductors. It appears in the low-frequency region below the superconducting gap at the zero magnetic fields. We have attributed this component of the photocurrent to the JDOS $\tilde{J}_{\mathrm{B2(ii)}}$ by decomposing the JDOS to the two components, $\tilde{J}_{\mathrm{B2(i)}}$ and $\tilde{J}_{\mathrm{B2(ii)}}$. 
Although the JDOS component $\tilde{J}_{\mathrm{B2(ii)}}$ is tiny, this low-frequency photocurrent is significantly enhanced by the quantum geometry of Bogoliubov quasiparticles and is dominant in some tensor elements of the photocurrent conductivity. 
It is known that the photocurrent generation in the normal state is closely related to the 
quantum geometric property of Dirac fermions~\cite{Hosur2011,Lyanda-Geller2015} and Weyl fermions~\cite{deJuan2017,Rees2020,Ma2017}. 
In this paper, the concept of quantum-geometrically enhanced optical responses is extended to superconductors.


One of the candidate superconductors for observing this unique photocurrent generation is FeSe thin films grown on $\mathrm{SrTiO_{3}}$(001)~\cite{Huang2017}, where an electron pocket appears at M point~\cite{He2013, Tan2013}. Although the bulk FeSe is centrosymmetric~\cite{Shibauchi_FeSe_review}, the inversion symmetry is broken in thin films due to the effects of the substrate. Indeed, a large spin-orbit coupling with the inversion symmetry breaking has been reported~\cite{Zakeri2023}. Significant quantum geometry of normal electrons has also been pointed out in FeSe~\cite{Kitamura2021,Kitamura2022a,Kitamura2022b}. Thus, it is expected that unique behaviors of photocurrent conductivity 
can be found in the superconducting state of FeSe/$\mathrm{SrTiO_{3}}$ and related materials.
The interface of superconductors and topological insulators is another intriguing platform, where the interplay between superconductivity and surface Dirac/Weyl states has been recognized~\cite{Yasuda2019,Tobias2012, Erratum_Tobias2012}. 
Transition metal dichalcogenides~\cite{Saito2016, Lu2015, Xi2016, delaBarrera2018, Fatemi2018, Sajadi2018} are also expected to show giant superconducting nonlinear optical responses due to pronounced quantum geometry.
In particular, monolayer $\mathrm{WTe_{2}}$ is a centrosymmetric topological insulator~\cite{Wu2018, Fei2017}, which can be gate-tuned to the superconducting state~\cite{Fatemi2018,Sajadi2018}. The electrically induced nonlinear response has been demonstrated in the topological insulating state~\cite{Xu2018}, while the effect of nontrivial topology and quantum geometry in the superconducting state has not been uncovered.
The evaluation of linear and nonlinear optical responses in these fascinating superconductors will be a future work of interest.

Finally, we comment on the vertex correction for the optical conductivity in the superconductors. Vertex correction is a key step in maintaining the gauge invariance and consistency with the gap equation~\cite{Huang2023, Papaj2022, Dai2017, Nambu1960}. The vertex correction method has been adopted in the research on the linear and nonlinear responses of superconductors, where the $\mathcal{P}$ symmetry is broken by the finite Cooper pairs' momentum~\cite{Huang2023, Papaj2022, Dai2017}. As a result of the vertex correction, the magnitude of the second-order responses is reduced, and some components reverse the sign, while the frequency dependence is qualitatively unchanged~\cite{Huang2023}. Therefore, the characteristic frequency dependence of our results is expected to appear even after the vertex correction. The contribution from collective modes is included in the vertex correction. How the contribution of collective modes changes before and after the topological transition will be interesting future work.


\section{Summary}
\label{sec:summary}
This paper elaborated on the second-order nonlinear optical responses of noncentrosymmetric superconductors in the magnetic field. We showed that the magnetic field is essential for the nonlinear responses in broad classes of superconductors. In particular, $\mathcal{T}$-symmetry breaking nullifies the selection rule forbidding the optical transition between electron and hole bands in $\mathcal{T}$-symmetric $s$-wave superconductors. Therefore, the nonlinear responses due to Bogoliubov quasiparticles can occur in 
conventional superconductors under magnetic fields. In addition, we found a new component of superconducting photocurrent generation that emerges under the magnetic field and is closely related to the Dirac point of normal electrons. We ascribed the characteristic dependence on magnetic fields and chemical potentials to this component. This photocurrent characterizes low-carrier-density superconductors which are candidates for class D topological superconductors. 



The low-carrier-density Rashba superconductors have been intensively studied for the realization and detection of topological superconductivity. 
However, it has been challenging to search for signatures in bulk properties because the DOS and JDOS resulting from gap closing are tiny and difficult to observe.
However, we have shown that the characteristic photocurrent generation is significantly enhanced by quantum geometry arising from the ${\bm k}$-space around the Dirac point.
The band-resolved quantum metric of Bogoliubov quasiparticles inherits peculiar geometric properties of Dirac electrons. 
The photocurrent originating from the low-energy quasiparticles due to gap closing can be much larger than other components. 
In this sense, the optical response is a bulk property sensitive to the topological transition in superconductors. 
Furthermore, the sign reversal of the photocurrent conductivity occurs associated with the gap closing and band inversion at the topological transition. The longitudinal component of linear conductivity does not show the sign reversal and this characteristic behavior is unique to the nonlinear responses.
Therefore, the nonlinear superconducting optics can provide a bulk probe for the detection of topological superconductivity in Rashba systems. 
The nonlinear responses yield rich information on attractive materials. Exploring other superconducting nonlinear responses will also be a promising route for future research.

\begin{acknowledgments}
This work is supported by JSPS KAKENHI (Grants Numbers JP21K18145,
JP22H01181,  JP22H04933, JP23K17353).
H.T. is supported by JSPS KAKENHI (Grants Number JP23KJ1344).
H.W. is supported by JSPS KAKENHI (Grants Number JP23K13058).
\end{acknowledgments}

\appendix
\onecolumngrid

\section{Finite-momentum Cooper pairs in the superconducting state}
\label{app:finite_momentum}

In the calculation of the electric current under the irradiation of light, we need to determine the momentum of Cooper pairs in the equilibrium state. 
In the setup of our model, both $\mathcal{T}$ and $\mathcal{P}$ symmetries are broken, and therefore, the Cooper pairs may acquire total momentum $2\bm{q}$, which is determined by minimizing the free energy $F$ in Eq.~\eqref{eq:free_energy}. To this end, we calculate the $q_{x}$ dependence of the free energy with the fixed chemical potential $\mu$ and magnetic field $\bm{h}$. The free energy $F$ is symmetric with respect to $q_{y}\leftrightarrow -q_{y}$ because of the $\mathcal{T}\mathrm{m_{y}}$ symmetry at $h_{x}=0$. Thus, the free energy $F$ is expected to have a minimum at $q_{y}=0$ when the higher order terms of $q_{y}$ are suppressed.  In our numerical results, $F$ actually has the minimum at $q_{y}=0$. We substitute the obtained $\bm{q}$ to Eq.~\eqref{eq:tot_Res} and numerically calculate the photocurrent conductivity and second harmonic generation coefficient.

\section{Energy of Bogoliubov quasiparticles and asymmetry}
\label{app:energy_appro_asym}

When the condition $\bm{g_{k}}\cdot \bm{h}=0$ is satisfied, the eigenenergies of the Bogoliubov-de Gennes Hamiltonian are exactly obtained as 
\begin{align}
E_{\bm{k}} = \pm\sqrt{{\xi_{\bm{k}}}^{2} + {g_{\bm{k}}}^{2} + h^{2} + \psi^{2} \pm^{\prime} 2\sqrt{{\xi_{\bm{k}}}^{2}({g_{\bm{k}}}^{2} + h^{2}) + \psi^{2}h^{2}}}.
\label{eq:exact_E}
\end{align}
The detailed calculation is presented in Appendix \ref{app:appro_JDOS}. This formula satisfies the relation $E_{a\bm{k}}=E_{a-\bm{k}}$. However, this symmetry is broken when the inner product of $\bm{g}$ and $\bm{h}$ is finite. 
By neglecting the asymmetry, we approximate the energy dispersion by Eq.~\eqref{eq:exact_E} in the main text for the discussion of JDOS [Sec.~\ref{sec:approximate_JDOS}]. Equation~\eqref{eq:exact_E} is equivalent to Eq.~\eqref{eq:approximate_energy}. Here, we show the correction term resulting from $\bm{g_{k}}\cdot \bm{h}$ and discuss its irrelevance to the optical response.

In the normal state, the eigenenergies $\epsilon_{\bm{k}}$ are given as
\begin{align}
    \epsilon_{\bm{k}} = \xi_{\bm{k}} \pm |\bm{g_{k}} + \bm{h}|.
\end{align}
We define the asymmetric component of $\epsilon_{\bm{k}}$ as 
\begin{align}
    \Delta \epsilon_{\bm{k}} = \frac{1}{2}\left(\epsilon_{\bm{k}} - \epsilon_{-\bm{k}}\right) = \frac{1}{2}\left(|\bm{g_{k}} + \bm{h}|-|-\bm{g_{k}} + \bm{h}|\right).
\end{align}
When the g-vector $\bm{g_{\bm{k}}}$ is perpendicular to the magnetic field $\bm{h}$, $\Delta \epsilon_{\bm{k}}$ is zero. Therefore, $\Delta \epsilon_{\bm{k}}$ is expected to be the source of the correction term in the Bogoliubov quasiparticles' energies.

We focus on the eigenenergies $E_{2\bm{k}}$ and $E_{3\bm{k}}$ because the transition between the $E_{2\bm{k}}$ and $E_{3\bm{k}}$ bands give dominant nonlinear responses in the low-frequency region. For finite $\bm{g_{k}}\cdot \bm{h}$, we can approximate the energy bands as $E_{2\bm{k}} \simeq \tilde{E}_{2\bm{k}}$ and $E_{3\bm{k}} \simeq \tilde{E}_{3\bm{k}}$ with
\begin{align}
    \tilde{E}_{2\bm{k}} &= -\sqrt{{\xi_{\bm{k}}}^{2} + {g_{\bm{k}}}^{2} + h^{2} + \psi^{2} - 2\sqrt{{\xi_{\bm{k}}}^{2}({g_{\bm{k}}}^{2} + h^{2}) + \psi^{2}h^{2}}} - \frac{\xi_{\bm{k}}}{\sqrt{{\xi_{\bm{k}}}^{2} + \psi^{2}}}\Delta \epsilon_{\bm{k}},
\label{eq:apendix_app_energy2}
    \\
    \tilde{E}_{3\bm{k}} &= \sqrt{{\xi_{\bm{k}}}^{2} + {g_{\bm{k}}}^{2} + h^{2} + \psi^{2} - 2\sqrt{{\xi_{\bm{k}}}^{2}({g_{\bm{k}}}^{2} + h^{2}) + \psi^{2}h^{2}}} - \frac{\xi_{\bm{k}}}{\sqrt{{\xi_{\bm{k}}}^{2} + \psi^{2}}}\Delta \epsilon_{\bm{k}}.
\label{eq:apendix_app_energy3}
\end{align}
Note that $\tilde{E}_{2\bm{k}}$ and $\tilde{E}_{3\bm{k}}$ are related to each other under the charge-conjugation transformation 
$\tilde{E}_{2\bm{k}} = -\tilde{E}_{3-\bm{k}}$.


\begin{figure*}
\includegraphics[width=\linewidth]{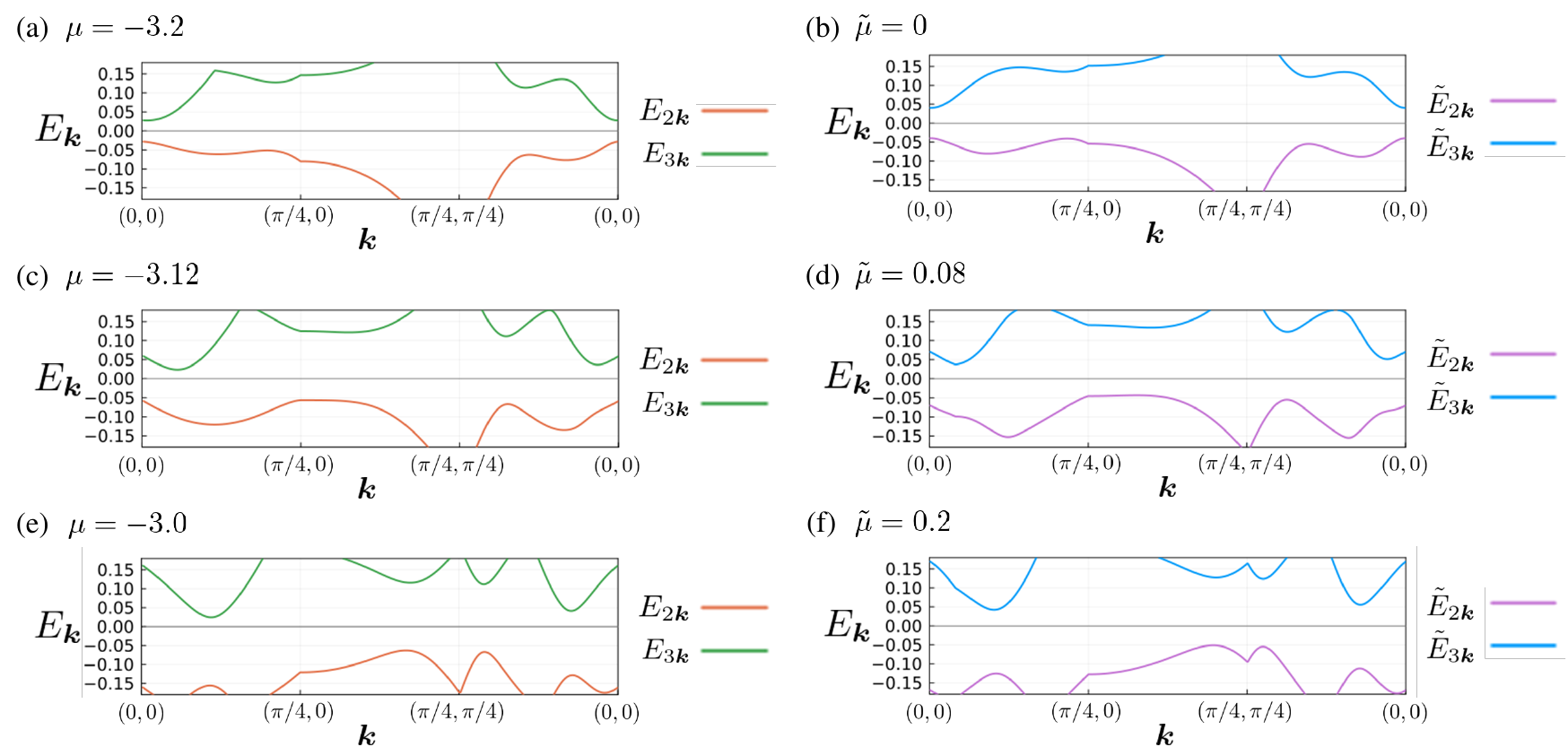}
\caption{(Left panels) The energy bands $E_{2\bm{k}}$ and $E_{3\bm{k}}$ around $\bm{k}=\bm{0}$ for the chemical potential (a) $\mu =-3.2$, (c) $-3.12$, and (e) $-3.0$. (Right panels) The approximated energy bands $\tilde{E}_{2\bm{k}}$ and $\tilde{E}_{3\bm{k}}$ around $\bm{k}=\bm{0}$. The effective chemical potential in (b) $\tilde{\mu}=0$, (d) $0.08$, and (f) $0.2$ correspond to (a) $\mu=-3.2$, (c) $-3.12$, and (e) $-3.0$, respectively. We adopt Eqs.~\eqref{eq:simple_xi_2} and \eqref{eq:simple_g_2} and set $t=0.6$ consistent with the tight-binding parameters $t_{1}=1$ and $t_{2}=0.2$.}
\label{fig:bands_E2_E3}
\end{figure*}

Next, we compare the exact eigenenergies and the approximation formulae, Eqs.~\eqref{eq:apendix_app_energy2} and \eqref{eq:apendix_app_energy3}. Figures \ref{fig:bands_E2_E3}(a), \ref{fig:bands_E2_E3}(c), and \ref{fig:bands_E2_E3}(e) show the numerically obtained exact eigenenergies $E_{2\bm{k}}$ and $E_{3\bm{k}}$ around $\bm{k}=\bm{0}$. The approximate formulae $\tilde{E}_{2\bm{k}}$ and $\tilde{E}_{3\bm{k}}$ reproduce the energy bands, as shown in Figs.~\ref{fig:bands_E2_E3}(b), \ref{fig:bands_E2_E3}(d), and \ref{fig:bands_E2_E3}(f) which correspond to Figs.~\ref{fig:bands_E2_E3}(a), \ref{fig:bands_E2_E3}(c), and \ref{fig:bands_E2_E3}(e), respectively. Note that we assume Eqs.~\eqref{eq:simple_xi_2} and \eqref{eq:simple_g_2} in the calculation of $\tilde{E}_{2\bm{k}}$ and $\tilde{E}_{3\bm{k}}$.

In the main text, we focus on the optical transition between $E_{2\bm{k}}$ and $E_{3\bm{k}}$, which we call the B2 transition. The JDOS corresponding to the B2 transition is defined as
\begin{align}
    J_{\mathrm{B2}}(\Omega) = \sum_{\bm{k}} \delta(\Omega - E_{3\bm{k}} + E_{2\bm{k}}).
\end{align}
Although the correction term shifts the energy bands $E_{2\bm{k}}$ and $E_{3\bm{k}}$, the difference $E_{3\bm{k}} - E_{2\bm{k}}$ is not affected by the correction, since the correction terms are equivalent between Eqs.~\eqref{eq:apendix_app_energy2} and \eqref{eq:apendix_app_energy3}. Thus, the effect of energy asymmetry 
is negligible when we analyze the JDOS $J_{\mathrm{B2}}(\Omega)$.

\section{Calculation of the JDOS}
\label{app:appro_JDOS}

\subsection{Formula of the Bogoliubov quasiparticles' energy}
In this Appendix, we discuss the JDOS with approximate calculations. First, we approximately derive the analytic form of energy eigenvalues. We ignore the total momentum of Cooper pairs $2\bm{q}$ for simplicity. The BdG Hamiltonian is given by
\begin{align}
H_{\mathrm{BdG}}(\bm{k})=
    \begin{pmatrix}
    H_{\mathrm{N}}(\bm{k}) & \psi(i\sigma_{y}) \\
    \psi(i\sigma_{y})^{\top} & -H_{\mathrm{N}}(-\bm{k})^{\top}
    \end{pmatrix}
    ,
\end{align}
where $H_{\mathrm{N}}(\bm{k}) = \xi_{\bm{k}} + \left(\bm{g}_{\bm{k}}+\bm{h}\right)\cdot\bm{\sigma}$ is the Hamiltonian in the normal state, and $\psi$ is the pair potential of the spin-singlet pairing. We obtain ${H_{\mathrm{BdG}}}^{2}$ as
\begin{align}
\label{eq:square_BdG}
{\mathcal{H}_{\mathrm{BdG}}}^{2} = {\xi_{\bm{k}}}^{2} + {g_{\bm{k}}}^{2} + \bm{h}^{2} + 2\bm{g_{k}}\cdot\bm{h}\tau_{\mathrm{z}} + \psi^{2} + 2 
\begin{pmatrix}
\xi_{\bm{k}}(\bm{g_{k}}+\bm{h})\cdot\bm{\sigma} & \psi(\bm{h}\cdot\bm{\sigma})i\sigma_{\mathrm{y}} \\*
\psi(-i\sigma_{\mathrm{y}})(\bm{h}\cdot\bm{\sigma}) & -\xi_{\bm{k}}(\bm{\tilde{g}_{k}}-\tilde{\bm{h}})\cdot\bm{\sigma}
\end{pmatrix},
\end{align}
where the Pauli matrices $\tau_{\alpha}$ represent the particle-hole degree of freedom in the Nambu space and the notation $\tilde{\bm{a}}$ means $\Tilde{\bm{a}}=(a_{x}, -a_{y}, a_{z})^{\top}$. Here, we define $A$, $B$, and $C$ as $A=\xi_{\bm{k}}(\bm{g_{k}}+\bm{h})\cdot\bm{\sigma}$, $B=\psi(\bm{h}\cdot\bm{\sigma})i\sigma_{\mathrm{y}}$, and $C=-\xi_{\bm{k}}(\tilde{\bm{g_{k}}}-\tilde{\bm{h}})\cdot\bm{\sigma}$ and consider the square of the last term in Eq.~\eqref{eq:square_BdG}, 
\begin{align}
\left({\mathcal{H}_{\mathrm{BdG}}}^{2} - {\xi_{\bm{k}}}^{2} - {g_{\bm{k}}}^{2} - \bm{h}^{2} - 2\bm{g_{k}}\cdot\bm{h}\tau_{\mathrm{z}} - \psi^{2}\right)^{2} 
= 4
\begin{pmatrix}
A^{2} + BB^{\dagger} & AB+BC \\*
B^{\dagger}A+CB^{\dagger} & B^{\dagger}B + C^{2}
\end{pmatrix}.
\end{align}
The matrix elements are obtained as
\begin{align}
A^{2} + BB^{\dagger} = {\xi_{\bm{k}}}^{2}(\bm{g_{k}}+\bm{h})^{2} + \psi^{2}h^{2}, \\
C^{2} + B^{\dagger}B = {\xi_{\bm{k}}}^{2}(\bm{g_{k}}-\bm{h})^{2} + \psi^{2}h^{2}, \\
AB + BC = 2\xi_{\bm{k}}\psi(\bm{g_{k}}\cdot\bm{h})i\sigma_{\mathrm{y}}.
\end{align}

In the following discussion, we approximate $\bm{g_{k}}\cdot \bm{h}\sim 0$. Then, we obtain the relation
\begin{align}
\left({\mathcal{H}_{\mathrm{BdG}}}^{2} - {\xi_{\bm{k}}}^{2} - {g_{\bm{k}}}^{2} - \bm{h}^{2} - \psi^{2}\right)^{2} 
= 4\left[{\xi_{\bm{k}}}^{2}({\bm{g}_{\bm{k}}}^{2}+\bm{h}^{2})+\psi^{2}h^{2}\right],
\end{align}
and the energy eigenvalues 
as 
\begin{align}
E_{\bm{k}} = \pm\sqrt{{\xi_{\bm{k}}}^{2} + {g_{\bm{k}}}^{2} + h^{2} + \psi^{2} \pm^{\prime} 2\sqrt{{\xi_{\bm{k}}}^{2}({g_{\bm{k}}}^{2} + h^{2}) + \psi^{2}h^{2}}},
\label{eq:appro_E}
\end{align}
which is equivalent to Eq.~\eqref{eq:exact_E}.
When the g-vector of spin-orbit coupling is perpendicular to the magnetic field, $\bm{g_{k}}\cdot \bm{h}=0$, the formula 
Eq.~\eqref{eq:appro_E} is exact. 
The correction term resulting from $\bm{g_{k}}\cdot \bm{h}$ has been discussed in Appendix~\ref{app:energy_appro_asym} and shown to be unimportant for the optical responses.
Below, we discuss two simple cases where Eq.~\eqref{eq:appro_E} gives exact results.

\subsubsection{Zero magnetic field}
When the magnetic field is absent, Eq.~\eqref{eq:appro_E} is rewritten as
\begin{align}
E_{\bm{k}} = \pm\sqrt{(|\xi_{\bm{k}}|\pm^{\prime}g_{\bm{k}})^{2} + \psi^{2} } = \pm\sqrt{(\xi_{\bm{k}}\pm^{\prime}g_{\bm{k}})^{2} + \psi^{2} }.
\label{eq:appendix_zeromagneticfield}
\end{align}
where we redefine $\pm^{\prime}$ with corresponding to the sign of $\xi_{\bm{k}}$ at the second equality. Because $\bm{g_{k}}\cdot \bm{h}$ is zero due to the absence of the magnetic field $\bm{h}=\bm{0}$, the energy eigenvalues $E_{\bm{k}}$ are exact in this case. Indeed, the formula Eq.~\eqref{eq:appendix_zeromagneticfield} is known as Bogoliubov bands in the noncentrosymmetric spin-singlet superconductors~\cite{Bauer2012}.

\subsubsection{Spin-orbit coupling free systems}
When the anti-symmetric spin-orbit coupling is absent $\bm{g_{k}}=\bm{0}$, Eq.~\eqref{eq:appro_E} is rewritten as 
\begin{align}
E_{\bm{k}} &= \pm\sqrt{{\xi_{\bm{k}}}^{2} + {h}^{2} + \psi^{2} \pm^{\prime} 2h\sqrt{{\xi_{\bm{k}}}^{2}+\psi^{2}}} \\*
&=\pm\left|\sqrt{{\xi_{\bm{k}}}^{2}+\psi^{2}}\pm^{\prime} h\right|.
\label{eq:appendix_spinorbitcouplingfree}
\end{align}
Because $\bm{g_{k}}\cdot \bm{h}$ vanishes due to the absence of the anti-symmetric spin-orbit coupling $\bm{g_{k}}=\bm{0}$, the energy eigenvalues $E_{\bm{k}}$ are exact. Indeed, Eq.~\eqref{eq:appendix_spinorbitcouplingfree} is the energy of Bogoliubov quasiparticles in the spin-orbit coupling free systems.
It should be noticed that the energy of Bogoliubov quasiparticles at $\bm{k}=\bm{0}$ is exactly given by Eq.~\eqref{eq:appro_E} because $\bm{g_{k=0}}=\bm{0}$.

\subsection{Approximation for analytic calculations of the JDOS}

Here, we calculate the JDOS $J_{\mathrm{B2}}(\Omega)$, which is essential for the superconducting nonlinear optical responses. The definition is given by 
\begin{align}
    J_{\mathrm{B2}}(\Omega)=\sum_{\bm{k}}\delta(\Omega - E_{3\bm{k}} + E_{2\bm{k}}),
\end{align}
where $E_{1\bm{k}}$, $E_{2\bm{k}}$, $E_{3\bm{k}}$, and $E_{4\bm{k}}$ are eigenenergies of Bogoliubov quasiparticles $(E_{1\bm{k}}\leq E_{2\bm{k}} \leq E_{3\bm{k}} \leq E_{4\bm{k}})$ in the superconducting state.
We start from Eq.~\eqref{eq:appro_E} which is not exact when $\bm{g_k}\cdot\bm{h} \ne 0$. However, when we calculate the difference between $E_{2\bm{k}}$ and $E_{3\bm{k}}$, we can ignore the effect of $\bm{g_k}\cdot\bm{h}$ because the effect cancels each other out between $E_{2\bm{k}}$ and $E_{3\bm{k}}$. The detailed discussion has been presented in Appendix~\ref{app:energy_appro_asym}.
Here, we adopt further approximation for the Bogoliubov quasiparticle bands by considering the cases of ${\xi_{\bm{k}}}^{2}{g_{\bm{k}}}^{2} \gg \psi^{2}h^{2}$ and $\psi^{2}h^{2} \gg {\xi_{\bm{k}}}^{2}{g_{\bm{k}}}^{2}$. 

First, we discuss the former case. Under the condition ${\xi_{\bm{k}}}^{2}{g_{\bm{k}}}^{2} \gg \psi^{2}h^{2}$, we can evaluate
\begin{align}
2\sqrt{{\xi_{\bm{k}}}^{2}({g_{\bm{k}}}^{2} + h^{2}) + \psi^{2}h^{2}} \simeq 2|\xi_{\bm{k}}|\sqrt{{g_{\bm{k}}}^{2} + h^{2}},
\end{align}
and thus, obtain the approximate energies $\tilde{E}_{\bm{k}}$ as 
\begin{align}
\tilde{E}_{\pm\bm{k}} \simeq& \pm\sqrt{\left(|\xi_{\bm{k}}|\pm^{\prime}\sqrt{{g_{\bm{k}}}^{2} + h^{2}}\right)^{2}+\psi^{2}}.
\end{align}
Under the charge-conjugation transformation, the energy of the band $\tilde{E}_{\bm{k}}$ changes to $-\tilde{E}_{-\bm{k}}$.
Here, we consider the Bogoliubov quasiparticle band of $E_{3\bm{k}}$
\begin{align}
\tilde{E}_{3\bm{k}} = \sqrt{\left(|\xi_{\bm{k}}|-\sqrt{{g_{\bm{k}}}^{2} + h^{2}}\right)^{2}+\psi^{2}}.
\end{align}
Because of the charge-conjugation relation, we have $E_{3\bm{k}} =  -E_{2-\bm{k}}$, and thus, the energy difference in the transition between the particle and hole bands is given by 
\begin{align}
\Delta \tilde{E}_{\bm{k}}  
\equiv \tilde{E}_{3\bm{k}} - \tilde{E}_{2\bm{k}}
= 2\sqrt{\left(|\xi_{\bm{k}}|-\sqrt{{g_{\bm{k}}}^{2} + h^{2}}\right)^{2}+\psi^{2}}.
\label{eq:appro_energy_diff_1}
\end{align}

Next, we evaluate the energy eigenvalues in another case, $\psi^{2}h^{2} \gg {\xi_{\bm{k}}}^{2}{g_{\bm{k}}}^{2}$. In this case, we have 
\begin{align}
2\sqrt{{\xi_{\bm{k}}}^{2}({g_{\bm{k}}}^{2} + h^{2}) + \psi^{2}h^{2}} \simeq 2h\sqrt{{\xi_{\bm{k}}}^{2} + \psi^{2}},
\end{align}
and therefore, we estimate $\tilde{E}_{\bm{k}}$ as
\begin{align}
\tilde{E}_{\pm\bm{k}} \simeq  \pm\sqrt{{\xi_{\bm{k}}}^{2} + {g_{{\bm k}}}^{2} + h^{2} + \psi^{2} \pm^{\prime}2h\sqrt{{\xi_{\bm{k}}}^{2} + \psi^{2}}}.
\end{align}
For the particle band of $E_{3\bm{k}}$, the energy is approximated as 
\begin{align}
\tilde{E}_{3\bm{k}} \simeq& \sqrt{{\xi_{\bm{k}}}^{2} + {g_{{\bm k}}}^{2} + h^{2} + \psi^{2} - 2h\sqrt{{\xi_{\bm{k}}}^{2} + \psi^{2}}}.
\end{align}
The energy difference in the transition between the particle and hole bands is given by
\begin{align}
\Delta \tilde{E}_{\bm{k}} 
\equiv \tilde{E}_{3\bm{k}} - \tilde{E}_{2\bm{k}}
= \tilde{E}_{3\bm{k}}+\left(-\tilde{E}_{3-\bm{k}}\right) = 2\sqrt{{\xi_{\bm{k}}}^{2} + {g_{\bm k}}^{2} + h^{2} + \psi^{2} - 2h\sqrt{{\xi_{\bm{k}}}^{2} + \psi^{2}}}.
\label{eq:appro_energy_diff_2}
\end{align}

\subsection{Approximate JDOS}

We introduce the approximate JDOS based on the approximate energy difference as
\begin{align}
    \tilde{J}_{\mathrm{B2(i)}}(\Omega) &= \sum_{\bm{k}} \delta \left(\Omega - 2\sqrt{\left(|\xi_{\bm{k}}|-\sqrt{{g_{\bm{k}}}^{2} + h^{2}}\right)^{2}+\psi^{2}}\right), \label{eq:appro_B_i}\\
    \tilde{J}_{\mathrm{B2(ii)}}(\Omega) &= \sum_{\bm{k}} \delta \left(\Omega - 2\sqrt{{\xi_{\bm{k}}}^{2} + {g_{\bm k}}^{2} + h^{2} + \psi^{2} - 2h\sqrt{{\xi_{\bm{k}}}^{2} + \psi^{2}}}\right), \label{eq:appro_B_ii_appendix}
\end{align}
which correspond to Eqs.~\eqref{eq:appro_energy_diff_1} and \eqref{eq:appro_energy_diff_2}, respectively. We have introduced the assumption ${\xi_{\bm{k}}}^{2}{g_{\bm{k}}}^{2} \gg \psi^{2}h^{2}$ and $\psi^{2}h^{2} \gg {\xi_{\bm{k}}}^{2}{g_{\bm{k}}}^{2}$ for derivation of the energy difference between transition bands.   
These assumptions are satisfied in a part of the Brillouin zone. 
While in the main text the integration is carried out in the appropriate momentum space satisfying the assumption, here we do not restrict the momentum for the integration. 
However, the following evaluation of $\tilde{J}_{\mathrm{B2(i)}}$ and $\tilde{J}_{\mathrm{B2(ii)}}$ takes into account the essential contributions.
The approximate JDOS $\tilde{J}_{\mathrm{B2(i)}}$ and $\tilde{J}_{\mathrm{B2(ii)}}$ are helpful to discuss the behavior of the photocurrent conductivity (Sec.~\ref{sec:Numerical_results}). Below, we derive analytic formulae of these contributions and explain their characteristic behaviors.

\subsubsection{Analytical formula of B2(i) contribution}

\begin{figure*}
\includegraphics[width=\linewidth]{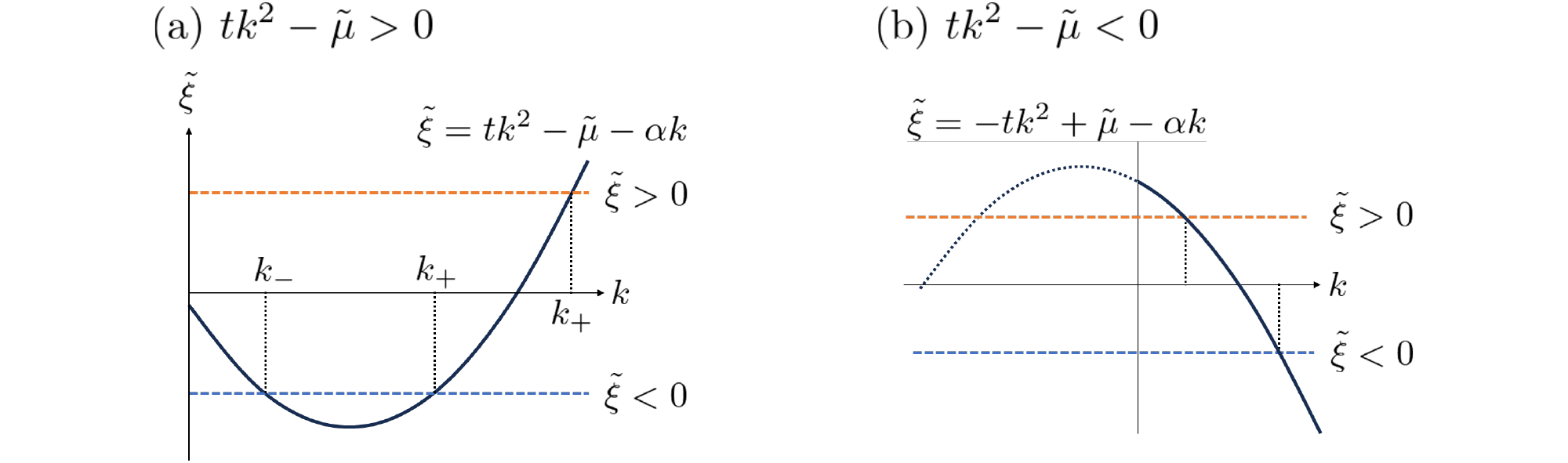}
\caption{Schematic illustration of the $k = |\bm{k}|$ which contribute to $\tilde{J}_{\mathrm{B2(i)}}(\Omega)$ when the magnetic field is sufficiently small. We assume that $\tilde{\mu}$ is positive. (a) The case of $tk^{2} -\tilde{\mu} > 0$. We here consider the positive and negative solutions of $\tilde{\xi}$, namely, $\tilde{\xi}=\pm\sqrt{(\Omega/2)^{2}+\psi^{2}}$. For the negative $\tilde{\xi}<0$, the equation $\tilde{\xi} = tk^{2} - \tilde{\mu} -\alpha k$ has the two solutions $k=k_{+}$ and $k=k_{-}$, 
which correspond to $D_{\tilde{\xi}_{+} < 0}(\Omega)$ and $D_{\tilde{\xi}_{-} < 0}(\Omega)$. For the positive $\tilde{\xi}>0$, there is only one solution $k=k_{+}$, which corresponds to $D_{\tilde{\xi}_{+} > 0}(\Omega)$. (b) The solution of $k$ for positive and negative $\tilde{\xi}$ when the condition $tk^{2} -\tilde{\mu} < 0$ is satisfied. There is only one solution of $k$ for a given $\tilde{\xi}$. $G_{\tilde{\xi} > 0}$ and $G_{\tilde{\xi} < 0}$ correspond to the positive case $\tilde{\xi}>0$ and the negative case $\tilde{\xi}<0$, respectively.}
\end{figure*}

First, we discuss the JDOS component $\tilde{J}_{\mathrm{B2(i)}}$. 
From the similarity to the JDOS in conventional superconductors, we expect that $\tilde{J}_{\mathrm{B2(i)}}(\Omega)$ has a peak around $\Omega =2\psi$. Indeed, the JDOS $J_{\mathrm{B2}}$ and the approximate JDOS $\tilde{J}_{\mathrm{B2(i)}}$ share the peak around $\Omega =2\psi$ in Fig.~\ref{fig:JDOS2}.
To analyze $\tilde{J}_{\mathrm{B2(i)}}$ further, we introduce an additional assumption $g_{\bm{k}} \gg h$, by which we have
\begin{align}
    \tilde{J}_{\mathrm{B2(i)}}(\Omega) \simeq \sum_{\bm{k}}\delta\left(\Omega -2\sqrt{\left(|\xi_{\bm{k}}|-g_{\bm{k}}\right)^{2}+\psi^{2}}\right).
\end{align}
Below, we discuss the behaviors of $\tilde{J}_{\mathrm{B2(i)}}(\Omega)$ by analytic calculations. We approximate the normal Hamiltonian with a focus on the momentum space around the $\Gamma$ point ($\bm{k}=\bm{0}$) as
\begin{align}
    \xi(\bm{k}) &= t({k_{x}}^{2} + {k_{y}}^{2}) - \tilde{\mu} \quad (t>0, \tilde{\mu}\geq 0), \label{eq:simple_xi}\\
    \bm{g}(\bm{k}) &= (\alpha k_{y}, -\alpha k_{x}, 0) \quad (\alpha >0). \label{eq:simple_g}
\end{align}
The energy difference is obtained as 
\begin{align}
    \Delta \tilde{E}_{\bm{k}} = 2\sqrt{{\tilde{\xi}_{\bm{k}}}^{2} + \psi^{2}},
\end{align}
where we define 
$\tilde{\xi}_{\bm{k}} \equiv |tk^{2} - \tilde{\mu}| -\alpha k$. 

First, we assume $tk^{2} -\tilde{\mu} >0$ and $\tilde{\xi}_{\bm{k}} = tk^{2} - \tilde{\mu} -\alpha k$.
We have the relation between the infinitesimal changes $d\tilde{\xi}$ and $d\tilde{E}$ 
\begin{align}
    d\Delta\tilde{E} = \frac{2|\tilde{\xi}|}{\sqrt{{\tilde{\xi}}^{2} + \psi^{2}}}d\tilde{\xi} = \frac{2\sqrt{(\Delta\tilde{E})^{2} - (2\psi)^{2}}}{\Delta \tilde{E}} d\tilde{\xi}.
\end{align}
Solutions of the equation $\tilde{\xi} = tk^{2} - \tilde{\mu} -\alpha k$ for $k$ are given by 
\begin{align}
    k_{\pm} = \frac{\alpha \pm \sqrt{\alpha^{2} + 4t(\tilde{\mu} + \tilde{\xi})}}{2t}.
\end{align}
The infinitesimal changes of $\tilde{\xi}$ at $k=k_\pm$ are denoted as 
$d\tilde{\xi}_{\pm}$ and given by
\begin{align}
    d\tilde{\xi}_{+} = \frac{\sqrt{\alpha^{2} + 4t(\tilde{\mu} + \tilde{\xi})}}{\alpha + \sqrt{\alpha^{2} + 4t(\tilde{\mu} + \tilde{\xi})}}\cdot 2tk_{+}dk_{+}, \\
    d\tilde{\xi}_{-} = \frac{\sqrt{\alpha^{2} + 4t(\tilde{\mu} + \tilde{\xi})}}{\alpha - \sqrt{\alpha^{2} + 4t(\tilde{\mu} + \tilde{\xi})}}\cdot 2tk_{-}dk_{-}.
\end{align}
Then, $dk_{\pm}$ satisfy the relations
\begin{align}
    2\pi k_{+}dk_{+} = \frac{\pi}{2t}\frac{\alpha + \sqrt{\alpha^{2} + 4t(\tilde{\mu} + \tilde{\xi}_{+})}}{\sqrt{\alpha^{2} + 4t(\tilde{\mu} + \tilde{\xi}_{+})}}\cdot \frac{\Delta \tilde{E}}{\sqrt{(\Delta \tilde{E})^{2} - (2\psi)^{2}}} d\Delta \tilde{E}, \\
    2\pi k_{-}dk_{-} = \frac{\pi}{2t}\frac{\alpha - \sqrt{\alpha^{2} + 4t(\tilde{\mu} + \tilde{\xi}_{-})}}{\sqrt{\alpha^{2} + 4t(\tilde{\mu} + \tilde{\xi}_{-})}}\cdot \frac{\Delta \tilde{E}}{\sqrt{(\Delta \tilde{E})^{2} - (2\psi)^{2}}} d\Delta \tilde{E}.
\end{align}
Given $tk^{2}-\tilde{\mu}>0$, we obtain the range of $k$ as $k>\sqrt{\frac{\tilde{\mu}}{t}}$. Thus, the range of $\tilde{\xi}_{\pm}$ is
\begin{align}
    &\begin{cases}
        \tilde{\xi}_{+} \geq -\frac{\alpha^{2}}{4t} -\tilde{\mu} & (0 \leq \tilde{\mu} \leq \frac{\alpha^{2}}{4t}) \\
        \tilde{\xi}_{+} \geq -\alpha\sqrt{\frac{\tilde{\mu}}{t}} & (\frac{\alpha^{2}}{4t} < \tilde{\mu})
    \end{cases}, \\
    &\begin{cases}
        -\frac{\alpha^{2}}{4t} -\tilde{\mu} \leq \tilde{\xi}_{-} \leq -\alpha\sqrt{\frac{\tilde{\mu}}{t}} & (0 \leq \tilde{\mu} \leq \frac{\alpha^{2}}{4t}) \\
        \text{no }\tilde{\xi}_{-} \text{ satisfies the conditions.} & (\frac{\alpha^{2}}{4t} < \tilde{\mu})
    \end{cases}.\\
\end{align}
We define $\Delta \tilde{E}_{\pm} = 2\sqrt{{\tilde{\xi}_{\pm}}^{2} + \psi^{2}}$ and obtain the range of $\Delta \tilde{E}_{\pm}$ as
\begin{align}
    &\Delta \tilde{E}_{+} \geq 2\psi, \\
    &\begin{cases}
      2\sqrt{\frac{\alpha^{2}\tilde{\mu}}{t} + \psi^{2}} \leq \Delta {E}_{-} \leq 2\sqrt{\left(-\frac{\alpha^{2}}{4t}-\tilde{\mu}\right)^{2} +\psi^{2}} & (0 \leq \tilde{\mu} \leq \frac{\alpha^{2}}{4t}) \\
      \text{no } \Delta \tilde{E}_{-} \text{satisfies the conditions.} & (\frac{\alpha^{2}}{4t} < \tilde{\mu})
    \end{cases}.
\end{align}

Here, we estimate $\tilde{J}_{\mathrm{B2(i)}}^{(\xi>0)}(\Omega)$ defined as the contribution to $\tilde{J}_{\mathrm{B2(i)}}(\Omega)$ from the region where the condition $tk^{2} -\tilde{\mu} > 0$ is satisfied. When 
$0 \leq \tilde{\mu} \leq \frac{\alpha^{2}}{4t}$,
it is obtained as 
\begin{align}
\tilde{J}_{\mathrm{B2(i)}}^{(\xi>0)}(\Omega) =
    \begin{cases}
    0 & (0\leq \Omega <2\psi)    \\
    \left(D_{\tilde{\xi}_{+} >0}(\Omega) + D_{\tilde{\xi}_{+}<0}(\Omega)\right)\frac{\Omega}{8\pi t\sqrt{\Omega^{2} - (2\psi)^{2}}} & \left(2\psi\leq \Omega <2\sqrt{\frac{\alpha^{2}\tilde{\mu}}{t} + \psi^{2}}\right) \\
    \left(D_{\tilde{\xi}_{+} >0}(\Omega) + D_{\tilde{\xi}_{+}<0}(\Omega) + D_{\tilde{\xi}_{-}<0}(\Omega)\right)\frac{\Omega}{8\pi t\sqrt{\Omega^{2} - (2\psi)^{2}}} & \left(2\sqrt{\frac{\alpha^{2}\tilde{\mu}}{t} + \psi^{2}}\leq \Omega <2\sqrt{\left(-\frac{\alpha^{2}}{4t}-\tilde{\mu}\right)^{2} + \psi^{2}}\right) \\
    D_{\tilde{\xi}_{+} >0}(\Omega)\frac{\Omega}{8\pi t\sqrt{\Omega^{2} - (2\psi)^{2}}} & \left(2\sqrt{\left(-\frac{\alpha^{2}}{4t}-\tilde{\mu}\right)^{2} + \psi^{2}} \leq \Omega\right) 
    \end{cases}.
\end{align}
We have defined $D_{\tilde{\xi}_{+} >0}$, $D_{\tilde{\xi}_{+}<0}$, and $D_{\tilde{\xi}_{-}<0}$ by
\begin{align}
    D_{\tilde{\xi}_{+} > 0}(\Omega) = \frac{\alpha + \sqrt{\alpha^{2} + 4t(\tilde{\mu}+\tilde{\xi})}}{\sqrt{\alpha^{2} + 4t(\tilde{\mu}+\tilde{\xi})}}, \\
    D_{\tilde{\xi}_{+} < 0}(\Omega) = \frac{\alpha + \sqrt{\alpha^{2} + 4t(\tilde{\mu}-\tilde{\xi})}}{\sqrt{\alpha^{2} + 4t(\tilde{\mu}-\tilde{\xi})}}, \\
    D_{\tilde{\xi}_{-} < 0}(\Omega) = \frac{\alpha - \sqrt{\alpha^{2} + 4t(\tilde{\mu}-\tilde{\xi})}}{\sqrt{\alpha^{2} + 4t(\tilde{\mu}-\tilde{\xi})}},    
\end{align}
with $\tilde{\xi} \equiv \frac{1}{2}\sqrt{\Omega^{2} - (2\psi)^{2}}$. These factors $D_{\tilde{\xi}_{+} >0}$, $D_{\tilde{\xi}_{+}<0}$, and $D_{\tilde{\xi}_{-}<0}$ correspond to the contributions from the region where $\tilde{\xi}_{+} = \frac{1}{2}\sqrt{\Omega^{2} - (2\psi)^{2}}$, $\tilde{\xi}_{+} = -\frac{1}{2}\sqrt{\Omega^{2} - (2\psi)^{2}}$, and $\tilde{\xi}_{-} = -\frac{1}{2}\sqrt{\Omega^{2} - (2\psi)^{2}}$, respectively.

When 
$\frac{\alpha^{2}}{4t} < \tilde{\mu}$, 
no $\tilde{E}_{-}$ satisfies the conditions. Thus, we obtain $\tilde{J}_{\mathrm{B2(i)}}^{(\xi>0)}$ as
\begin{align}
\tilde{J}_{\mathrm{B2(i)}}^{(\xi>0)}(\Omega) =
    \begin{cases}
    0 & (0\leq \Omega <2\psi)    \\
    \left(D_{\tilde{\xi}_{+} >0}(\Omega) + D_{\tilde{\xi}_{+}<0}(\Omega)\right)\frac{\Omega}{8\pi t\sqrt{\Omega^{2} - (2\psi)^{2}}} & \left(2\psi\leq \Omega <2\sqrt{\frac{\alpha^{2}\tilde{\mu}}{t} + \psi^{2}}\right) \\
    D_{\tilde{\xi}_{+} >0}(\Omega)\frac{\Omega}{8\pi t\sqrt{\Omega^{2} - (2\psi)^{2}}} & \left(2\sqrt{\frac{\alpha^{2}\tilde{\mu}}{t} + \psi^{2}} \leq \Omega\right) 
    \end{cases}.
    \label{eq:J_Bi_xi_p}
\end{align}

Next, we consider the case $tk^{2} - \tilde{\mu} < 0$ and $\tilde{\xi}_{\bm{k}}=-tk^{2} + \tilde{\mu} - \alpha k$. 
Solutions of the equation $\tilde{\xi} = -tk^{2} + \tilde{\mu} -\alpha k$ are given by
\begin{align}
    k = \frac{-\alpha + \sqrt{\alpha^{2} - 4t(\tilde{\xi} -\tilde{\mu})}}{2t},
\end{align}
where we use $k\geq 0$. We obtain $d\tilde{\xi}$ as
\begin{align}
    d\tilde{\xi} = \frac{\sqrt{\alpha^{2} - 4t(\tilde{\xi} -\tilde{\mu})}}{-\alpha + \sqrt{\alpha^{2} - 4t(\tilde{\xi} -\tilde{\mu})}}\cdot 2tkdk.
\end{align}
Then, 
$dk$ is given by
\begin{align}
    2\pi kdk = \frac{\pi}{2t}\frac{-\alpha + \sqrt{\alpha^{2} - 4t(\tilde{\xi} -\tilde{\mu})}}{\sqrt{\alpha^{2} - 4t(\tilde{\xi} -\tilde{\mu})}}\cdot\frac{\Delta \tilde{E}}{\sqrt{(\Delta \tilde{E})^{2} - (2\psi)^{2}}} d\Delta \tilde{E}.
\end{align}
Given $tk^{2} - \tilde{\mu} < 0$, we obtain the range of $k$ as $0<k<\frac{\tilde{\mu}}{t}$. Thus, the range of $\tilde{\xi}$ is 
\begin{align}
    -\alpha\sqrt{\frac{\tilde{\mu}}{t}} \leq \tilde{\xi} \leq \tilde{\mu}.
\end{align}
When $\tilde{\mu}$ is smaller (larger) than $\frac{\alpha^{2}}{t}$, the relation $\tilde{\mu}<\alpha\sqrt{\frac{\tilde{\mu}}{t}}$ $\left(\tilde{\mu}\geq\alpha\sqrt{\frac{\tilde{\mu}}{t}}\right)$ is given. Thus, we obtain the range of $\Delta\tilde{E}$
\begin{align}
    \begin{cases}
        2\psi \leq \Delta\tilde{E} \leq 2\sqrt{\frac{\alpha^{2}\tilde{\mu}}{t} + \psi^{2}} & (0\leq \tilde{\mu} < \frac{\alpha^{2}}{t}) \\
        2\psi \leq \Delta\tilde{E} \leq 2\sqrt{{\tilde{\mu}}^{2} + \psi^{2}} & (\frac{\alpha^{2}}{t} \leq \mu)
    \end{cases}.
\end{align}
Let us consider the case $\tilde{\mu} < \frac{\alpha^{2}}{t}$ because the parameters of Fig.~\ref{fig:JDOS2} satisfy the condition. We obtain $\tilde{J}_{\mathrm{B2(i)}}^{(\xi<0)}$, the contribution to the approximate JDOS $\tilde{J}_{\mathrm{B2(i)}}$ as
\begin{align}
    \tilde{J}_{\mathrm{B2(i)}}^{(\xi<0)}(\Omega) =
    \begin{cases}
        0 & (0 \leq \Omega <2\psi) \\
        \left(G_{\tilde{\xi} > 0}(\Omega) + G_{\tilde{\xi} < 0}(\Omega)\right)\frac{\Omega}{8\pi t\sqrt{\Omega^{2} - (2\psi)^{2}}} & \left(2\psi \leq \Omega < 2\sqrt{{\tilde{\mu}}^{2} + \psi^{2}}\right) \\
        G_{\tilde{\xi} < 0}(\Omega)\frac{\Omega}{8\pi t\sqrt{\Omega^{2} - (2\psi)^{2}}} & \left(2\sqrt{{\tilde{\mu}}^{2} + \psi^{2}} \leq \Omega < 2\sqrt{\frac{\alpha^{2}\tilde{\mu}}{t} + \psi^{2}}\right) \\
        0 & \left(2\sqrt{\frac{\alpha^{2}\tilde{\mu}}{t} + \psi^{2}} \leq \Omega\right)
    \end{cases},
    \label{eq:J_Bi_xi_n}
\end{align}
where we define $G_{\tilde{\xi} > 0}$ and $G_{\tilde{\xi} < 0}$ as 
\begin{align}
    G_{\tilde{\xi} > 0} = \frac{-\alpha + \sqrt{\alpha^{2} - 4t(\bar{\xi} -\tilde{\mu})}}{\sqrt{\alpha^{2} + 4t(\tilde{\mu}-\bar{\xi}) }}, \\
    G_{\tilde{\xi} < 0} = \frac{-\alpha + \sqrt{\alpha^{2} + 4t(\bar{\xi} +\tilde{\mu})}}{\sqrt{\alpha^{2} + 4t(\bar{\xi} +\tilde{\mu})}},
\end{align}
with $\bar{\xi}=\frac{1}{2}\sqrt{\Omega^{2} - (2\psi)^{2}}$. The factors $G_{\tilde{\xi} > 0}$ and $G_{\tilde{\xi} < 0}$ correspond to the contributions from the region where 
$\tilde{\xi}=\frac{1}{2}\sqrt{\Omega^{2} - (2\psi)^{2}}$ and $\tilde{\xi}=-\frac{1}{2}\sqrt{\Omega^{2} - (2\psi)^{2}}$, respectively.

\begin{figure*}
\includegraphics[width=0.45\linewidth]{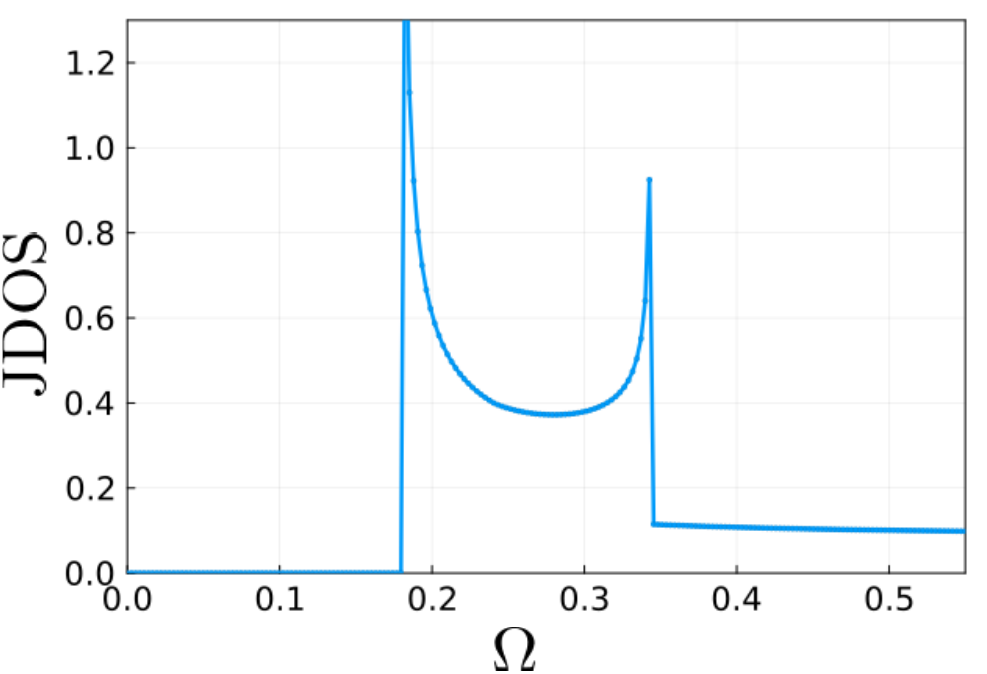}
\caption{
The analytic formula of the approximate
JDOS} $\tilde{J}_{\mathrm{B2(i)}}^{(\xi>0)}(\Omega)+\tilde{J}_{\mathrm{B2(i)}}^{(\xi<0)}(\Omega)$ is plotted. We set $\tilde{\mu}=0.08$, $t=0.6$, $\alpha =0.4$, and $\psi =0.09$. These parameters correspond to Figs.~\ref{fig:JDOS2}(b) and \ref{fig:JDOS2}(e).
\label{fig:appro_J_Bii}
\end{figure*}

The approximate JDOS $\tilde{J}_{\mathrm{B2(i)}}$ is given by the sum of $\tilde{J}_{\mathrm{B2(i)}}^{(\xi>0)}$ and $\tilde{J}_{\mathrm{B2(i)}}^{(\xi<0)}$. 
As shown in Fig.~\ref{fig:appro_J_Bii}, it shows the divergent behavior at the characteristic frequencies $\Omega=2\psi$ and $\Omega =2\sqrt{\left(-\frac{\alpha^{2}}{4t}-\tilde{\mu}\right)^{2} + \psi^{2}}$. Thus, the formula obtained in this section reproduces the behaviors of $\tilde{J}_{\mathrm{B2(i)}}(\Omega)$ in Figs.~\ref{fig:JDOS2}(b) and \ref{fig:JDOS2}(e). 
At $\Omega=2\psi$
the denominator $\sqrt{\Omega^{2} - (2\psi)^{2}}$ of $\tilde{J}_{\mathrm{B2(i)}}^{(\xi>0)}$ and $\tilde{J}_{\mathrm{B2(i)}}^{(\xi<0)}$ cause the divergent behavior. 
The divergence at $\Omega =2\sqrt{\left(-\frac{\alpha^{2}}{4t}-\tilde{\mu}\right)^{2} + \psi^{2}}$
is due to the denominator $\sqrt{\alpha^{2} + 4t(\tilde{\mu}-\tilde{\xi})}$ of $D_{\tilde{\xi}_{+} < 0}$ and $D_{\tilde{\xi}_{-} < 0}$, where we defined $\tilde{\xi}=\frac{1}{2}\sqrt{\Omega^{2} - (2\psi)^{2}}$. 

Finally, we identify the ${\bm k}$ space which contributes to the peak at $\Omega =2\psi$. When we set $\Omega = 2\psi$, $\tilde{\xi}$ must be zero. Thus, the two solutions of $k$ are obtained as
\begin{align}
    k_{1} = \frac{\alpha + \sqrt{\alpha^{2} + 4t\tilde{\mu}}}{2t}, \quad k_{2} = \frac{-\alpha + \sqrt{\alpha^{2} + 4t\tilde{\mu}}}{2t}.
\end{align}
In particular, when we set $\tilde{\mu}=0$, we obtain the solutions $k_{1}=\alpha/t$ and $k_{2}=0$. However, the contribution originating from the region around $\bm{k}=\bm{0}$ is unphysical because the condition ${\xi_{\bm{k}}}^{2}{g_{\bm{k}}}^{2} \gg \psi^{2}h^{2}$ is not satisfied at the $\Gamma$ point. Actually, the contribution around the $\Gamma$ point is negligible as shown in Fig.~\ref{fig:JDOS_k_profile}(d).
In this case, the contribution from the momentum around $k_1$ gives the peak of the JDOS.

\subsubsection{Analytical formula of B2(ii) contribution}

The approximate formula of the JDOS $\tilde{J}_{\mathrm{B2(ii)}}(\Omega)$ [Eq.~\eqref{eq:appro_B_ii_appendix}] is justified around the $\Gamma$ point ($\bm{k}\sim \bm{0}$) because the conditions $\psi^{2}h^{2} \gg {\xi_{\bm{k}}}^{2}{g_{\bm{k}}}^{2}$ and $\bm{h}\cdot \bm{g_{k}} \sim 0$ are satisfied there. We evaluate $\tilde{J}_{\mathrm{B2(ii)}}(\Omega)$ and show that it gives contribution in the frequency region $\Omega \geq 2(\sqrt{{\tilde{\mu}}^{2} + \psi^{2}} -h)$. In the region $\bm{k}\sim \bm{0}$, the energy difference $\Delta \tilde{E}_{\bm k}$ can be further approximated by
\begin{align}
    \Delta \tilde{E}_{\bm k} = 2\sqrt{{\xi_{\bm{k}}}^{2} + {g_{\bm k}}^{2} + h^{2} + \psi^{2} - 2h\sqrt{{\xi_{\bm{k}}}^{2} + \psi^{2}}}\simeq 2\sqrt{{\tilde{\mu}}^{2} + \psi^{2} + h^{2} - 2h\sqrt{{\tilde{\mu}}^{2} + \psi^{2}} + {\tilde{\alpha}}^{2}k^{2}},
\end{align}
where we define $\tilde{\alpha}$ as
\begin{align}
    \tilde{\alpha} = \sqrt{\alpha^{2} -2t\tilde{\mu} + \frac{2t\tilde{\mu}h}{\sqrt{{\tilde{\mu}}^{2} + \psi^{2}}}}.
\end{align}
We obtain $dk$ as
\begin{align}
    2\pi kdk = \frac{\pi\Delta \tilde{E}}{2\tilde{\alpha}^{2}}d\Delta \tilde{E}.
\end{align}
Thus, the approximate JDOS $\tilde{J}_{\mathrm{B2(ii)}}(\Omega)$ is given by
\begin{align}
    \tilde{J}_{\mathrm{B2(ii)}}(\Omega) = 
    \begin{cases}
        0 & \left(0\leq \Omega <2|\sqrt{{\tilde{\mu}}^{2} + \psi^{2}}-h| \right) \\
        \frac{\Omega}{8\pi{\tilde{\alpha}}^{2}} & \left(2|\sqrt{{\tilde{\mu}}^{2} + \psi^{2}}-h| \leq \Omega\right)
    \end{cases}.
\end{align}
When $|\tilde{\mu}|$ is sufficiently small and the Fermi level is in the vicinity of the Dirac point, this contribution gives a finite JDOS at frequencies lower than $\Omega = 2\psi$.

The $\tilde{J}_{\mathrm{B(ii)}}(\Omega)$ is contributed from the ${\bm k}$ space where $k$ satisfies the condition
\begin{align}
    k = \frac{\sqrt{(\Omega/2)^{2} - (h-\sqrt{{\tilde{\mu}}^{2} + \psi^{2}})^{2}}}{\tilde{\alpha}}.
\end{align}
When we set $\tilde{\mu}=0$ corresponding to $\mu=-3.2$ for our choice of parameters, the above $k$ is given by $k_{\Omega=0.12}=3.56\pi\times 10^{-2}$ for $\Omega=0.12$ and $k_{\Omega=0.18}=6.42\pi\times 10^{-2}$ for $\Omega=0.18$. In agreement with these estimations, we see the finite contribution to the JDOS from the regions around $k=k_{\Omega=0.12}$ and $k=k_{\Omega=0.18}$ in Figs.~\ref{fig:JDOS_k_profile}(a) and \ref{fig:JDOS_k_profile}(d), respectively.

\section{Classification of photocurrent conductivity}
\label{app:general_property}

\begin{table}[htbp]
\caption{Classification of the normal and anomalous photocurrent responses. Parities under the $\mathcal{T}-$ and $\mathcal{PT}-$symmetry operations are denoted by $\pm$. "Linear" and "Circular" indicate the polarization of light. Electric conductivity derivative term $\sigma_{\mathrm{ECD}}$ and nonreciprocal superfluid density term $\sigma_{\mathrm{NRSF}}$ show the divergent behavior in the low-frequency regime. We introduce the electric and magnetic conductivity derivative terms,  $\sigma_{\mathrm{ECD}}$ and $\sigma_{\mathrm{MCD}}$, as the imaginary and real part of the conductivity derivative term derived in Ref.~\onlinecite{Watanabe2022}.}
\centering
\begin{tabular}{lccccc}
\hline \hline
Mechanism & Notation & $\mathcal{T}$ & $\mathcal{PT}$ & Polarization & Low-frequency divergence\\
\hline
(Normal photocurrent) &&&&\\
Injection current (electric) & $\sigma_{\mathrm{Einj}}$& + & - & Circular &  \\
Injection current (magnetic) & $\sigma_{\mathrm{Minj}}$& - & + & Linear & \\
Shift current & $\sigma_{\mathrm{shift}}$& + & - & Linear & \\
Gyration current & $\sigma_{\mathrm{gyro}}$& - & + & Circular & \\
&&&&\\
(Anomalous photocurrent) &&&\\
Conductivity derivative (electric) & $\sigma_{\mathrm{ECD}}$& + & - & Circular & $O(\Omega^{-1})$ \\
Conductivity derivative (magnetic) & $\sigma_{\mathrm{MCD}}$& - & + & Linear & \\
Nonreciprocal superfluid density & $\sigma_{\mathrm{NRSF}}$& - & + & Linear & $O(\Omega^{-2})$\\
\hline\hline
\end{tabular}
\label{tab:mechanism_class}
\end{table}


\begin{figure*}[htbp] 
\includegraphics[width=\linewidth]{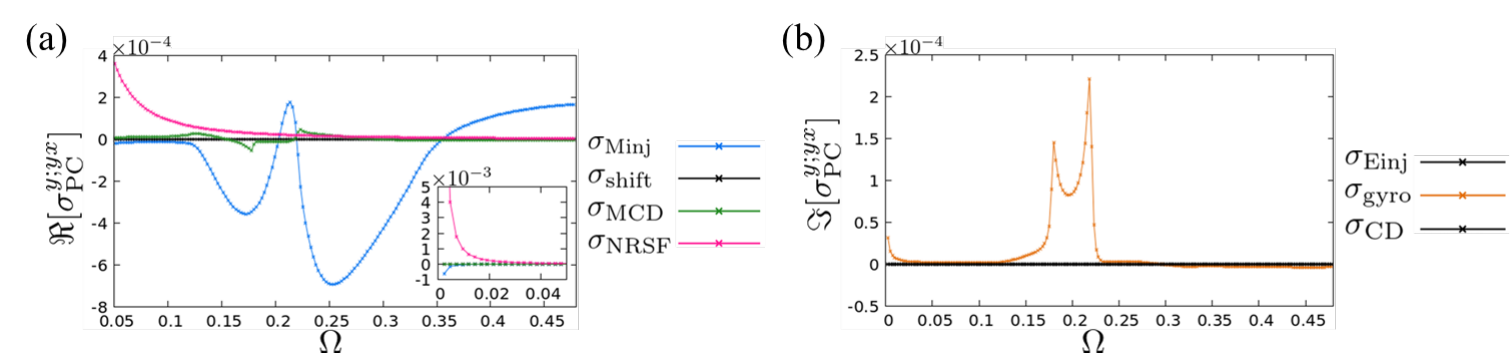}
 \caption{Each component of the photocurrent conductivity $\sigma^{y;yx}_{\mathrm{PC}}$ in the in-plane magnetic field. (a) and (b) plot the real part and the imaginary part, respectively. The calculation is performed with $h=0.025$ and $\mu =-3.2$, which are the same as Fig.~\ref{fig:general2.pdf}. The $\mathcal{T}$-odd components are colored, while the $\mathcal{T}$-even components are shown in black and revealed to disappear.}
 \label{fig:general3.pdf}
\end{figure*}
\begin{figure*}[htbp]
\includegraphics[width=\linewidth]{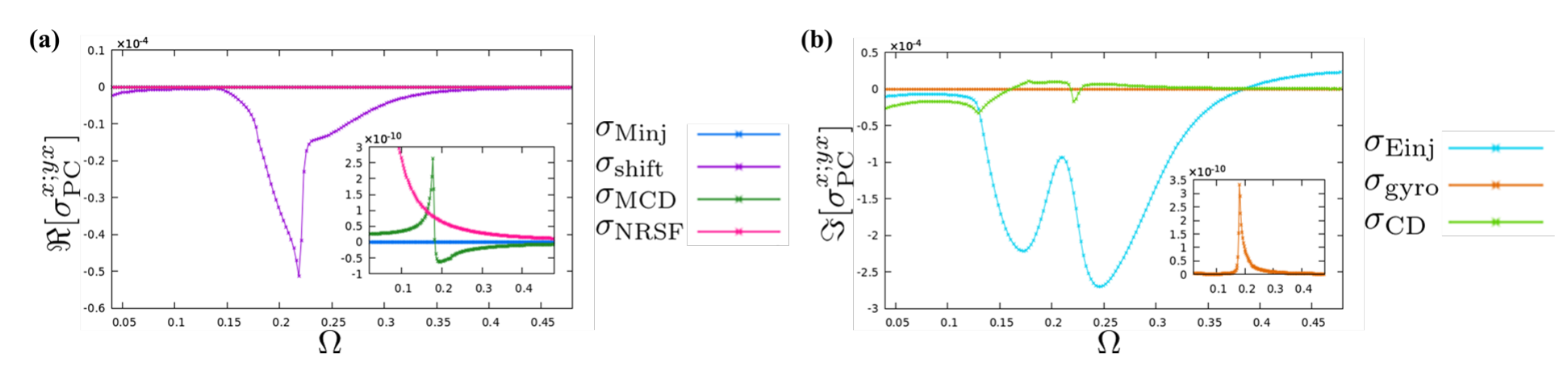}
 \caption{Each component of the photocurrent conductivity $\sigma^{x;yx}_{\mathrm{PC}}$ in the tilted magnetic field ($\theta =\ang{45}$). (a) real part and (b) imaginary part. The parameters other than the angle $\theta$ are the same as Figs.~\ref{fig:general2.pdf} and \ref{fig:general3.pdf}. 
 The $\mathcal{T}$-even components are dominant in contrast to $\sigma^{y;yx}_{\mathrm{PC}}$ (Fig.~\ref{fig:general3.pdf}).}
 \label{fig:various_mechanism}
\end{figure*}

Classification of the photocurrent conductivity based on the mechanism and parity under the  $\mathcal{T}$- and $\mathcal{PT}$-operations is summarized in Table~\ref{tab:mechanism_class}.
Figure \ref{fig:general3.pdf} plots the contributions stemming from each photocurrent generation mechanism under the in-plane magnetic field $\theta =\ang{0}$. We see that the magnetic injection current and gyration current dominate the real and imaginary parts of the photocurrent conductivity, respectively. These dominant contributions are $\mathcal{T}$-odd terms. 
In all the components with hyperscripts $\alpha,\beta,\gamma$, the $\mathcal{T}$-even contributions vanish consistent with the symmetry constraints in Table~\ref{tab:constraints}. On the other hand, the $\mathcal{T}$-even contributions are allowed to be finite in some components when the magnetic field is tilted ($\ang{0} < \theta < \ang{90}$). Figure~\ref{fig:various_mechanism} shows that the $\mathcal{T}$-even mechanisms, such as the shift current and electric injection current, give significant contributions to $\sigma^{x;yx}_{\mathrm{PC}}$. Note that $\sigma^{x;xx}$, $\sigma^{x;yy}$, and $\sigma^{y;yx}$ do not have finite $\mathcal{T}$-even contributions even for the field angle $\ang{0} < \theta < \ang{90}$ due to the symmetry constraint. Thus, the constraints in Table~\ref{tab:constraints} are useful for the identification of leading photocurrent generation mechanisms.
\section{Momentum-resolved contribution of photocurrent conductivity and origin of sign reversal}
\label{app:model_with_mu}
The discussion in the main text is mainly focused on the model for the chemical potential $\mu\simeq -3.2$. We also showed the results for $\mu =-0.8$ in the subsection \ref{subsec:comparison_normal_SC}. In both cases for $\mu=-3.2$ and $\mu=-0.8$, the Fermi level lies on the Dirac point in the normal state. However, the position and number of Dirac points are different between the two cases. For the chemical potential $\mu =-3.2$, the Dirac point at $\bm{k}=\bm{0}$ is on the Fermi level and relevant for the superconducting nonlinear optical responses. On the other hand, in the case with $\mu =-0.8$, Dirac points on the Brillouin zone boundary, $\bm{k} =(\pi, 0)$ and $(0, \pi)$, play essential roles. In this Appendix, we explain the characteristic properties of the model for $\mu =-0.8$ through the comparison with the case of $\mu =-3.2$.

\begin{figure*}[tbp]
 \includegraphics[width=\linewidth]{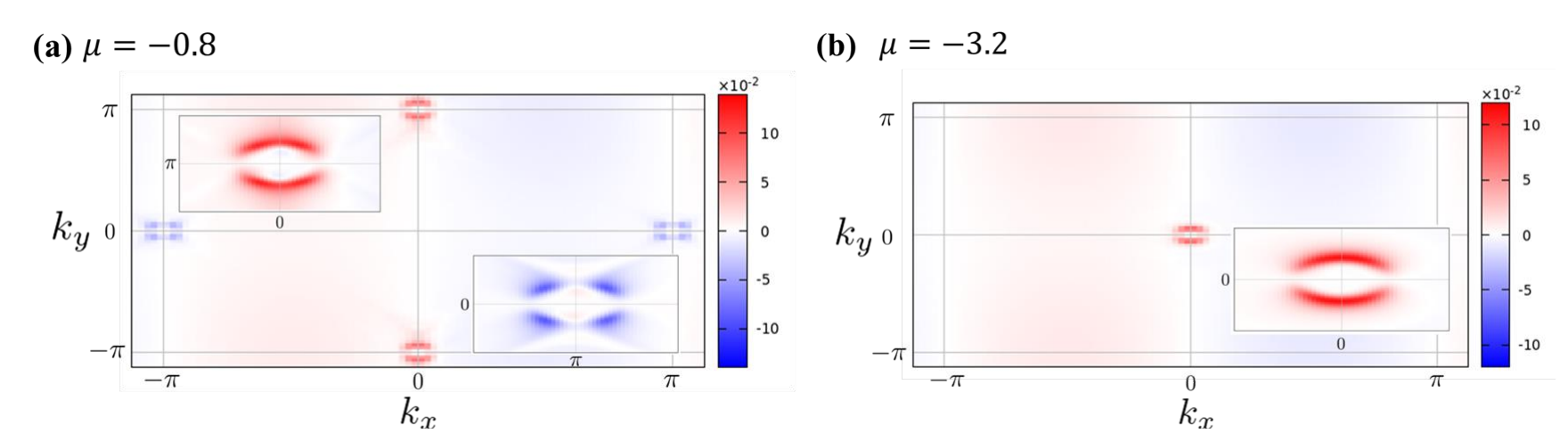}
 \caption{Distribution of the photocurrent conductivity $\Re\left[\sigma^{x;xx}_{\mathrm{PC}}\right]$ in the ${\bm k}$-space. We fix $\theta =\ang{30}$, $h=0.025$, and $\psi =0.09$ and set (a) $\mu=-0.8$ and (b) $\mu=-3.2$. Note that the distribution has periodicity due to the translation symmetry in the crystal structure. The inset shows the enlarged figures.}
 \label{fig:k_profile}
\end{figure*}

First, we show where in the ${\bm k}$-space the nonlinear conductivity mainly arises. Figure~\ref{fig:k_profile} shows the distribution of the photocurrent conductivity $\sigma^{x;xx}_{\mathrm{PC}}$ in the ${\bm k}$-space. We set the chemical potential as 
$\mu =-0.8$ and $\mu =-3.2$ in Fig.~\ref{fig:k_profile}(a) and Fig.~\ref{fig:k_profile}(b), respectively. The frequency of incident light is assumed as $\Omega =2\psi$. For $\mu=-3.2$, a large positive contribution to the photocurrent conductivity arises from $\bm{k} \sim (0,0)$, while for $\mu=-0.8$ the contributions from $(\pi,0)$ and $(0,\pi)$ are opposite as shown in Fig.~\ref{fig:k_profile}(b). Thus, it is expected that the nonlinear optical responses show more complicated behaviors for $\mu =-0.8$ than for $\mu =-3.2$.

\begin{figure*}[tbp]
 \includegraphics[width=\linewidth]{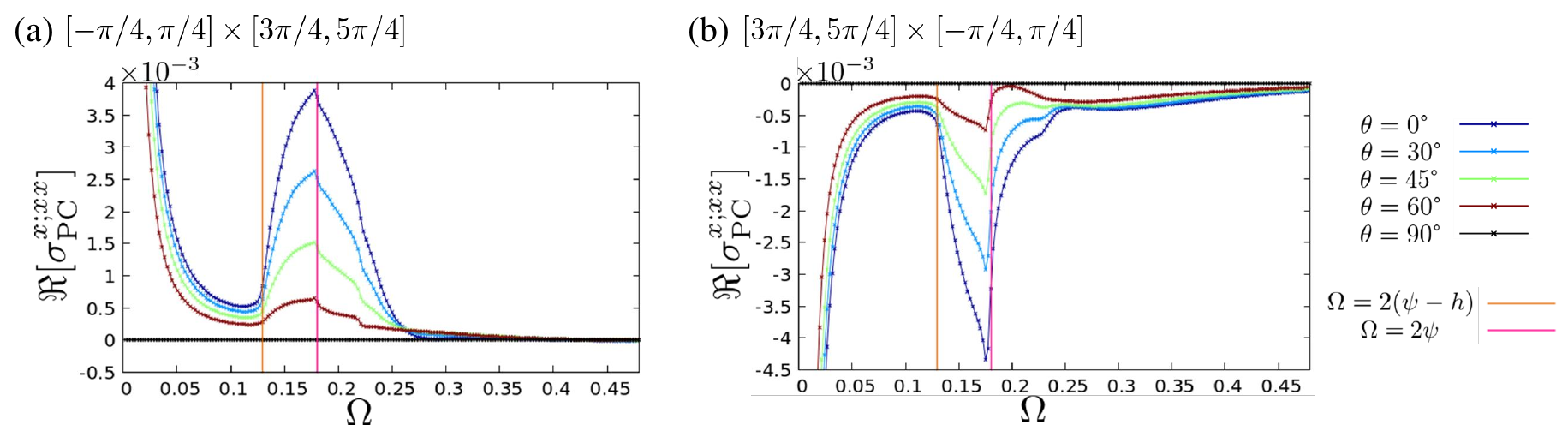}
 \caption{Partial contribution of the photocurrent conductivity $\Re\left[\sigma^{x;xx}_{\mathrm{PC}}\right]$ from the ${\bm k}$-space region (a) $\bm{k}\in [-\pi/4, \pi/4]\times [3\pi/4, 5\pi/4]$ and (b)$[3\pi/4, 5\pi/4]\times [-\pi/4, \pi/4]$.
 We vary the angle $\theta$ of the magnetic field and adopt the other parameters as in Fig.~\ref{fig:k_profile}. 
 }
 \label{fig:k_decompose}
\end{figure*}

Next, we focus on the frequency dependence of the contribution from $\bm{k} \sim (\pi,0)$ and $(0,\pi)$ in the case of $\mu =-0.8$. Figures~\ref{fig:k_decompose}(a) and \ref{fig:k_decompose}(b) plot the real part of the photocurrent conductivity $\Re\left[\sigma^{x;xx}_{\mathrm{PC}}\right]$ contributed from the momentum space around $(\pi,0)$ and $(0,\pi)$, respectively. Both contributions show the enhancement maximized at $\Omega =2\psi$ with a relatively simple frequency dependence. However, because the signs of the contributions are opposite to each other, the frequency dependence of the total photocurrent conductivity is complex as shown in Fig.~\ref{fig:SC_normal}(b). In particular, we notice that the sign reversal of the photocurrent conductivity at $\Omega =2\psi$ is caused by the partial cancellation of opposite contributions from ${\bm k} \sim (\pi,0)$ and $(0,\pi)$.

\section{Diagonal elements of paramagnetic current operator and group velocities}
\label{app:diagonal_elements}
In this Appendix, we discuss the relationship between the diagonal elements of the paramagnetic current and the group velocities. In the normal state, the group velocity $v^{\alpha}_{a}$ is closely related to the paramagnetic current operator $J^{\alpha}$. With the Hellmann-Feynman theorem, the group velocity is obtained as
\begin{align}
v^{\alpha}_{a} = \frac{\partial}{\partial k_{\alpha}}E_{a} = J^{\alpha}_{aa}.
\label{eq:gp_velocity}
\end{align}

In contrast to the normal state, the Hellmann-Feynman relation fails in the superconducting state. In the Bogoliubov de-Gennes formalism, particles with opposite charges, that is electron and hole, are treated on equal footing. Thus, a naive treatment based on the minimal coupling $\bm{p}\rightarrow \bm{p}-q\bm{A}$ is not justified. However, the group velocity of electrons and holes is closely related to the diagonal elements of the paramagnetic current operator.

\subsection{Diagonal elements of paramagnetic current operator}
\label{app:diag_ele_para_current_op}
First, we derive the diagonal elements of the paramagnetic current operator $J^{\alpha}_{aa}$ at $\bm{k}=\bm{0}$. In the main text, we showed that the momentum space around $\bm{k}=\bm{0}$ gives a dominant contribution to the photocurrent conductivity in the low-frequency regime when the Fermi level is close to the Dirac point. In this region,  $J^{\alpha}_{aa}(\bm{k})$ can be approximated by $J^{\alpha}_{aa}(\bm{k}=\bm{0})$. 
The BdG Hamiltonian is given by 
\begin{align}
    H(\bm{k}) = 
    \begin{pmatrix}
    H_{\mathrm{N}}(\bm{k}) & \psi(i\sigma_{y}) \\
    \psi(i\sigma_{y})^{\dagger} & -H_{\mathrm{N}}(-\bm{k})^{\top}
    \end{pmatrix},
    \label{eq:BdG_Hamiltonian}
\end{align}
where $H_{\mathrm{N}}(\bm{k}) = \xi_{\bm{k}} + \bm{g_{k}}\cdot\bm{\sigma} + \bm{h}\cdot\bm{\sigma}$ is the Hamiltonian in the normal state, and $\psi(i\sigma_{y})$ is the pair potential of superconductivity. 
Under the appropriate gauge, $\psi$ is real-valued.  We ignore the total momentum of the Cooper pairs for simplicity.
At $\bm{k}=\bm{0}$, the spin-orbit coupling $\bm{g_{k}}\cdot\bm{\sigma}$ vanishes. We define $H_{0}$ as $H_{0}\equiv H(\bm{k}=\bm{0})$.
The paramagnetic current operator $J^{\alpha}$ is given by
\begin{align}
J^{\alpha} =
\begin{pmatrix}
\partial_{\alpha}H_{\mathrm{N}}(\bm{k}) & 0 \\
0 & \partial_{\alpha}[{H_{\mathrm{N}}(-\bm{k})}^{\top}]
\end{pmatrix}.
\end{align}

We diagonalize $H_{\mathrm{N}}(\bm{k}=\bm{0})$ and $-H_{\mathrm{N}}(-\bm{k}=\bm{0})^{\top}$. For this purpose, we introduce unitary matrices $U$ and $\tilde{U}$ as
\begin{align}
U = \frac{1}{\sqrt{2h(h+h_{z})}}(h_{x}, h_{y}, h+h_{z})\cdot\bm{\sigma}, \\
\tilde{U} = \frac{1}{\sqrt{2h(h+h_{z})}}(h_{x}, -h_{y}, h+h_{z})\cdot\bm{\sigma}.
\end{align}
Under the unitary transformation, the BdG Hamiltonian at ${\bm k}={\bm 0}$ is transformed as
\begin{align}
&
\begin{pmatrix}
U^{\dagger} & 0 \\
0 & \tilde{U}^{\dagger}
\end{pmatrix}
\begin{pmatrix}
    H_{\mathrm{N}} & \psi(i\sigma_{y}) \\
    \psi(i\sigma_{y})^{\dagger} & -H_{\mathrm{N}}^{\top}
\end{pmatrix}
\begin{pmatrix}
U & 0 \\
0 & \tilde{U}
\end{pmatrix} \\
=&
\begin{pmatrix}
U^{\dagger}H_{\mathrm{N}}U & U^{\dagger}\psi(i\sigma_{y})\tilde{U} \\
\tilde{U}^{\dagger}\psi(i\sigma_{y})^{\dagger}U & -\tilde{U}^{\dagger}H_{\mathrm{N}}^{\top}\tilde{U}
\end{pmatrix}
=
\begin{pmatrix}
\xi +h\sigma_{z} & -\psi i\sigma_{y} \\
-(\psi i\sigma_{y})^{\dagger} & -(\xi +h\sigma_{z})
\end{pmatrix}.
\end{align}
Under the same unitary transformation, the paramagnetic current operator is given by
\begin{align}
J^{\alpha} = 
\begin{pmatrix}
U^{\dagger}\partial_{\alpha}H_{\mathrm{N}}(\bm{k})U & 0 \\
0 & \tilde{U}^{\dagger}\partial_{\alpha}[{H_{\mathrm{N}}(-\bm{k})}^{\top}]\tilde{U}
\end{pmatrix}
=
\begin{pmatrix}
\partial_{\alpha}\xi + \bm{\eta}^{\alpha}\cdot\bm{\sigma} & 0 \\
0 & \partial_{\alpha}\xi -\tilde{\bm{\eta}}^{\alpha}\cdot\bm{\sigma}
\end{pmatrix}, \label{eq:para_current_op_normal_diag}
\end{align}
where we define $\bm{\eta}^{\alpha} \equiv (\eta^{\alpha}_{x}, \eta^{\alpha}_{y}, \eta^{\alpha}_{z})$ as
\begin{align}
&
\begin{pmatrix}
\eta^{\alpha}_{x} \\
\eta^{\alpha}_{y}
\end{pmatrix}
=-
\begin{pmatrix}
\partial_{\alpha}g_{x} \\
\partial_{\alpha}g_{y}
\end{pmatrix}
+\frac{\partial_{\alpha}\bm{g}\cdot\bm{h} + h\partial_{\alpha}g_{z}}{h(h + h_{z})}
\begin{pmatrix}
h_{x} \\
h_{y}
\end{pmatrix}
, \quad \eta^{\alpha}_{z} = \partial_{\alpha}\bm{g}\cdot\hat{\bm{h}}, \label{eq:alpha_def} 
\end{align}
and $\tilde{\bm{\eta}}^{\alpha}= (\eta^{\alpha}_{x}, -\eta^{\alpha}_{y}, \eta^{\alpha}_{z})$. 
We have introduced the unit vector $\hat{\bm{h}}=(h_{x}, h_{y}, h_{z})^{\top}/h$.

By an appropriate permutation of the bases, the BdG Hamiltonian and the paramagnetic current operator are represented by 
\begin{align}
H_{0}=
\begin{pmatrix}
\xi -h & \psi & 0 & 0 \\
\psi & -\xi -h & 0 & 0 \\
0 & 0 & \xi + h & -\psi \\
0 & 0 & -\psi & -\xi + h
\end{pmatrix}
= 
\begin{pmatrix}
A & 0 \\
0 & B
\end{pmatrix},
\end{align}
and
\begin{align}
J^{\alpha}=
\begin{pmatrix}
\partial_{\alpha}\xi - \eta^{\alpha}_{z} & (\eta^{\alpha}_{x}+ i\eta^{\alpha}_{y})\sigma_{z} \\
(\eta^{\alpha}_{x}- i\eta^{\alpha}_{y})\sigma_{z} & \partial_{\alpha}\xi + \eta^{\alpha}_{z}
\end{pmatrix}.
\label{eq:J_alpha_element}
\end{align}
We diagonalize $A$ and $B$ by the unitary matrices $U_{A}$ and $U_{B}$ defined as
\begin{align}
U_{A} &= \frac{1}{\sqrt{2u(u+\xi)}}(\psi, 0, u+\xi)\cdot \bm{\sigma}, \\
U_{B} &= \frac{1}{\sqrt{2u(u+\xi)}}(-\psi, 0, u+\xi)\cdot \bm{\sigma}, 
\end{align}
with $u = \sqrt{\xi^{2} + \psi^{2}}$.
The unitary transformation of the Hamiltonian is written as
\begin{align}
H_{0} = 
\begin{pmatrix}
U_{A}^{\dagger} & 0 \\
0 & U_{B}^{\dagger}
\end{pmatrix}
\begin{pmatrix}
A & 0 \\
0 & B
\end{pmatrix}
\begin{pmatrix}
U_{A} & 0 \\
0 & U_{B}
\end{pmatrix}
=
\begin{pmatrix}
-h +u\sigma_{z} & 0 \\
0 & h + u\sigma_{z}
\end{pmatrix}.
\label{eq:H_o_elements}
\end{align}
The diagonal elements of the paramagnetic current operator do not change under the unitary transformation $U_{A}$ and $U_{B}$.

From Eqs.~\eqref{eq:J_alpha_element} and \eqref{eq:H_o_elements}, we obtain the eigenenergies of Bogoliubov quasiparticles $E_{a}$ ($E_{1}\leq E_{2}\leq E_{3}\leq E_{4}$) and the diagonal elements of the paramagnetic current operator $J^{\alpha}_{aa}$,
\begin{align}
(E_{1}, E_{2}, E_{3}, E_{4})&=
\begin{cases}
(-u-h, -u+h, u-h, u+h) & \left(\sqrt{\xi^{2}+\psi^{2}}>h\right) \\
(-u-h, u-h, -u+h, u+h) & \left(\sqrt{\xi^{2}+\psi^{2}}<h\right) 
\end{cases}, \\
(J^{\alpha}_{11}, J^{\alpha}_{22}, J^{\alpha}_{33}, J^{\alpha}_{44})&=
\begin{cases}
(\partial_{\alpha}\xi - \eta^{\alpha}_{z}, \partial_{\alpha}\xi+ \eta^{\alpha}_{z}, \partial_{\alpha}\xi- \eta^{\alpha}_{z}, \partial_{\alpha}\xi+ \eta^{\alpha}_{z}) & \left(\sqrt{\xi^{2}+\psi^{2}}>h\right) \\
(\partial_{\alpha}\xi- \eta^{\alpha}_{z}, \partial_{\alpha}\xi- \eta^{\alpha}_{z}, \partial_{\alpha}\xi+ \eta^{\alpha}_{z}, \partial_{\alpha}\xi+ \eta^{\alpha}_{z}) & \left(\sqrt{\xi^{2}+\psi^{2}}<h\right) \\
\end{cases}.
\end{align}

As we explained in the main text, $J^{x}_{22}-J^{x}_{33}$ is an important factor for the magnetic injection current in the low-frequency regime. Therefore, we explicitly show the formula of $J^{x}_{22}-J^{x}_{33}$ 
\begin{align}
J^{x}_{22}-J^{x}_{33}=
\begin{cases}
2(\partial_{\alpha}\bm{g}\cdot\hat{\bm{h}}) & \left(\sqrt{\xi^{2}+\psi^{2}}>h\right) \\
-2(\partial_{\alpha}\bm{g}\cdot\hat{\bm{h}}) & \left(\sqrt{\xi^{2}+\psi^{2}}<h\right)
\end{cases}.
\end{align}
The condition $\sqrt{{\xi_{\bm{k}=\bm{0}}}^{2}+\psi^{2}} < h$ is closely related to the topological superconductivity. Indeed, we show that around the topological transition the photocurrent conductivity drastically changes accompanied by the sign reversal due to the sign change of $J^{x}_{22}-J^{x}_{33}$ [Sec.~\ref{sec:photocurrent conductivity around topological transition}]. From this property, we propose that the photocurrent conductivity is a bulk probe of topological superconductivity.

\subsection{Group velocity}
Next, we discuss the relation between the paramagnetic current operator in the superconducting state and the group velocity in the normal state. 
In the normal state, we approximately obtain the energies of hole and electron bands around $\bm{k}=\bm{0}$
\begin{align}
(\epsilon^{\mathrm{e+}}_{\bm{k}}, \epsilon^{\mathrm{e-}}_{\bm{k}}, \epsilon^{\mathrm{h+}}_{\bm{k}}, \epsilon^{\mathrm{h-}}_{\bm{k}})=(\xi_{\bm{k}}+h+\bm{g_{k}}\cdot\hat{\bm{h}}, \xi_{\bm{k}}-h-\bm{g_{k}}\cdot\hat{\bm{h}}, -\xi_{\bm{k}}-h+\bm{g_{k}}\cdot\hat{\bm{h}}, -\xi_{\bm{k}}+h-\bm{g_{k}}\cdot\hat{\bm{h}}),
\end{align}
where we use $U$ and $\tilde{U}$ for the diagonalization of Hamiltonian. The group velocity is obtained as
\begin{align}
(v^{\alpha}_{\mathrm{e+}}, v^{\alpha}_{\mathrm{e-}}, v^{\alpha}_{\mathrm{h+}}, v^{\alpha}_{\mathrm{h-}})=(\partial_{\alpha}\xi_{\bm{k}} +\partial_{\alpha}\bm{g_{k}}\cdot\hat{\bm{h}}, \partial_{\alpha}\xi_{\bm{k}} -\partial_{\alpha}\bm{g_{k}}\cdot\hat{\bm{h}}, -\partial_{\alpha}\xi_{\bm{k}} +\partial_{\alpha}\bm{g_{k}}\cdot\hat{\bm{h}}, -\partial_{\alpha}\xi_{\bm{k}} -\partial_{\alpha}\bm{g_{k}}\cdot\hat{\bm{h}}).
\label{eq:group_velocity}
\end{align}
The electron and hole have an opposite charge and we set the unit charge $\mathrm{e}=1$. Thus, the current of the quasiparticle is given by
\begin{align}
(j^{\alpha}_{\mathrm{e+}}, j^{\alpha}_{\mathrm{e-}}, j^{\alpha}_{\mathrm{h+}}, j^{\alpha}_{\mathrm{h-}}) = (v^{\alpha}_{\mathrm{e+}}, v^{\alpha}_{\mathrm{e-}}, -v^{\alpha}_{\mathrm{h+}}, -v^{\alpha}_{\mathrm{h-}}).
\end{align}
These currents of quasiparticles correspond to the diagonal elements of the paramagnetic current operator in Eq.~\eqref{eq:para_current_op_normal_diag}. 

Here we consider an approximate current operator $\tilde{J}^{\alpha}$ defined as
\begin{align}
\tilde{J}^{\alpha} = \mathrm{diag}(v^{\alpha}_{\mathrm{e+}}, v^{\alpha}_{\mathrm{e-}}, -v^{\alpha}_{\mathrm{h+}}, -v^{\alpha}_{\mathrm{h-}}),
\end{align}
where we ignore the off-diagonal elements of the paramagnetic current operator in Eq.~\eqref{eq:para_current_op_normal_diag}.
In the superconducting state, the mixing of electrons and holes influences the effective current of Bogoliubov quasiparticles. To treat the effect of the band mixing, we carry out the unitary transformation $U_{A}$, $U_{B}$, and the permutation of the bases. On this basis, the approximate current operator is represented as,
\begin{align}
\tilde{J}^{\alpha}=
\begin{pmatrix}
\beta^{\alpha}_{0} + \bm{\beta}^{\alpha}\cdot\bm{\sigma} & 0 \\
0 & \gamma^{\alpha}_{0} + \bm{\gamma}^{\alpha}\cdot\bm{\sigma}
\end{pmatrix},
\end{align}
where we define $\beta^{\alpha}$ and $\gamma^{\alpha}$ as
\begin{align}
\beta^{\alpha}_{0} &=\frac{1}{2}\left(v^{\alpha}_{\mathrm{e-}} - v^{\alpha}_{\mathrm{h+}}\right),\quad \beta^{\alpha}_{x}=\frac{\psi}{2u}\left(v^{\alpha}_{\mathrm{e-}} + v^{\alpha}_{\mathrm{h+}}\right),\quad \beta^{\alpha}_{y}=0,\quad \beta^{\alpha}_{z}=\frac{\xi}{2u}\left(v^{\alpha}_{\mathrm{e-}} + v^{\alpha}_{\mathrm{h+}}\right), \\
\gamma^{\alpha}_{0} &=\frac{1}{2}\left(v^{\alpha}_{\mathrm{e+}} - v^{\alpha}_{\mathrm{h-}}\right), \quad \gamma^{\alpha}_{x}=-\frac{\psi}{2u}\left(v^{\alpha}_{\mathrm{e+}} + v^{\alpha}_{\mathrm{h-}}\right), \quad \gamma^{\alpha}_{y}=0, \quad \gamma^{\alpha}_{z}=\frac{\xi}{2u}\left(v^{\alpha}_{\mathrm{e+}} + v^{\alpha}_{\mathrm{h-}}\right).
\end{align}

We define the effective current of Bogoliubov quasiparticles with energies $E_{a}$ by the approximate paramagnetic current operator as $j^{\alpha}_{a} \equiv \tilde{J}^{\alpha}_{aa}$. Here, we focus on $j^{\alpha}_{2}$ and $j^{\alpha}_{3}$ because the transition between the $E_{2}$ and $E_{3}$ bands dominantly contributes to the photocurrent conductivity in the low-frequency regime. The effective currents $j^{\alpha}_{2}$ and $j^{\alpha}_{3}$ are obtained as
\begin{align}
(j^{\alpha}_{2}, j^{\alpha}_{3})&=
\begin{cases}
(w_{\mathrm{-}}v^{\alpha}_{\mathrm{e+}}-w_{\mathrm{+}}v^{\alpha}_{\mathrm{h-}},\quad  w_{\mathrm{+}}v^{\alpha}_{\mathrm{e-}}-w_{\mathrm{-}}v^{\alpha}_{\mathrm{h+}}) & \left(\sqrt{\xi^{2} + \psi^{2}}>h\right) \\
( w_{\mathrm{+}}v^{\alpha}_{\mathrm{e-}}-w_{\mathrm{-}}v^{\alpha}_{\mathrm{h+}},\quad w_{\mathrm{-}}v^{\alpha}_{\mathrm{e+}}-w_{\mathrm{+}}v^{\alpha}_{\mathrm{h-}}) & \left(\sqrt{\xi^{2} + \psi^{2}}<h\right)
\end{cases}, \\
w_{\mathrm{+}}&=\frac{u + \xi}{2u}, \quad w_{\mathrm{-}}=\frac{u - \xi}{2u}.
\label{eq:j_2_j_3_formula}
\end{align}
With Eq.~\eqref{eq:group_velocity}, the effective  current $j^{\alpha}_{a}$ is coincident with $J^{\alpha}_{aa}$ at $\bm{k}=\bm{0}$.
Under the charge conjugate transformation $(\mathrm{e\pm}\leftrightarrow \mathrm{h\pm})$, the effective current $j^{\alpha}_{2}$ and $j^{\alpha}_{3}$ changes to each other. The exchange of formula between $j^{\alpha}_{2}$ and $j^{\alpha}_{3}$ at $h=\sqrt{\xi^{2}+\psi^{2}}$ in Eq.~\eqref{eq:j_2_j_3_formula} corresponds to the band inversion between electrons and holes.

Finally, we clarify the condition for the finite difference between $j^{\alpha}_{2}$ and $j^{\alpha}_{3}$. The difference is obtained as
\begin{align}
j^{\alpha}_{2} - j^{\alpha}_{3} =
\begin{cases}
w_{\mathrm{-}}(v^{\alpha}_{\mathrm{e+}}+v^{\alpha}_{\mathrm{h+}})-w_{\mathrm{+}}(v^{\alpha}_{\mathrm{e-}}+v^{\alpha}_{\mathrm{h-}}) & \left(\sqrt{\xi^{2} + \psi^{2}}>h\right) \\
-w_{\mathrm{-}}(v^{\alpha}_{\mathrm{e+}}+v^{\alpha}_{\mathrm{h+}})+w_{\mathrm{+}}(v^{\alpha}_{\mathrm{e-}}+v^{\alpha}_{\mathrm{h-}}) & \left(\sqrt{\xi^{2} + \psi^{2}}<h\right)
\end{cases}.
\label{eq:j_2_j_3}
\end{align}
When the energy bands $\epsilon_{\bm{k}}$ has the symmetry $\epsilon_{\bm{k}}=\epsilon_{\bm{-k}}$ in the normal state, the group velocity of the hole bands $-\epsilon_{-\bm{k}}$ is given by
\begin{align}
v^{\alpha}_{\mathrm{h}}=\partial_{\alpha}(-\epsilon_{-\bm{k}})=-\partial_{\alpha}\epsilon_{\bm{k}}=-v^{\alpha}_{\mathrm{e}}.
\end{align}
With Eqs.~\eqref{eq:j_2_j_3_formula} and \eqref{eq:j_2_j_3}, the asymmetry $\epsilon^{\mathrm{e\pm}}_{\bm{k}}\neq\epsilon^{\mathrm{e\pm}}_{-\bm{k}}$ is necessary for the finite $j^{\alpha}_{2} - j^{\alpha}_{3}$, indicating that the nonreciprocity in the normal state band structure is essential for the magnetic injection current and its sign-reversal at the topological transition.

Although the Hellmann-Feynman relation breaks down in the superconducting state, the diagonal elements of the paramagnetic current operator $J^{\alpha}_{aa}$ are closely related to the group velocity of electron and hole bands in the normal state. Thus, the diagonal elements $J^{\alpha}_{22}$ and $J^{\alpha}_{33}$ are sensitive to the band inversion between the electron and hole bands, and this property is useful for detecting the topological transition in the superconducting state. In the photocurrent conductivity, the difference $J^{\alpha}_{22}-J^{\alpha}_{33}$ gives a profound effect on the magnetic injection current, which would be a probe of topological superconductivity. The asymmetric dispersion $\epsilon^{\mathrm{e\pm}}_{\bm{k}}\neq\epsilon^{\mathrm{e\pm}}_{-\bm{k}}$ in the normal states is essential for the finite difference $J^{\alpha}_{22}-J^{\alpha}_{33}$.

\section{Condition for the finite quantum metric of the B2 transition}
In this Appendix, we elucidate the condition for the finite quantum metric $g^{\alpha\beta}_{32}$. We show that both the magnetic field and spin-orbit coupling are needed for the quantum metric $g^{\alpha\beta}_{32}$ to be finite. The $g^{\alpha\beta}_{32}$ is given by
\begin{align}
g^{\alpha\beta}_{32} = \frac{1}{{E_{23}}^{2}}\Re[J^{\alpha}_{32}J^{\beta}_{23}].
\end{align}
Thus, we have only to evaluate the matrix element of the paramagnetic current operator $J^{\alpha}_{32}$. When the $J^{\alpha}_{32}$ is zero, the quantum metric $g^{\alpha\beta}_{32}$ must be zero. 

\subsection{Quantum metric in the absence of spin-orbit coupling} 
First, we investigate the quantum metric $g^{\alpha\beta}_{32}$ at the $\Gamma$ point ($\bm{k}=\bm{0}$) where the spin-orbit coupling vanishes. We have already diagonalized the BdG Hamiltonian $H(\bm{k}=\bm{0})$ in Appendix~\ref{app:diag_ele_para_current_op}. After the appropriate unitary transformation, we obtain the diagonalized Hamiltonian
\begin{align}
H(\bm{k}=\bm{0}) &= \text{diag}\left(\sqrt{{\xi}^{2}+\psi^{2}}-h, -\sqrt{{\xi}^{2}+\psi^{2}}-h, \sqrt{{\xi}^{2}+\psi^{2}}+h, -\sqrt{{\xi}^{2}+\psi^{2}}+h\right),\\
&=
\begin{cases}
(E_{3}, E_{1}, E_{4}, E_{2}) & (\sqrt{{\xi}^{2}+\psi^{2}}> h) \\
(E_{2}, E_{1}, E_{4}, E_{3}) & (\sqrt{{\xi}^{2}+\psi^{2}}< h)
\end{cases},
\label{eq:eignvalue_position}
\end{align}
where $\xi$ means $\xi_{\bm{k}=\bm{0}}$ for short hand notation. Note that the subscript of $E_{a}$ is defined as $E_{1\bm{k}}\leq E_{2\bm{k}}\leq E_{3\bm{k}}\leq E_{4\bm{k}}$. Under the same transformation as the diagonalization of the Hamiltonian, the paramagnetic current operator is represented as
\begin{align}
J^{\alpha}=
\begin{pmatrix}
\partial_{\alpha}\xi -\eta^{\alpha}_{z} & (\eta^{\alpha}_{x}+i\eta^{\alpha}_{y})\sigma_{z} \\
(\eta^{\alpha}_{x}-i\eta^{\alpha}_{y})\sigma_{z} & \partial_{\alpha}\xi +\eta^{\alpha}_{z}
\end{pmatrix},
\label{eq:J_alpha_App_quantum_metric}
\end{align}
with $\eta^{\alpha}_{x}$, $\eta^{\alpha}_{y}$, and $\eta^{\alpha}_{z}$ introduced in Eq.~\eqref{eq:alpha_def}. With Eqs.~\eqref{eq:eignvalue_position} and \eqref{eq:J_alpha_App_quantum_metric}, the matrix element $J^{\alpha}_{23}$ is zero. Thus, the quantum metric $g^{\alpha\beta}_{32}$ vanishes at the $\Gamma$ point ($\bm{k}=\bm{0}$). This result indicates that the g-vector of spin-orbit coupling is essential for the finite quantum metric $g^{\alpha\beta}_{32}$ even when the derivative of g-vector $\partial_{\alpha}\bm{g}$ is finite.

\subsection{Quantum metric at the zero magnetic field}
\label{app:q_metric_without_MF}
Next, we evaluate the quantum metric of Rashba superconductors at the zero magnetic field. The BdG Hamiltonian in the coordinate $\hat{z}\parallel \bm{g}$ has the form 
\begin{align}
H(\bm{k}) =
\begin{pmatrix}
\xi + g & 0 & 0 & \psi \\
0 & \xi - g & -\psi & 0 \\
0 & -\psi & -\xi +g & 0 \\
\psi & 0 & 0 & -\xi -g
\end{pmatrix}.
\end{align}
Then, the paramagnetic current operator is obtained as
\begin{align}
J^{\alpha}=
\begin{pmatrix}
\partial_{\alpha}\xi + \partial_{\alpha}g_{z} & \partial_{\alpha}g_{x}-i\partial_{\alpha}g_{y} & 0 & 0\\
\partial_{\alpha}g_{x}+i\partial_{\alpha}g_{y} & \partial_{\alpha}\xi - \partial_{\alpha}g_{z} & 0 & 0\\
0 & 0 & \partial_{\alpha}\xi - \partial_{\alpha}g_{z} & -\partial_{\alpha}g_{x}-i\partial_{\alpha}g_{y} \\
0 & 0 & -\partial_{\alpha}g_{x}+i\partial_{\alpha}g_{y} & \partial_{\alpha}\xi + \partial_{\alpha}g_{z}
\end{pmatrix}.
\end{align}
By an appropriate permutation of the bases, the BdG Hamiltonian and the paramagnetic current operator are rewritten as
\begin{align}
H(\bm{k}) &=
\begin{pmatrix}
\xi + g & \psi & 0 & 0 \\
\psi & -\xi - g & 0 & 0 \\
0 & 0 & \xi -g & -\psi \\
0 & 0 & -\psi & -\xi +g
\end{pmatrix}
=
\begin{pmatrix}
C & 0 \\
0 & D
\end{pmatrix}, \\
J^{\alpha} &=
\begin{pmatrix}
(\partial_{\alpha}\xi + \partial_{\alpha}g_{z})\sigma_{0} & (\partial_{\alpha}g_{x}-i\partial_{\alpha}g_{y})\sigma_{z} \\
(\partial_{\alpha}g_{x}+i\partial_{\alpha}g_{y})\sigma_{z} & (\partial_{\alpha}\xi - \partial_{\alpha}g_{z})\sigma_{0}
\end{pmatrix}.
\end{align}
We can diagonalize the diagonal block $C$ and $D$ by using unitary operators $U_{C}$ and $U_{D}$, respectively. We introduce $U_{C}$ and $U_{D}$ as
\begin{align}
U_{C} = \frac{1}{\sqrt{2u_{C}(u_{C} + \xi +g)}}(\psi, 0, u_{C} + \xi +g)\cdot \bm{\sigma}, \quad u_{C}=\sqrt{(\xi +g)^{2} + \psi^{2}}, \\
U_{D} = \frac{1}{\sqrt{2u_{D}(u_{D} + \xi -g)}}(-\psi, 0, u_{D} + \xi -g)\cdot \bm{\sigma}, \quad u_{D}=\sqrt{(\xi -g)^{2} + \psi^{2}}. 
\end{align}
Under the unitary transformation, we obtain the Hamiltonian as
\begin{align}
H(\bm{k}) &= \text{diag}\left(\sqrt{(\xi+g)^{2} + \psi^{2}},-\sqrt{(\xi+g)^{2} + \psi^{2}},\sqrt{(\xi-g)^{2} + \psi^{2}}, -\sqrt{(\xi+g)^{2} + \psi^{2}} \right) \\
&=
\begin{cases}
(E_{4}, E_{1}, E_{3}, E_{2}) & (\xi > 0) \\
(E_{3}, E_{2}, E_{4}, E_{1}) & (\xi < 0) 
\end{cases}. \label{eq:eigenenergies_no_MF}
\end{align}
Under the same unitary transformation, the paramagnetic current operator is given by
\begin{align}
J^{\alpha} &=
\begin{pmatrix}
(\partial_{\alpha}\xi + \partial_{\alpha}g_{z})\sigma_{0} & (\partial_{\alpha}g_{x}-i\partial_{\alpha}g_{y}){U_{C}}^{\dagger}\sigma_{z}U_{D} \\
(\partial_{\alpha}g_{x}+i\partial_{\alpha}g_{y}){U_{D}}^{\dagger}\sigma_{z}U_{C} & (\partial_{\alpha}\xi - \partial_{\alpha}g_{z})\sigma_{0}
\end{pmatrix}. \label{eq:paramag_current_op_no_MF}
\end{align}
With Eqs.~\eqref{eq:eigenenergies_no_MF} and \eqref{eq:paramag_current_op_no_MF}, we find $J^{\alpha}_{23}=0$ and $g^{\alpha\beta}_{32}=0$. Therefore, the magnetic field is essential for a finite matrix element of the quantum metric $g^{\alpha\beta}_{32}$.


\bibliographystyle{apsrev4-1}
\bibliography{reference}

\begin{thebibliography}{89}%
\makeatletter
\providecommand \@ifxundefined [1]{%
 \@ifx{#1\undefined}
}%
\providecommand \@ifnum [1]{%
 \ifnum #1\expandafter \@firstoftwo
 \else \expandafter \@secondoftwo
 \fi
}%
\providecommand \@ifx [1]{%
 \ifx #1\expandafter \@firstoftwo
 \else \expandafter \@secondoftwo
 \fi
}%
\providecommand \natexlab [1]{#1}%
\providecommand \enquote  [1]{``#1''}%
\providecommand \bibnamefont  [1]{#1}%
\providecommand \bibfnamefont [1]{#1}%
\providecommand \citenamefont [1]{#1}%
\providecommand \href@noop [0]{\@secondoftwo}%
\providecommand \href [0]{\begingroup \@sanitize@url \@href}%
\providecommand \@href[1]{\@@startlink{#1}\@@href}%
\providecommand \@@href[1]{\endgroup#1\@@endlink}%
\providecommand \@sanitize@url [0]{\catcode `\\12\catcode `\$12\catcode
  `\&12\catcode `\#12\catcode `\^12\catcode `\_12\catcode `\%12\relax}%
\providecommand \@@startlink[1]{}%
\providecommand \@@endlink[0]{}%
\providecommand \url  [0]{\begingroup\@sanitize@url \@url }%
\providecommand \@url [1]{\endgroup\@href {#1}{\urlprefix }}%
\providecommand \urlprefix  [0]{URL }%
\providecommand \Eprint [0]{\href }%
\providecommand \doibase [0]{http://dx.doi.org/}%
\providecommand \selectlanguage [0]{\@gobble}%
\providecommand \bibinfo  [0]{\@secondoftwo}%
\providecommand \bibfield  [0]{\@secondoftwo}%
\providecommand \translation [1]{[#1]}%
\providecommand \BibitemOpen [0]{}%
\providecommand \bibitemStop [0]{}%
\providecommand \bibitemNoStop [0]{.\EOS\space}%
\providecommand \EOS [0]{\spacefactor3000\relax}%
\providecommand \BibitemShut  [1]{\csname bibitem#1\endcsname}%
\let\auto@bib@innerbib\@empty
\bibitem [{\citenamefont {Ideue}\ and\ \citenamefont
  {Iwasa}(2021)}]{Ideue2021}%
  \BibitemOpen
  \bibfield  {author} {\bibinfo {author} {\bibfnamefont {T.}~\bibnamefont
  {Ideue}}\ and\ \bibinfo {author} {\bibfnamefont {Y.}~\bibnamefont {Iwasa}},\
  }\href {\doibase 10.1146/annurev-conmatphys-060220-100347} {\bibfield
  {journal} {\bibinfo  {journal} {Annual Review of Condensed Matter Physics}\
  }\textbf {\bibinfo {volume} {12}},\ \bibinfo {pages} {201} (\bibinfo {year}
  {2021})},\ \Eprint
  {http://arxiv.org/abs/https://doi.org/10.1146/annurev-conmatphys-060220-100347}
  {https://doi.org/10.1146/annurev-conmatphys-060220-100347} \BibitemShut
  {NoStop}%
\bibitem [{\citenamefont {Tokura}\ and\ \citenamefont
  {Nagaosa}(2018)}]{Tokura2018}%
  \BibitemOpen
  \bibfield  {author} {\bibinfo {author} {\bibfnamefont {Y.}~\bibnamefont
  {Tokura}}\ and\ \bibinfo {author} {\bibfnamefont {N.}~\bibnamefont
  {Nagaosa}},\ }\href {\doibase 10.1038/s41467-018-05759-4} {\bibfield
  {journal} {\bibinfo  {journal} {Nature Communications}\ }\textbf {\bibinfo
  {volume} {9}},\ \bibinfo {pages} {3740} (\bibinfo {year} {2018})}\BibitemShut
  {NoStop}%
\bibitem [{\citenamefont {Rikken}\ \emph {et~al.}(2001)\citenamefont {Rikken},
  \citenamefont {F\"olling},\ and\ \citenamefont {Wyder}}]{Rikken2001}%
  \BibitemOpen
  \bibfield  {author} {\bibinfo {author} {\bibfnamefont {G.~L. J.~A.}\
  \bibnamefont {Rikken}}, \bibinfo {author} {\bibfnamefont {J.}~\bibnamefont
  {F\"olling}}, \ and\ \bibinfo {author} {\bibfnamefont {P.}~\bibnamefont
  {Wyder}},\ }\href {\doibase 10.1103/PhysRevLett.87.236602} {\bibfield
  {journal} {\bibinfo  {journal} {Phys. Rev. Lett.}\ }\textbf {\bibinfo
  {volume} {87}},\ \bibinfo {pages} {236602} (\bibinfo {year}
  {2001})}\BibitemShut {NoStop}%
\bibitem [{\citenamefont {Krstić}\ \emph {et~al.}(2002)\citenamefont
  {Krstić}, \citenamefont {Roth}, \citenamefont {Burghard}, \citenamefont
  {Kern},\ and\ \citenamefont {Rikken}}]{Krstic2002}%
  \BibitemOpen
  \bibfield  {author} {\bibinfo {author} {\bibfnamefont {V.}~\bibnamefont
  {Krstić}}, \bibinfo {author} {\bibfnamefont {S.}~\bibnamefont {Roth}},
  \bibinfo {author} {\bibfnamefont {M.}~\bibnamefont {Burghard}}, \bibinfo
  {author} {\bibfnamefont {K.}~\bibnamefont {Kern}}, \ and\ \bibinfo {author}
  {\bibfnamefont {G.~L. J.~A.}\ \bibnamefont {Rikken}},\ }\href {\doibase
  10.1063/1.1523895} {\bibfield  {journal} {\bibinfo  {journal} {The Journal of
  Chemical Physics}\ }\textbf {\bibinfo {volume} {117}},\ \bibinfo {pages}
  {11315} (\bibinfo {year} {2002})},\ \Eprint
  {http://arxiv.org/abs/https://pubs.aip.org/aip/jcp/article-pdf/117/24/11315/10844217/11315\_1\_online.pdf}
  {https://pubs.aip.org/aip/jcp/article-pdf/117/24/11315/10844217/11315\_1\_online.pdf}
  \BibitemShut {NoStop}%
\bibitem [{\citenamefont {Pop}\ \emph {et~al.}(2014)\citenamefont {Pop},
  \citenamefont {Auban-Senzier}, \citenamefont {Canadell}, \citenamefont
  {Rikken},\ and\ \citenamefont {Avarvari}}]{Pop2014}%
  \BibitemOpen
  \bibfield  {author} {\bibinfo {author} {\bibfnamefont {F.}~\bibnamefont
  {Pop}}, \bibinfo {author} {\bibfnamefont {P.}~\bibnamefont {Auban-Senzier}},
  \bibinfo {author} {\bibfnamefont {E.}~\bibnamefont {Canadell}}, \bibinfo
  {author} {\bibfnamefont {G.~L. J.~A.}\ \bibnamefont {Rikken}}, \ and\
  \bibinfo {author} {\bibfnamefont {N.}~\bibnamefont {Avarvari}},\ }\href
  {\doibase 10.1038/ncomms4757} {\bibfield  {journal} {\bibinfo  {journal}
  {Nature Communications}\ }\textbf {\bibinfo {volume} {5}},\ \bibinfo {pages}
  {3757} (\bibinfo {year} {2014})}\BibitemShut {NoStop}%
\bibitem [{\citenamefont {Rikken}\ and\ \citenamefont
  {Wyder}(2005)}]{Rikken2005}%
  \BibitemOpen
  \bibfield  {author} {\bibinfo {author} {\bibfnamefont {G.~L. J.~A.}\
  \bibnamefont {Rikken}}\ and\ \bibinfo {author} {\bibfnamefont
  {P.}~\bibnamefont {Wyder}},\ }\href {\doibase 10.1103/PhysRevLett.94.016601}
  {\bibfield  {journal} {\bibinfo  {journal} {Phys. Rev. Lett.}\ }\textbf
  {\bibinfo {volume} {94}},\ \bibinfo {pages} {016601} (\bibinfo {year}
  {2005})}\BibitemShut {NoStop}%
\bibitem [{\citenamefont {Ideue}\ \emph {et~al.}(2017)\citenamefont {Ideue},
  \citenamefont {Hamamoto}, \citenamefont {Koshikawa}, \citenamefont {Ezawa},
  \citenamefont {Shimizu}, \citenamefont {Kaneko}, \citenamefont {Tokura},
  \citenamefont {Nagaosa},\ and\ \citenamefont {Iwasa}}]{Ideue2017}%
  \BibitemOpen
  \bibfield  {author} {\bibinfo {author} {\bibfnamefont {T.}~\bibnamefont
  {Ideue}}, \bibinfo {author} {\bibfnamefont {K.}~\bibnamefont {Hamamoto}},
  \bibinfo {author} {\bibfnamefont {S.}~\bibnamefont {Koshikawa}}, \bibinfo
  {author} {\bibfnamefont {M.}~\bibnamefont {Ezawa}}, \bibinfo {author}
  {\bibfnamefont {S.}~\bibnamefont {Shimizu}}, \bibinfo {author} {\bibfnamefont
  {Y.}~\bibnamefont {Kaneko}}, \bibinfo {author} {\bibfnamefont
  {Y.}~\bibnamefont {Tokura}}, \bibinfo {author} {\bibfnamefont
  {N.}~\bibnamefont {Nagaosa}}, \ and\ \bibinfo {author} {\bibfnamefont
  {Y.}~\bibnamefont {Iwasa}},\ }\href {\doibase 10.1038/nphys4056} {\bibfield
  {journal} {\bibinfo  {journal} {Nature Physics}\ }\textbf {\bibinfo {volume}
  {13}},\ \bibinfo {pages} {578} (\bibinfo {year} {2017})}\BibitemShut
  {NoStop}%
\bibitem [{\citenamefont {Wakatsuki}\ and\ \citenamefont
  {Nagaosa}(2018)}]{Wakatsuki2018}%
  \BibitemOpen
  \bibfield  {author} {\bibinfo {author} {\bibfnamefont {R.}~\bibnamefont
  {Wakatsuki}}\ and\ \bibinfo {author} {\bibfnamefont {N.}~\bibnamefont
  {Nagaosa}},\ }\href {\doibase 10.1103/PhysRevLett.121.026601} {\bibfield
  {journal} {\bibinfo  {journal} {Phys. Rev. Lett.}\ }\textbf {\bibinfo
  {volume} {121}},\ \bibinfo {pages} {026601} (\bibinfo {year}
  {2018})}\BibitemShut {NoStop}%
\bibitem [{\citenamefont {Hoshino}\ \emph {et~al.}(2018)\citenamefont
  {Hoshino}, \citenamefont {Wakatsuki}, \citenamefont {Hamamoto},\ and\
  \citenamefont {Nagaosa}}]{Hoshino2018}%
  \BibitemOpen
  \bibfield  {author} {\bibinfo {author} {\bibfnamefont {S.}~\bibnamefont
  {Hoshino}}, \bibinfo {author} {\bibfnamefont {R.}~\bibnamefont {Wakatsuki}},
  \bibinfo {author} {\bibfnamefont {K.}~\bibnamefont {Hamamoto}}, \ and\
  \bibinfo {author} {\bibfnamefont {N.}~\bibnamefont {Nagaosa}},\ }\href
  {\doibase 10.1103/PhysRevB.98.054510} {\bibfield  {journal} {\bibinfo
  {journal} {Phys. Rev. B}\ }\textbf {\bibinfo {volume} {98}},\ \bibinfo
  {pages} {054510} (\bibinfo {year} {2018})}\BibitemShut {NoStop}%
\bibitem [{\citenamefont {Wakatsuki}\ \emph {et~al.}(2017)\citenamefont
  {Wakatsuki}, \citenamefont {Saito}, \citenamefont {Hoshino}, \citenamefont
  {Itahashi}, \citenamefont {Ideue}, \citenamefont {Ezawa}, \citenamefont
  {Iwasa},\ and\ \citenamefont {Nagaosa}}]{Wakatsuki2017}%
  \BibitemOpen
  \bibfield  {author} {\bibinfo {author} {\bibfnamefont {R.}~\bibnamefont
  {Wakatsuki}}, \bibinfo {author} {\bibfnamefont {Y.}~\bibnamefont {Saito}},
  \bibinfo {author} {\bibfnamefont {S.}~\bibnamefont {Hoshino}}, \bibinfo
  {author} {\bibfnamefont {Y.~M.}\ \bibnamefont {Itahashi}}, \bibinfo {author}
  {\bibfnamefont {T.}~\bibnamefont {Ideue}}, \bibinfo {author} {\bibfnamefont
  {M.}~\bibnamefont {Ezawa}}, \bibinfo {author} {\bibfnamefont
  {Y.}~\bibnamefont {Iwasa}}, \ and\ \bibinfo {author} {\bibfnamefont
  {N.}~\bibnamefont {Nagaosa}},\ }\href {\doibase 10.1126/sciadv.1602390}
  {\bibfield  {journal} {\bibinfo  {journal} {Science Advances}\ }\textbf
  {\bibinfo {volume} {3}},\ \bibinfo {pages} {e1602390} (\bibinfo {year}
  {2017})},\ \Eprint
  {http://arxiv.org/abs/https://www.science.org/doi/pdf/10.1126/sciadv.1602390}
  {https://www.science.org/doi/pdf/10.1126/sciadv.1602390} \BibitemShut
  {NoStop}%
\bibitem [{\citenamefont {Qin}\ \emph {et~al.}(2017)\citenamefont {Qin},
  \citenamefont {Shi}, \citenamefont {Ideue}, \citenamefont {Yoshida},
  \citenamefont {Zak}, \citenamefont {Tenne}, \citenamefont {Kikitsu},
  \citenamefont {Inoue}, \citenamefont {Hashizume},\ and\ \citenamefont
  {Iwasa}}]{Qin2017}%
  \BibitemOpen
  \bibfield  {author} {\bibinfo {author} {\bibfnamefont {F.}~\bibnamefont
  {Qin}}, \bibinfo {author} {\bibfnamefont {W.}~\bibnamefont {Shi}}, \bibinfo
  {author} {\bibfnamefont {T.}~\bibnamefont {Ideue}}, \bibinfo {author}
  {\bibfnamefont {M.}~\bibnamefont {Yoshida}}, \bibinfo {author} {\bibfnamefont
  {A.}~\bibnamefont {Zak}}, \bibinfo {author} {\bibfnamefont {R.}~\bibnamefont
  {Tenne}}, \bibinfo {author} {\bibfnamefont {T.}~\bibnamefont {Kikitsu}},
  \bibinfo {author} {\bibfnamefont {D.}~\bibnamefont {Inoue}}, \bibinfo
  {author} {\bibfnamefont {D.}~\bibnamefont {Hashizume}}, \ and\ \bibinfo
  {author} {\bibfnamefont {Y.}~\bibnamefont {Iwasa}},\ }\href {\doibase
  10.1038/ncomms14465} {\bibfield  {journal} {\bibinfo  {journal} {Nature
  Communications}\ }\textbf {\bibinfo {volume} {8}},\ \bibinfo {pages} {14465}
  (\bibinfo {year} {2017})}\BibitemShut {NoStop}%
\bibitem [{\citenamefont {Yasuda}\ \emph {et~al.}(2019)\citenamefont {Yasuda},
  \citenamefont {Yasuda}, \citenamefont {Liang}, \citenamefont {Yoshimi},
  \citenamefont {Tsukazaki}, \citenamefont {Takahashi}, \citenamefont
  {Nagaosa}, \citenamefont {Kawasaki},\ and\ \citenamefont
  {Tokura}}]{Yasuda2019}%
  \BibitemOpen
  \bibfield  {author} {\bibinfo {author} {\bibfnamefont {K.}~\bibnamefont
  {Yasuda}}, \bibinfo {author} {\bibfnamefont {H.}~\bibnamefont {Yasuda}},
  \bibinfo {author} {\bibfnamefont {T.}~\bibnamefont {Liang}}, \bibinfo
  {author} {\bibfnamefont {R.}~\bibnamefont {Yoshimi}}, \bibinfo {author}
  {\bibfnamefont {A.}~\bibnamefont {Tsukazaki}}, \bibinfo {author}
  {\bibfnamefont {K.~S.}\ \bibnamefont {Takahashi}}, \bibinfo {author}
  {\bibfnamefont {N.}~\bibnamefont {Nagaosa}}, \bibinfo {author} {\bibfnamefont
  {M.}~\bibnamefont {Kawasaki}}, \ and\ \bibinfo {author} {\bibfnamefont
  {Y.}~\bibnamefont {Tokura}},\ }\href {\doibase 10.1038/s41467-019-10658-3}
  {\bibfield  {journal} {\bibinfo  {journal} {Nature Communications}\ }\textbf
  {\bibinfo {volume} {10}},\ \bibinfo {pages} {2734} (\bibinfo {year}
  {2019})}\BibitemShut {NoStop}%
\bibitem [{\citenamefont {Ideue}\ \emph {et~al.}(2020)\citenamefont {Ideue},
  \citenamefont {Koshikawa}, \citenamefont {Namiki}, \citenamefont {Sasagawa},\
  and\ \citenamefont {Iwasa}}]{Ideue2020}%
  \BibitemOpen
  \bibfield  {author} {\bibinfo {author} {\bibfnamefont {T.}~\bibnamefont
  {Ideue}}, \bibinfo {author} {\bibfnamefont {S.}~\bibnamefont {Koshikawa}},
  \bibinfo {author} {\bibfnamefont {H.}~\bibnamefont {Namiki}}, \bibinfo
  {author} {\bibfnamefont {T.}~\bibnamefont {Sasagawa}}, \ and\ \bibinfo
  {author} {\bibfnamefont {Y.}~\bibnamefont {Iwasa}},\ }\href {\doibase
  10.1103/PhysRevResearch.2.042046} {\bibfield  {journal} {\bibinfo  {journal}
  {Phys. Rev. Res.}\ }\textbf {\bibinfo {volume} {2}},\ \bibinfo {pages}
  {042046} (\bibinfo {year} {2020})}\BibitemShut {NoStop}%
\bibitem [{\citenamefont {Itahashi}\ \emph {et~al.}(2020)\citenamefont
  {Itahashi}, \citenamefont {Ideue}, \citenamefont {Saito}, \citenamefont
  {Shimizu}, \citenamefont {Ouchi}, \citenamefont {Nojima},\ and\ \citenamefont
  {Iwasa}}]{Itahashi2020}%
  \BibitemOpen
  \bibfield  {author} {\bibinfo {author} {\bibfnamefont {Y.~M.}\ \bibnamefont
  {Itahashi}}, \bibinfo {author} {\bibfnamefont {T.}~\bibnamefont {Ideue}},
  \bibinfo {author} {\bibfnamefont {Y.}~\bibnamefont {Saito}}, \bibinfo
  {author} {\bibfnamefont {S.}~\bibnamefont {Shimizu}}, \bibinfo {author}
  {\bibfnamefont {T.}~\bibnamefont {Ouchi}}, \bibinfo {author} {\bibfnamefont
  {T.}~\bibnamefont {Nojima}}, \ and\ \bibinfo {author} {\bibfnamefont
  {Y.}~\bibnamefont {Iwasa}},\ }\href {\doibase 10.1126/sciadv.aay9120}
  {\bibfield  {journal} {\bibinfo  {journal} {Science Advances}\ }\textbf
  {\bibinfo {volume} {6}},\ \bibinfo {pages} {eaay9120} (\bibinfo {year}
  {2020})},\ \Eprint
  {http://arxiv.org/abs/https://www.science.org/doi/pdf/10.1126/sciadv.aay9120}
  {https://www.science.org/doi/pdf/10.1126/sciadv.aay9120} \BibitemShut
  {NoStop}%
\bibitem [{\citenamefont {Orenstein}\ \emph {et~al.}(2021)\citenamefont
  {Orenstein}, \citenamefont {Moore}, \citenamefont {Morimoto}, \citenamefont
  {Torchinsky}, \citenamefont {Harter},\ and\ \citenamefont
  {Hsieh}}]{Orenstein2021}%
  \BibitemOpen
  \bibfield  {author} {\bibinfo {author} {\bibfnamefont {J.}~\bibnamefont
  {Orenstein}}, \bibinfo {author} {\bibfnamefont {J.}~\bibnamefont {Moore}},
  \bibinfo {author} {\bibfnamefont {T.}~\bibnamefont {Morimoto}}, \bibinfo
  {author} {\bibfnamefont {D.}~\bibnamefont {Torchinsky}}, \bibinfo {author}
  {\bibfnamefont {J.}~\bibnamefont {Harter}}, \ and\ \bibinfo {author}
  {\bibfnamefont {D.}~\bibnamefont {Hsieh}},\ }\href {\doibase
  10.1146/annurev-conmatphys-031218-013712} {\bibfield  {journal} {\bibinfo
  {journal} {Annual Review of Condensed Matter Physics}\ }\textbf {\bibinfo
  {volume} {12}},\ \bibinfo {pages} {247} (\bibinfo {year} {2021})},\ \Eprint
  {http://arxiv.org/abs/https://doi.org/10.1146/annurev-conmatphys-031218-013712}
  {https://doi.org/10.1146/annurev-conmatphys-031218-013712} \BibitemShut
  {NoStop}%
\bibitem [{\citenamefont {Zhao}\ \emph {et~al.}(2017)\citenamefont {Zhao},
  \citenamefont {Belvin}, \citenamefont {Liang}, \citenamefont {Bonn},
  \citenamefont {Hardy}, \citenamefont {Armitage},\ and\ \citenamefont
  {Hsieh}}]{Zhao2017}%
  \BibitemOpen
  \bibfield  {author} {\bibinfo {author} {\bibfnamefont {L.}~\bibnamefont
  {Zhao}}, \bibinfo {author} {\bibfnamefont {C.~A.}\ \bibnamefont {Belvin}},
  \bibinfo {author} {\bibfnamefont {R.}~\bibnamefont {Liang}}, \bibinfo
  {author} {\bibfnamefont {D.~A.}\ \bibnamefont {Bonn}}, \bibinfo {author}
  {\bibfnamefont {W.~N.}\ \bibnamefont {Hardy}}, \bibinfo {author}
  {\bibfnamefont {N.~P.}\ \bibnamefont {Armitage}}, \ and\ \bibinfo {author}
  {\bibfnamefont {D.}~\bibnamefont {Hsieh}},\ }\href {\doibase
  10.1038/nphys3962} {\bibfield  {journal} {\bibinfo  {journal} {Nature
  Physics}\ }\textbf {\bibinfo {volume} {13}},\ \bibinfo {pages} {250}
  (\bibinfo {year} {2017})}\BibitemShut {NoStop}%
\bibitem [{\citenamefont {Torre}\ \emph {et~al.}(2021)\citenamefont {Torre},
  \citenamefont {Seyler}, \citenamefont {Zhao}, \citenamefont {Matteo},
  \citenamefont {Scheurer}, \citenamefont {Li}, \citenamefont {Yu},
  \citenamefont {Greven}, \citenamefont {Sachdev}, \citenamefont {Norman},\
  and\ \citenamefont {Hsieh}}]{Torre2021}%
  \BibitemOpen
  \bibfield  {author} {\bibinfo {author} {\bibfnamefont {A.~d.~l.}\
  \bibnamefont {Torre}}, \bibinfo {author} {\bibfnamefont {K.~L.}\ \bibnamefont
  {Seyler}}, \bibinfo {author} {\bibfnamefont {L.}~\bibnamefont {Zhao}},
  \bibinfo {author} {\bibfnamefont {S.~D.}\ \bibnamefont {Matteo}}, \bibinfo
  {author} {\bibfnamefont {M.~S.}\ \bibnamefont {Scheurer}}, \bibinfo {author}
  {\bibfnamefont {Y.}~\bibnamefont {Li}}, \bibinfo {author} {\bibfnamefont
  {B.}~\bibnamefont {Yu}}, \bibinfo {author} {\bibfnamefont {M.}~\bibnamefont
  {Greven}}, \bibinfo {author} {\bibfnamefont {S.}~\bibnamefont {Sachdev}},
  \bibinfo {author} {\bibfnamefont {M.~R.}\ \bibnamefont {Norman}}, \ and\
  \bibinfo {author} {\bibfnamefont {D.}~\bibnamefont {Hsieh}},\ }\href
  {\doibase 10.1038/s41567-021-01210-6} {\bibfield  {journal} {\bibinfo
  {journal} {Nature Physics}\ }\textbf {\bibinfo {volume} {17}},\ \bibinfo
  {pages} {777} (\bibinfo {year} {2021})}\BibitemShut {NoStop}%
\bibitem [{\citenamefont {Fiebig}\ \emph {et~al.}(2005)\citenamefont {Fiebig},
  \citenamefont {Pavlov},\ and\ \citenamefont {Pisarev}}]{Fiebig2005}%
  \BibitemOpen
  \bibfield  {author} {\bibinfo {author} {\bibfnamefont {M.}~\bibnamefont
  {Fiebig}}, \bibinfo {author} {\bibfnamefont {V.~V.}\ \bibnamefont {Pavlov}},
  \ and\ \bibinfo {author} {\bibfnamefont {R.~V.}\ \bibnamefont {Pisarev}},\
  }\href {\doibase 10.1364/JOSAB.22.000096} {\bibfield  {journal} {\bibinfo
  {journal} {J. Opt. Soc. Am. B}\ }\textbf {\bibinfo {volume} {22}},\ \bibinfo
  {pages} {96} (\bibinfo {year} {2005})}\BibitemShut {NoStop}%
\bibitem [{\citenamefont {Liu}\ \emph {et~al.}(2020)\citenamefont {Liu},
  \citenamefont {Xia}, \citenamefont {Xiao}, \citenamefont {Garc{\'i}a~de
  Abajo},\ and\ \citenamefont {Sun}}]{Liu2020}%
  \BibitemOpen
  \bibfield  {author} {\bibinfo {author} {\bibfnamefont {J.}~\bibnamefont
  {Liu}}, \bibinfo {author} {\bibfnamefont {F.}~\bibnamefont {Xia}}, \bibinfo
  {author} {\bibfnamefont {D.}~\bibnamefont {Xiao}}, \bibinfo {author}
  {\bibfnamefont {F.~J.}\ \bibnamefont {Garc{\'i}a~de Abajo}}, \ and\ \bibinfo
  {author} {\bibfnamefont {D.}~\bibnamefont {Sun}},\ }\href {\doibase
  10.1038/s41563-020-0715-7} {\bibfield  {journal} {\bibinfo  {journal} {Nature
  Materials}\ }\textbf {\bibinfo {volume} {19}},\ \bibinfo {pages} {830}
  (\bibinfo {year} {2020})}\BibitemShut {NoStop}%
\bibitem [{\citenamefont {Ogawa}\ \emph {et~al.}(2016)\citenamefont {Ogawa},
  \citenamefont {Yoshimi}, \citenamefont {Yasuda}, \citenamefont {Tsukazaki},
  \citenamefont {Kawasaki},\ and\ \citenamefont {Tokura}}]{Ogawa2016}%
  \BibitemOpen
  \bibfield  {author} {\bibinfo {author} {\bibfnamefont {N.}~\bibnamefont
  {Ogawa}}, \bibinfo {author} {\bibfnamefont {R.}~\bibnamefont {Yoshimi}},
  \bibinfo {author} {\bibfnamefont {K.}~\bibnamefont {Yasuda}}, \bibinfo
  {author} {\bibfnamefont {A.}~\bibnamefont {Tsukazaki}}, \bibinfo {author}
  {\bibfnamefont {M.}~\bibnamefont {Kawasaki}}, \ and\ \bibinfo {author}
  {\bibfnamefont {Y.}~\bibnamefont {Tokura}},\ }\href {\doibase
  10.1038/ncomms12246} {\bibfield  {journal} {\bibinfo  {journal} {Nature
  Communications}\ }\textbf {\bibinfo {volume} {7}},\ \bibinfo {pages} {12246}
  (\bibinfo {year} {2016})}\BibitemShut {NoStop}%
\bibitem [{\citenamefont {Burger}\ \emph {et~al.}(2020)\citenamefont {Burger},
  \citenamefont {Gao}, \citenamefont {Agarwal}, \citenamefont {Aprelev},
  \citenamefont {Spanier}, \citenamefont {Rappe},\ and\ \citenamefont
  {Fridkin}}]{Burger2020}%
  \BibitemOpen
  \bibfield  {author} {\bibinfo {author} {\bibfnamefont {A.~M.}\ \bibnamefont
  {Burger}}, \bibinfo {author} {\bibfnamefont {L.}~\bibnamefont {Gao}},
  \bibinfo {author} {\bibfnamefont {R.}~\bibnamefont {Agarwal}}, \bibinfo
  {author} {\bibfnamefont {A.}~\bibnamefont {Aprelev}}, \bibinfo {author}
  {\bibfnamefont {J.~E.}\ \bibnamefont {Spanier}}, \bibinfo {author}
  {\bibfnamefont {A.~M.}\ \bibnamefont {Rappe}}, \ and\ \bibinfo {author}
  {\bibfnamefont {V.~M.}\ \bibnamefont {Fridkin}},\ }\href {\doibase
  10.1103/PhysRevB.102.081113} {\bibfield  {journal} {\bibinfo  {journal}
  {Phys. Rev. B}\ }\textbf {\bibinfo {volume} {102}},\ \bibinfo {pages}
  {081113} (\bibinfo {year} {2020})}\BibitemShut {NoStop}%
\bibitem [{\citenamefont {Matsunaga}\ \emph {et~al.}(2014)\citenamefont
  {Matsunaga}, \citenamefont {Tsuji}, \citenamefont {Fujita}, \citenamefont
  {Sugioka}, \citenamefont {Makise}, \citenamefont {Uzawa}, \citenamefont
  {Terai}, \citenamefont {Wang}, \citenamefont {Aoki},\ and\ \citenamefont
  {Shimano}}]{Matsunaga2014}%
  \BibitemOpen
  \bibfield  {author} {\bibinfo {author} {\bibfnamefont {R.}~\bibnamefont
  {Matsunaga}}, \bibinfo {author} {\bibfnamefont {N.}~\bibnamefont {Tsuji}},
  \bibinfo {author} {\bibfnamefont {H.}~\bibnamefont {Fujita}}, \bibinfo
  {author} {\bibfnamefont {A.}~\bibnamefont {Sugioka}}, \bibinfo {author}
  {\bibfnamefont {K.}~\bibnamefont {Makise}}, \bibinfo {author} {\bibfnamefont
  {Y.}~\bibnamefont {Uzawa}}, \bibinfo {author} {\bibfnamefont
  {H.}~\bibnamefont {Terai}}, \bibinfo {author} {\bibfnamefont
  {Z.}~\bibnamefont {Wang}}, \bibinfo {author} {\bibfnamefont {H.}~\bibnamefont
  {Aoki}}, \ and\ \bibinfo {author} {\bibfnamefont {R.}~\bibnamefont
  {Shimano}},\ }\href {\doibase 10.1126/science.1254697} {\bibfield  {journal}
  {\bibinfo  {journal} {Science}\ }\textbf {\bibinfo {volume} {345}},\ \bibinfo
  {pages} {1145} (\bibinfo {year} {2014})},\ \Eprint
  {http://arxiv.org/abs/https://www.science.org/doi/pdf/10.1126/science.1254697}
  {https://www.science.org/doi/pdf/10.1126/science.1254697} \BibitemShut
  {NoStop}%
\bibitem [{\citenamefont {Cea}\ \emph {et~al.}(2016)\citenamefont {Cea},
  \citenamefont {Castellani},\ and\ \citenamefont {Benfatto}}]{Cea2016}%
  \BibitemOpen
  \bibfield  {author} {\bibinfo {author} {\bibfnamefont {T.}~\bibnamefont
  {Cea}}, \bibinfo {author} {\bibfnamefont {C.}~\bibnamefont {Castellani}}, \
  and\ \bibinfo {author} {\bibfnamefont {L.}~\bibnamefont {Benfatto}},\ }\href
  {\doibase 10.1103/PhysRevB.93.180507} {\bibfield  {journal} {\bibinfo
  {journal} {Phys. Rev. B}\ }\textbf {\bibinfo {volume} {93}},\ \bibinfo
  {pages} {180507} (\bibinfo {year} {2016})}\BibitemShut {NoStop}%
\bibitem [{\citenamefont {Shimano}\ and\ \citenamefont
  {Tsuji}(2020)}]{Shimano-Tsuji}%
  \BibitemOpen
  \bibfield  {author} {\bibinfo {author} {\bibfnamefont {R.}~\bibnamefont
  {Shimano}}\ and\ \bibinfo {author} {\bibfnamefont {N.}~\bibnamefont
  {Tsuji}},\ }\href {\doibase 10.1146/annurev-conmatphys-031119-050813}
  {\bibfield  {journal} {\bibinfo  {journal} {Annual Review of Condensed Matter
  Physics}\ }\textbf {\bibinfo {volume} {11}},\ \bibinfo {pages} {103}
  (\bibinfo {year} {2020})},\ \Eprint
  {http://arxiv.org/abs/https://doi.org/10.1146/annurev-conmatphys-031119-050813}
  {https://doi.org/10.1146/annurev-conmatphys-031119-050813} \BibitemShut
  {NoStop}%
\bibitem [{\citenamefont {Gor'kov}\ and\ \citenamefont
  {Rashba}(2001)}]{Gorkov2001}%
  \BibitemOpen
  \bibfield  {author} {\bibinfo {author} {\bibfnamefont {L.~P.}\ \bibnamefont
  {Gor'kov}}\ and\ \bibinfo {author} {\bibfnamefont {E.~I.}\ \bibnamefont
  {Rashba}},\ }\href {\doibase 10.1103/PhysRevLett.87.037004} {\bibfield
  {journal} {\bibinfo  {journal} {Phys. Rev. Lett.}\ }\textbf {\bibinfo
  {volume} {87}},\ \bibinfo {pages} {037004} (\bibinfo {year}
  {2001})}\BibitemShut {NoStop}%
\bibitem [{\citenamefont {Frigeri}\ \emph {et~al.}(2004)\citenamefont
  {Frigeri}, \citenamefont {Agterberg}, \citenamefont {Koga},\ and\
  \citenamefont {Sigrist}}]{Frigeri2004}%
  \BibitemOpen
  \bibfield  {author} {\bibinfo {author} {\bibfnamefont {P.~A.}\ \bibnamefont
  {Frigeri}}, \bibinfo {author} {\bibfnamefont {D.~F.}\ \bibnamefont
  {Agterberg}}, \bibinfo {author} {\bibfnamefont {A.}~\bibnamefont {Koga}}, \
  and\ \bibinfo {author} {\bibfnamefont {M.}~\bibnamefont {Sigrist}},\ }\href
  {\doibase 10.1103/PhysRevLett.92.097001} {\bibfield  {journal} {\bibinfo
  {journal} {Phys. Rev. Lett.}\ }\textbf {\bibinfo {volume} {92}},\ \bibinfo
  {pages} {097001} (\bibinfo {year} {2004})}\BibitemShut {NoStop}%
\bibitem [{\citenamefont {Bauer}\ and\ \citenamefont
  {Sigrist}(2012)}]{Bauer2012}%
  \BibitemOpen
  \bibfield  {author} {\bibinfo {author} {\bibfnamefont {E.}~\bibnamefont
  {Bauer}}\ and\ \bibinfo {author} {\bibfnamefont {M.}~\bibnamefont
  {Sigrist}},\ }\href@noop {} {\emph {\bibinfo {title} {Non-centrosymmetric
  superconductors: introduction and overview}}}\ (\bibinfo  {publisher}
  {Springer Science \& Business Media},\ \bibinfo {year} {2012})\BibitemShut
  {NoStop}%
\bibitem [{\citenamefont {Smidman}\ \emph {et~al.}(2017)\citenamefont
  {Smidman}, \citenamefont {Salamon}, \citenamefont {Yuan},\ and\ \citenamefont
  {Agterberg}}]{Smidman2017}%
  \BibitemOpen
  \bibfield  {author} {\bibinfo {author} {\bibfnamefont {M.}~\bibnamefont
  {Smidman}}, \bibinfo {author} {\bibfnamefont {M.~B.}\ \bibnamefont
  {Salamon}}, \bibinfo {author} {\bibfnamefont {H.~Q.}\ \bibnamefont {Yuan}}, \
  and\ \bibinfo {author} {\bibfnamefont {D.~F.}\ \bibnamefont {Agterberg}},\
  }\href {\doibase 10.1088/1361-6633/80/3/036501} {\bibfield  {journal}
  {\bibinfo  {journal} {Reports on Progress in Physics}\ }\textbf {\bibinfo
  {volume} {80}},\ \bibinfo {pages} {036501} (\bibinfo {year}
  {2017})}\BibitemShut {NoStop}%
\bibitem [{\citenamefont {Alicea}(2012)}]{Alicea2012}%
  \BibitemOpen
  \bibfield  {author} {\bibinfo {author} {\bibfnamefont {J.}~\bibnamefont
  {Alicea}},\ }\href {\doibase 10.1088/0034-4885/75/7/076501} {\bibfield
  {journal} {\bibinfo  {journal} {Reports on Progress in Physics}\ }\textbf
  {\bibinfo {volume} {75}},\ \bibinfo {pages} {076501} (\bibinfo {year}
  {2012})}\BibitemShut {NoStop}%
\bibitem [{\citenamefont {Sato}\ and\ \citenamefont
  {Fujimoto}(2016)}]{Sato2016}%
  \BibitemOpen
  \bibfield  {author} {\bibinfo {author} {\bibfnamefont {M.}~\bibnamefont
  {Sato}}\ and\ \bibinfo {author} {\bibfnamefont {S.}~\bibnamefont
  {Fujimoto}},\ }\href {\doibase 10.7566/JPSJ.85.072001} {\bibfield  {journal}
  {\bibinfo  {journal} {Journal of the Physical Society of Japan}\ }\textbf
  {\bibinfo {volume} {85}},\ \bibinfo {pages} {072001} (\bibinfo {year}
  {2016})},\ \Eprint
  {http://arxiv.org/abs/https://doi.org/10.7566/JPSJ.85.072001}
  {https://doi.org/10.7566/JPSJ.85.072001} \BibitemShut {NoStop}%
\bibitem [{\citenamefont {Saito}\ \emph {et~al.}(2016)\citenamefont {Saito},
  \citenamefont {Nakamura}, \citenamefont {Bahramy}, \citenamefont {Kohama},
  \citenamefont {Ye}, \citenamefont {Kasahara}, \citenamefont {Nakagawa},
  \citenamefont {Onga}, \citenamefont {Tokunaga}, \citenamefont {Nojima},
  \citenamefont {Yanase},\ and\ \citenamefont {Iwasa}}]{Saito2016}%
  \BibitemOpen
  \bibfield  {author} {\bibinfo {author} {\bibfnamefont {Y.}~\bibnamefont
  {Saito}}, \bibinfo {author} {\bibfnamefont {Y.}~\bibnamefont {Nakamura}},
  \bibinfo {author} {\bibfnamefont {M.~S.}\ \bibnamefont {Bahramy}}, \bibinfo
  {author} {\bibfnamefont {Y.}~\bibnamefont {Kohama}}, \bibinfo {author}
  {\bibfnamefont {J.}~\bibnamefont {Ye}}, \bibinfo {author} {\bibfnamefont
  {Y.}~\bibnamefont {Kasahara}}, \bibinfo {author} {\bibfnamefont
  {Y.}~\bibnamefont {Nakagawa}}, \bibinfo {author} {\bibfnamefont
  {M.}~\bibnamefont {Onga}}, \bibinfo {author} {\bibfnamefont {M.}~\bibnamefont
  {Tokunaga}}, \bibinfo {author} {\bibfnamefont {T.}~\bibnamefont {Nojima}},
  \bibinfo {author} {\bibfnamefont {Y.}~\bibnamefont {Yanase}}, \ and\ \bibinfo
  {author} {\bibfnamefont {Y.}~\bibnamefont {Iwasa}},\ }\href {\doibase
  10.1038/nphys3580} {\bibfield  {journal} {\bibinfo  {journal} {Nature
  Physics}\ }\textbf {\bibinfo {volume} {12}},\ \bibinfo {pages} {144}
  (\bibinfo {year} {2016})}\BibitemShut {NoStop}%
\bibitem [{\citenamefont {Lu}\ \emph {et~al.}(2015)\citenamefont {Lu},
  \citenamefont {Zheliuk}, \citenamefont {Leermakers}, \citenamefont {Yuan},
  \citenamefont {Zeitler}, \citenamefont {Law},\ and\ \citenamefont
  {Ye}}]{Lu2015}%
  \BibitemOpen
  \bibfield  {author} {\bibinfo {author} {\bibfnamefont {J.~M.}\ \bibnamefont
  {Lu}}, \bibinfo {author} {\bibfnamefont {O.}~\bibnamefont {Zheliuk}},
  \bibinfo {author} {\bibfnamefont {I.}~\bibnamefont {Leermakers}}, \bibinfo
  {author} {\bibfnamefont {N.~F.~Q.}\ \bibnamefont {Yuan}}, \bibinfo {author}
  {\bibfnamefont {U.}~\bibnamefont {Zeitler}}, \bibinfo {author} {\bibfnamefont
  {K.~T.}\ \bibnamefont {Law}}, \ and\ \bibinfo {author} {\bibfnamefont
  {J.~T.}\ \bibnamefont {Ye}},\ }\href {\doibase 10.1126/science.aab2277}
  {\bibfield  {journal} {\bibinfo  {journal} {Science}\ }\textbf {\bibinfo
  {volume} {350}},\ \bibinfo {pages} {1353} (\bibinfo {year} {2015})},\ \Eprint
  {http://arxiv.org/abs/https://www.science.org/doi/pdf/10.1126/science.aab2277}
  {https://www.science.org/doi/pdf/10.1126/science.aab2277} \BibitemShut
  {NoStop}%
\bibitem [{\citenamefont {Xi}\ \emph {et~al.}(2016)\citenamefont {Xi},
  \citenamefont {Wang}, \citenamefont {Zhao}, \citenamefont {Park},
  \citenamefont {Law}, \citenamefont {Berger}, \citenamefont {Forr{\'o}},
  \citenamefont {Shan},\ and\ \citenamefont {Mak}}]{Xi2016}%
  \BibitemOpen
  \bibfield  {author} {\bibinfo {author} {\bibfnamefont {X.}~\bibnamefont
  {Xi}}, \bibinfo {author} {\bibfnamefont {Z.}~\bibnamefont {Wang}}, \bibinfo
  {author} {\bibfnamefont {W.}~\bibnamefont {Zhao}}, \bibinfo {author}
  {\bibfnamefont {J.-H.}\ \bibnamefont {Park}}, \bibinfo {author}
  {\bibfnamefont {K.~T.}\ \bibnamefont {Law}}, \bibinfo {author} {\bibfnamefont
  {H.}~\bibnamefont {Berger}}, \bibinfo {author} {\bibfnamefont
  {L.}~\bibnamefont {Forr{\'o}}}, \bibinfo {author} {\bibfnamefont
  {J.}~\bibnamefont {Shan}}, \ and\ \bibinfo {author} {\bibfnamefont {K.~F.}\
  \bibnamefont {Mak}},\ }\href {\doibase 10.1038/nphys3538} {\bibfield
  {journal} {\bibinfo  {journal} {Nature Physics}\ }\textbf {\bibinfo {volume}
  {12}},\ \bibinfo {pages} {139} (\bibinfo {year} {2016})}\BibitemShut
  {NoStop}%
\bibitem [{\citenamefont {de~la Barrera}\ \emph {et~al.}(2018)\citenamefont
  {de~la Barrera}, \citenamefont {Sinko}, \citenamefont {Gopalan},
  \citenamefont {Sivadas}, \citenamefont {Seyler}, \citenamefont {Watanabe},
  \citenamefont {Taniguchi}, \citenamefont {Tsen}, \citenamefont {Xu},
  \citenamefont {Xiao},\ and\ \citenamefont {Hunt}}]{delaBarrera2018}%
  \BibitemOpen
  \bibfield  {author} {\bibinfo {author} {\bibfnamefont {S.~C.}\ \bibnamefont
  {de~la Barrera}}, \bibinfo {author} {\bibfnamefont {M.~R.}\ \bibnamefont
  {Sinko}}, \bibinfo {author} {\bibfnamefont {D.~P.}\ \bibnamefont {Gopalan}},
  \bibinfo {author} {\bibfnamefont {N.}~\bibnamefont {Sivadas}}, \bibinfo
  {author} {\bibfnamefont {K.~L.}\ \bibnamefont {Seyler}}, \bibinfo {author}
  {\bibfnamefont {K.}~\bibnamefont {Watanabe}}, \bibinfo {author}
  {\bibfnamefont {T.}~\bibnamefont {Taniguchi}}, \bibinfo {author}
  {\bibfnamefont {A.~W.}\ \bibnamefont {Tsen}}, \bibinfo {author}
  {\bibfnamefont {X.}~\bibnamefont {Xu}}, \bibinfo {author} {\bibfnamefont
  {D.}~\bibnamefont {Xiao}}, \ and\ \bibinfo {author} {\bibfnamefont {B.~M.}\
  \bibnamefont {Hunt}},\ }\href {\doibase 10.1038/s41467-018-03888-4}
  {\bibfield  {journal} {\bibinfo  {journal} {Nature Communications}\ }\textbf
  {\bibinfo {volume} {9}},\ \bibinfo {pages} {1427} (\bibinfo {year}
  {2018})}\BibitemShut {NoStop}%
\bibitem [{\citenamefont {Ando}\ \emph {et~al.}(2020)\citenamefont {Ando},
  \citenamefont {Miyasaka}, \citenamefont {Li}, \citenamefont {Ishizuka},
  \citenamefont {Arakawa}, \citenamefont {Shiota}, \citenamefont {Moriyama},
  \citenamefont {Yanase},\ and\ \citenamefont {Ono}}]{Ando2020}%
  \BibitemOpen
  \bibfield  {author} {\bibinfo {author} {\bibfnamefont {F.}~\bibnamefont
  {Ando}}, \bibinfo {author} {\bibfnamefont {Y.}~\bibnamefont {Miyasaka}},
  \bibinfo {author} {\bibfnamefont {T.}~\bibnamefont {Li}}, \bibinfo {author}
  {\bibfnamefont {J.}~\bibnamefont {Ishizuka}}, \bibinfo {author}
  {\bibfnamefont {T.}~\bibnamefont {Arakawa}}, \bibinfo {author} {\bibfnamefont
  {Y.}~\bibnamefont {Shiota}}, \bibinfo {author} {\bibfnamefont
  {T.}~\bibnamefont {Moriyama}}, \bibinfo {author} {\bibfnamefont
  {Y.}~\bibnamefont {Yanase}}, \ and\ \bibinfo {author} {\bibfnamefont
  {T.}~\bibnamefont {Ono}},\ }\href {\doibase 10.1038/s41586-020-2590-4}
  {\bibfield  {journal} {\bibinfo  {journal} {Nature}\ }\textbf {\bibinfo
  {volume} {584}},\ \bibinfo {pages} {373} (\bibinfo {year}
  {2020})}\BibitemShut {NoStop}%
\bibitem [{\citenamefont {Nagaosa}\ and\ \citenamefont
  {Yanase}(2024)}]{Nagaosa-Yanase}%
  \BibitemOpen
  \bibfield  {author} {\bibinfo {author} {\bibfnamefont {N.}~\bibnamefont
  {Nagaosa}}\ and\ \bibinfo {author} {\bibfnamefont {Y.}~\bibnamefont
  {Yanase}},\ }\href {\doibase 10.1146/annurev-conmatphys-032822-033734}
  {\bibfield  {journal} {\bibinfo  {journal} {Annual Review of Condensed Matter
  Physics}\ }\textbf {\bibinfo {volume} {15}},\ \bibinfo {pages} {63} (\bibinfo
  {year} {2024})},\ \Eprint
  {http://arxiv.org/abs/https://doi.org/10.1146/annurev-conmatphys-032822-033734}
  {https://doi.org/10.1146/annurev-conmatphys-032822-033734} \BibitemShut
  {NoStop}%
\bibitem [{\citenamefont {Nadeem}\ \emph {et~al.}(2023)\citenamefont {Nadeem},
  \citenamefont {Fuhrer},\ and\ \citenamefont {Wang}}]{Nadeem2023}%
  \BibitemOpen
  \bibfield  {author} {\bibinfo {author} {\bibfnamefont {M.}~\bibnamefont
  {Nadeem}}, \bibinfo {author} {\bibfnamefont {M.~S.}\ \bibnamefont {Fuhrer}},
  \ and\ \bibinfo {author} {\bibfnamefont {X.}~\bibnamefont {Wang}},\ }\href
  {\doibase 10.1038/s42254-023-00632-w} {\bibfield  {journal} {\bibinfo
  {journal} {Nature Reviews Physics}\ }\textbf {\bibinfo {volume} {5}},\
  \bibinfo {pages} {558} (\bibinfo {year} {2023})}\BibitemShut {NoStop}%
\bibitem [{\citenamefont {Daido}\ \emph {et~al.}(2022)\citenamefont {Daido},
  \citenamefont {Ikeda},\ and\ \citenamefont {Yanase}}]{Daido2022}%
  \BibitemOpen
  \bibfield  {author} {\bibinfo {author} {\bibfnamefont {A.}~\bibnamefont
  {Daido}}, \bibinfo {author} {\bibfnamefont {Y.}~\bibnamefont {Ikeda}}, \ and\
  \bibinfo {author} {\bibfnamefont {Y.}~\bibnamefont {Yanase}},\ }\href
  {\doibase 10.1103/PhysRevLett.128.037001} {\bibfield  {journal} {\bibinfo
  {journal} {Phys. Rev. Lett.}\ }\textbf {\bibinfo {volume} {128}},\ \bibinfo
  {pages} {037001} (\bibinfo {year} {2022})}\BibitemShut {NoStop}%
\bibitem [{\citenamefont {He}\ \emph {et~al.}(2022)\citenamefont {He},
  \citenamefont {Tanaka},\ and\ \citenamefont {Nagaosa}}]{He2022}%
  \BibitemOpen
  \bibfield  {author} {\bibinfo {author} {\bibfnamefont {J.~J.}\ \bibnamefont
  {He}}, \bibinfo {author} {\bibfnamefont {Y.}~\bibnamefont {Tanaka}}, \ and\
  \bibinfo {author} {\bibfnamefont {N.}~\bibnamefont {Nagaosa}},\ }\href
  {\doibase 10.1088/1367-2630/ac6766} {\bibfield  {journal} {\bibinfo
  {journal} {New Journal of Physics}\ }\textbf {\bibinfo {volume} {24}},\
  \bibinfo {pages} {053014} (\bibinfo {year} {2022})}\BibitemShut {NoStop}%
\bibitem [{\citenamefont {Yuan}\ and\ \citenamefont {Fu}(2022)}]{Yuan2022}%
  \BibitemOpen
  \bibfield  {author} {\bibinfo {author} {\bibfnamefont {N.~F.~Q.}\
  \bibnamefont {Yuan}}\ and\ \bibinfo {author} {\bibfnamefont {L.}~\bibnamefont
  {Fu}},\ }\href {\doibase 10.1073/pnas.2119548119} {\bibfield  {journal}
  {\bibinfo  {journal} {Proceedings of the National Academy of Sciences}\
  }\textbf {\bibinfo {volume} {119}},\ \bibinfo {pages} {e2119548119} (\bibinfo
  {year} {2022})},\ \Eprint
  {http://arxiv.org/abs/https://www.pnas.org/doi/pdf/10.1073/pnas.2119548119}
  {https://www.pnas.org/doi/pdf/10.1073/pnas.2119548119} \BibitemShut {NoStop}%
\bibitem [{\citenamefont {Daido}\ and\ \citenamefont
  {Yanase}(2023)}]{daido2023rectification}%
  \BibitemOpen
  \bibfield  {author} {\bibinfo {author} {\bibfnamefont {A.}~\bibnamefont
  {Daido}}\ and\ \bibinfo {author} {\bibfnamefont {Y.}~\bibnamefont {Yanase}},\
  }\href@noop {} {\enquote {\bibinfo {title} {Rectification and nonlinear hall
  effect by fluctuating finite-momentum cooper pairs},}\ } (\bibinfo {year}
  {2023}),\ \Eprint {http://arxiv.org/abs/2302.10677} {arXiv:2302.10677
  [cond-mat.supr-con]} \BibitemShut {NoStop}%
\bibitem [{\citenamefont {Nakamura}\ \emph {et~al.}(2020)\citenamefont
  {Nakamura}, \citenamefont {Katsumi}, \citenamefont {Terai},\ and\
  \citenamefont {Shimano}}]{Nakamura2020}%
  \BibitemOpen
  \bibfield  {author} {\bibinfo {author} {\bibfnamefont {S.}~\bibnamefont
  {Nakamura}}, \bibinfo {author} {\bibfnamefont {K.}~\bibnamefont {Katsumi}},
  \bibinfo {author} {\bibfnamefont {H.}~\bibnamefont {Terai}}, \ and\ \bibinfo
  {author} {\bibfnamefont {R.}~\bibnamefont {Shimano}},\ }\href {\doibase
  10.1103/PhysRevLett.125.097004} {\bibfield  {journal} {\bibinfo  {journal}
  {Phys. Rev. Lett.}\ }\textbf {\bibinfo {volume} {125}},\ \bibinfo {pages}
  {097004} (\bibinfo {year} {2020})}\BibitemShut {NoStop}%
\bibitem [{\citenamefont {Vaswani}\ \emph {et~al.}(2020)\citenamefont
  {Vaswani}, \citenamefont {Mootz}, \citenamefont {Sundahl}, \citenamefont
  {Mudiyanselage}, \citenamefont {Kang}, \citenamefont {Yang}, \citenamefont
  {Cheng}, \citenamefont {Huang}, \citenamefont {Kim}, \citenamefont {Liu},
  \citenamefont {Luo}, \citenamefont {Perakis}, \citenamefont {Eom},\ and\
  \citenamefont {Wang}}]{Vaswani2020-ij}%
  \BibitemOpen
  \bibfield  {author} {\bibinfo {author} {\bibfnamefont {C.}~\bibnamefont
  {Vaswani}}, \bibinfo {author} {\bibfnamefont {M.}~\bibnamefont {Mootz}},
  \bibinfo {author} {\bibfnamefont {C.}~\bibnamefont {Sundahl}}, \bibinfo
  {author} {\bibfnamefont {D.~H.}\ \bibnamefont {Mudiyanselage}}, \bibinfo
  {author} {\bibfnamefont {J.~H.}\ \bibnamefont {Kang}}, \bibinfo {author}
  {\bibfnamefont {X.}~\bibnamefont {Yang}}, \bibinfo {author} {\bibfnamefont
  {D.}~\bibnamefont {Cheng}}, \bibinfo {author} {\bibfnamefont
  {C.}~\bibnamefont {Huang}}, \bibinfo {author} {\bibfnamefont {R.~H.~J.}\
  \bibnamefont {Kim}}, \bibinfo {author} {\bibfnamefont {Z.}~\bibnamefont
  {Liu}}, \bibinfo {author} {\bibfnamefont {L.}~\bibnamefont {Luo}}, \bibinfo
  {author} {\bibfnamefont {I.~E.}\ \bibnamefont {Perakis}}, \bibinfo {author}
  {\bibfnamefont {C.~B.}\ \bibnamefont {Eom}}, \ and\ \bibinfo {author}
  {\bibfnamefont {J.}~\bibnamefont {Wang}},\ }\href
  {http://dx.doi.org/10.1103/PhysRevLett.124.207003} {\bibfield  {journal}
  {\bibinfo  {journal} {Physical Review Letters}\ }\textbf {\bibinfo {volume}
  {124}},\ \bibinfo {pages} {207003} (\bibinfo {year} {2020})}\BibitemShut
  {NoStop}%
\bibitem [{\citenamefont {Xu}\ \emph {et~al.}(2019)\citenamefont {Xu},
  \citenamefont {Morimoto},\ and\ \citenamefont {Moore}}]{Xu2019}%
  \BibitemOpen
  \bibfield  {author} {\bibinfo {author} {\bibfnamefont {T.}~\bibnamefont
  {Xu}}, \bibinfo {author} {\bibfnamefont {T.}~\bibnamefont {Morimoto}}, \ and\
  \bibinfo {author} {\bibfnamefont {J.~E.}\ \bibnamefont {Moore}},\ }\href
  {\doibase 10.1103/PhysRevB.100.220501} {\bibfield  {journal} {\bibinfo
  {journal} {Phys. Rev. B}\ }\textbf {\bibinfo {volume} {100}},\ \bibinfo
  {pages} {220501} (\bibinfo {year} {2019})}\BibitemShut {NoStop}%
\bibitem [{\citenamefont {Watanabe}\ \emph {et~al.}(2022)\citenamefont
  {Watanabe}, \citenamefont {Daido},\ and\ \citenamefont
  {Yanase}}]{Watanabe2022}%
  \BibitemOpen
  \bibfield  {author} {\bibinfo {author} {\bibfnamefont {H.}~\bibnamefont
  {Watanabe}}, \bibinfo {author} {\bibfnamefont {A.}~\bibnamefont {Daido}}, \
  and\ \bibinfo {author} {\bibfnamefont {Y.}~\bibnamefont {Yanase}},\ }\href
  {\doibase 10.1103/PhysRevB.105.024308} {\bibfield  {journal} {\bibinfo
  {journal} {Phys. Rev. B}\ }\textbf {\bibinfo {volume} {105}},\ \bibinfo
  {pages} {024308} (\bibinfo {year} {2022})}\BibitemShut {NoStop}%
\bibitem [{\citenamefont {Tanaka}\ \emph {et~al.}(2023)\citenamefont {Tanaka},
  \citenamefont {Watanabe},\ and\ \citenamefont {Yanase}}]{Tanaka2023}%
  \BibitemOpen
  \bibfield  {author} {\bibinfo {author} {\bibfnamefont {H.}~\bibnamefont
  {Tanaka}}, \bibinfo {author} {\bibfnamefont {H.}~\bibnamefont {Watanabe}}, \
  and\ \bibinfo {author} {\bibfnamefont {Y.}~\bibnamefont {Yanase}},\ }\href
  {\doibase 10.1103/PhysRevB.107.024513} {\bibfield  {journal} {\bibinfo
  {journal} {Phys. Rev. B}\ }\textbf {\bibinfo {volume} {107}},\ \bibinfo
  {pages} {024513} (\bibinfo {year} {2023})}\BibitemShut {NoStop}%
\bibitem [{\citenamefont {Huang}\ and\ \citenamefont {Wang}(2023)}]{Huang2023}%
  \BibitemOpen
  \bibfield  {author} {\bibinfo {author} {\bibfnamefont {L.}~\bibnamefont
  {Huang}}\ and\ \bibinfo {author} {\bibfnamefont {J.}~\bibnamefont {Wang}},\
  }\href {\doibase 10.1103/PhysRevB.108.224516} {\bibfield  {journal} {\bibinfo
   {journal} {Phys. Rev. B}\ }\textbf {\bibinfo {volume} {108}},\ \bibinfo
  {pages} {224516} (\bibinfo {year} {2023})}\BibitemShut {NoStop}%
\bibitem [{\citenamefont {Watanabe}\ and\ \citenamefont
  {Yanase}(2021)}]{Watanabe2021}%
  \BibitemOpen
  \bibfield  {author} {\bibinfo {author} {\bibfnamefont {H.}~\bibnamefont
  {Watanabe}}\ and\ \bibinfo {author} {\bibfnamefont {Y.}~\bibnamefont
  {Yanase}},\ }\href {\doibase 10.1103/PhysRevX.11.011001} {\bibfield
  {journal} {\bibinfo  {journal} {Phys. Rev. X}\ }\textbf {\bibinfo {volume}
  {11}},\ \bibinfo {pages} {011001} (\bibinfo {year} {2021})}\BibitemShut
  {NoStop}%
\bibitem [{\citenamefont {Ahn}\ \emph {et~al.}(2020)\citenamefont {Ahn},
  \citenamefont {Guo},\ and\ \citenamefont {Nagaosa}}]{Ahn2020}%
  \BibitemOpen
  \bibfield  {author} {\bibinfo {author} {\bibfnamefont {J.}~\bibnamefont
  {Ahn}}, \bibinfo {author} {\bibfnamefont {G.-Y.}\ \bibnamefont {Guo}}, \ and\
  \bibinfo {author} {\bibfnamefont {N.}~\bibnamefont {Nagaosa}},\ }\href
  {\doibase 10.1103/PhysRevX.10.041041} {\bibfield  {journal} {\bibinfo
  {journal} {Phys. Rev. X}\ }\textbf {\bibinfo {volume} {10}},\ \bibinfo
  {pages} {041041} (\bibinfo {year} {2020})}\BibitemShut {NoStop}%
\bibitem [{\citenamefont {Resta}(2011)}]{Resta2011}%
  \BibitemOpen
  \bibfield  {author} {\bibinfo {author} {\bibfnamefont {R.}~\bibnamefont
  {Resta}},\ }\href {\doibase 10.1140/epjb/e2010-10874-4} {\bibfield  {journal}
  {\bibinfo  {journal} {The European Physical Journal B}\ }\textbf {\bibinfo
  {volume} {79}},\ \bibinfo {pages} {121} (\bibinfo {year} {2011})}\BibitemShut
  {NoStop}%
\bibitem [{\citenamefont {Sipe}\ and\ \citenamefont
  {Shkrebtii}(2000)}]{Sipe2000}%
  \BibitemOpen
  \bibfield  {author} {\bibinfo {author} {\bibfnamefont {J.~E.}\ \bibnamefont
  {Sipe}}\ and\ \bibinfo {author} {\bibfnamefont {A.~I.}\ \bibnamefont
  {Shkrebtii}},\ }\href {\doibase 10.1103/PhysRevB.61.5337} {\bibfield
  {journal} {\bibinfo  {journal} {Phys. Rev. B}\ }\textbf {\bibinfo {volume}
  {61}},\ \bibinfo {pages} {5337} (\bibinfo {year} {2000})}\BibitemShut
  {NoStop}%
\bibitem [{\citenamefont {Zhang}\ \emph {et~al.}(2019)\citenamefont {Zhang},
  \citenamefont {Holder}, \citenamefont {Ishizuka}, \citenamefont {de~Juan},
  \citenamefont {Nagaosa}, \citenamefont {Felser},\ and\ \citenamefont
  {Yan}}]{Zhang2019}%
  \BibitemOpen
  \bibfield  {author} {\bibinfo {author} {\bibfnamefont {Y.}~\bibnamefont
  {Zhang}}, \bibinfo {author} {\bibfnamefont {T.}~\bibnamefont {Holder}},
  \bibinfo {author} {\bibfnamefont {H.}~\bibnamefont {Ishizuka}}, \bibinfo
  {author} {\bibfnamefont {F.}~\bibnamefont {de~Juan}}, \bibinfo {author}
  {\bibfnamefont {N.}~\bibnamefont {Nagaosa}}, \bibinfo {author} {\bibfnamefont
  {C.}~\bibnamefont {Felser}}, \ and\ \bibinfo {author} {\bibfnamefont
  {B.}~\bibnamefont {Yan}},\ }\href {\doibase 10.1038/s41467-019-11832-3}
  {\bibfield  {journal} {\bibinfo  {journal} {Nature Communications}\ }\textbf
  {\bibinfo {volume} {10}},\ \bibinfo {pages} {3783} (\bibinfo {year}
  {2019})}\BibitemShut {NoStop}%
\bibitem [{\citenamefont {von Baltz}\ and\ \citenamefont
  {Kraut}(1981)}]{Von_Balz1981}%
  \BibitemOpen
  \bibfield  {author} {\bibinfo {author} {\bibfnamefont {R.}~\bibnamefont {von
  Baltz}}\ and\ \bibinfo {author} {\bibfnamefont {W.}~\bibnamefont {Kraut}},\
  }\href {\doibase 10.1103/PhysRevB.23.5590} {\bibfield  {journal} {\bibinfo
  {journal} {Phys. Rev. B}\ }\textbf {\bibinfo {volume} {23}},\ \bibinfo
  {pages} {5590} (\bibinfo {year} {1981})}\BibitemShut {NoStop}%
\bibitem [{\citenamefont {Okada}\ \emph {et~al.}(2016)\citenamefont {Okada},
  \citenamefont {Ogawa}, \citenamefont {Yoshimi}, \citenamefont {Tsukazaki},
  \citenamefont {Takahashi}, \citenamefont {Kawasaki},\ and\ \citenamefont
  {Tokura}}]{Okada2016}%
  \BibitemOpen
  \bibfield  {author} {\bibinfo {author} {\bibfnamefont {K.~N.}\ \bibnamefont
  {Okada}}, \bibinfo {author} {\bibfnamefont {N.}~\bibnamefont {Ogawa}},
  \bibinfo {author} {\bibfnamefont {R.}~\bibnamefont {Yoshimi}}, \bibinfo
  {author} {\bibfnamefont {A.}~\bibnamefont {Tsukazaki}}, \bibinfo {author}
  {\bibfnamefont {K.~S.}\ \bibnamefont {Takahashi}}, \bibinfo {author}
  {\bibfnamefont {M.}~\bibnamefont {Kawasaki}}, \ and\ \bibinfo {author}
  {\bibfnamefont {Y.}~\bibnamefont {Tokura}},\ }\href {\doibase
  10.1103/PhysRevB.93.081403} {\bibfield  {journal} {\bibinfo  {journal} {Phys.
  Rev. B}\ }\textbf {\bibinfo {volume} {93}},\ \bibinfo {pages} {081403}
  (\bibinfo {year} {2016})}\BibitemShut {NoStop}%
\bibitem [{\citenamefont {Morimoto}\ and\ \citenamefont
  {Nagaosa}(2016)}]{Morimoto2016}%
  \BibitemOpen
  \bibfield  {author} {\bibinfo {author} {\bibfnamefont {T.}~\bibnamefont
  {Morimoto}}\ and\ \bibinfo {author} {\bibfnamefont {N.}~\bibnamefont
  {Nagaosa}},\ }\href {\doibase 10.1103/PhysRevB.93.125125} {\bibfield
  {journal} {\bibinfo  {journal} {Phys. Rev. B}\ }\textbf {\bibinfo {volume}
  {93}},\ \bibinfo {pages} {125125} (\bibinfo {year} {2016})}\BibitemShut
  {NoStop}%
\bibitem [{\citenamefont {Hosur}(2011)}]{Hosur2011}%
  \BibitemOpen
  \bibfield  {author} {\bibinfo {author} {\bibfnamefont {P.}~\bibnamefont
  {Hosur}},\ }\href {\doibase 10.1103/PhysRevB.83.035309} {\bibfield  {journal}
  {\bibinfo  {journal} {Phys. Rev. B}\ }\textbf {\bibinfo {volume} {83}},\
  \bibinfo {pages} {035309} (\bibinfo {year} {2011})}\BibitemShut {NoStop}%
\bibitem [{\citenamefont {Sato}\ \emph {et~al.}(2009)\citenamefont {Sato},
  \citenamefont {Takahashi},\ and\ \citenamefont {Fujimoto}}]{Sato2009}%
  \BibitemOpen
  \bibfield  {author} {\bibinfo {author} {\bibfnamefont {M.}~\bibnamefont
  {Sato}}, \bibinfo {author} {\bibfnamefont {Y.}~\bibnamefont {Takahashi}}, \
  and\ \bibinfo {author} {\bibfnamefont {S.}~\bibnamefont {Fujimoto}},\ }\href
  {\doibase 10.1103/PhysRevLett.103.020401} {\bibfield  {journal} {\bibinfo
  {journal} {Phys. Rev. Lett.}\ }\textbf {\bibinfo {volume} {103}},\ \bibinfo
  {pages} {020401} (\bibinfo {year} {2009})}\BibitemShut {NoStop}%
\bibitem [{\citenamefont {Sato}\ \emph {et~al.}(2010)\citenamefont {Sato},
  \citenamefont {Takahashi},\ and\ \citenamefont {Fujimoto}}]{Sato2010}%
  \BibitemOpen
  \bibfield  {author} {\bibinfo {author} {\bibfnamefont {M.}~\bibnamefont
  {Sato}}, \bibinfo {author} {\bibfnamefont {Y.}~\bibnamefont {Takahashi}}, \
  and\ \bibinfo {author} {\bibfnamefont {S.}~\bibnamefont {Fujimoto}},\ }\href
  {\doibase 10.1103/PhysRevB.82.134521} {\bibfield  {journal} {\bibinfo
  {journal} {Phys. Rev. B}\ }\textbf {\bibinfo {volume} {82}},\ \bibinfo
  {pages} {134521} (\bibinfo {year} {2010})}\BibitemShut {NoStop}%
\bibitem [{\citenamefont {Sau}\ \emph {et~al.}(2010)\citenamefont {Sau},
  \citenamefont {Lutchyn}, \citenamefont {Tewari},\ and\ \citenamefont
  {Das~Sarma}}]{Sau2010}%
  \BibitemOpen
  \bibfield  {author} {\bibinfo {author} {\bibfnamefont {J.~D.}\ \bibnamefont
  {Sau}}, \bibinfo {author} {\bibfnamefont {R.~M.}\ \bibnamefont {Lutchyn}},
  \bibinfo {author} {\bibfnamefont {S.}~\bibnamefont {Tewari}}, \ and\ \bibinfo
  {author} {\bibfnamefont {S.}~\bibnamefont {Das~Sarma}},\ }\href {\doibase
  10.1103/PhysRevLett.104.040502} {\bibfield  {journal} {\bibinfo  {journal}
  {Phys. Rev. Lett.}\ }\textbf {\bibinfo {volume} {104}},\ \bibinfo {pages}
  {040502} (\bibinfo {year} {2010})}\BibitemShut {NoStop}%
\bibitem [{\citenamefont {He}\ \emph {et~al.}(2021)\citenamefont {He},
  \citenamefont {Tanaka},\ and\ \citenamefont {Nagaosa}}]{He2021PRL}%
  \BibitemOpen
  \bibfield  {author} {\bibinfo {author} {\bibfnamefont {J.~J.}\ \bibnamefont
  {He}}, \bibinfo {author} {\bibfnamefont {Y.}~\bibnamefont {Tanaka}}, \ and\
  \bibinfo {author} {\bibfnamefont {N.}~\bibnamefont {Nagaosa}},\ }\href
  {\doibase 10.1103/PhysRevLett.126.237002} {\bibfield  {journal} {\bibinfo
  {journal} {Phys. Rev. Lett.}\ }\textbf {\bibinfo {volume} {126}},\ \bibinfo
  {pages} {237002} (\bibinfo {year} {2021})}\BibitemShut {NoStop}%
\bibitem [{\citenamefont {He}\ and\ \citenamefont {Nagaosa}(2021)}]{He2021PRB}%
  \BibitemOpen
  \bibfield  {author} {\bibinfo {author} {\bibfnamefont {J.~J.}\ \bibnamefont
  {He}}\ and\ \bibinfo {author} {\bibfnamefont {N.}~\bibnamefont {Nagaosa}},\
  }\href {\doibase 10.1103/PhysRevB.103.L241109} {\bibfield  {journal}
  {\bibinfo  {journal} {Phys. Rev. B}\ }\textbf {\bibinfo {volume} {103}},\
  \bibinfo {pages} {L241109} (\bibinfo {year} {2021})}\BibitemShut {NoStop}%
\bibitem [{\citenamefont {Bi}\ and\ \citenamefont {He}(2023)}]{Bi2023}%
  \BibitemOpen
  \bibfield  {author} {\bibinfo {author} {\bibfnamefont {H.}~\bibnamefont
  {Bi}}\ and\ \bibinfo {author} {\bibfnamefont {J.~J.}\ \bibnamefont {He}},\
  }\href@noop {} {\enquote {\bibinfo {title} {Inter-band optical transitions of
  helical majorana edge modes in topological superconductors},}\ } (\bibinfo
  {year} {2023}),\ \Eprint {http://arxiv.org/abs/2310.17244} {arXiv:2310.17244
  [cond-mat.supr-con]} \BibitemShut {NoStop}%
\bibitem [{\citenamefont {Raj}\ \emph {et~al.}(2023)\citenamefont {Raj},
  \citenamefont {Postlewaite}, \citenamefont {Chaudhary},\ and\ \citenamefont
  {Fiete}}]{Raj2023}%
  \BibitemOpen
  \bibfield  {author} {\bibinfo {author} {\bibfnamefont {A.}~\bibnamefont
  {Raj}}, \bibinfo {author} {\bibfnamefont {A.}~\bibnamefont {Postlewaite}},
  \bibinfo {author} {\bibfnamefont {S.}~\bibnamefont {Chaudhary}}, \ and\
  \bibinfo {author} {\bibfnamefont {G.~A.}\ \bibnamefont {Fiete}},\ }\href@noop
  {} {\enquote {\bibinfo {title} {Nonlinear optical responses in multi-orbital
  topological superconductors},}\ } (\bibinfo {year} {2023}),\ \Eprint
  {http://arxiv.org/abs/2312.08638} {arXiv:2312.08638 [cond-mat.supr-con]}
  \BibitemShut {NoStop}%
\bibitem [{\citenamefont {Ye}\ \emph {et~al.}(2012)\citenamefont {Ye},
  \citenamefont {Zhang}, \citenamefont {Akashi}, \citenamefont {Bahramy},
  \citenamefont {Arita},\ and\ \citenamefont {Iwasa}}]{Ye2012}%
  \BibitemOpen
  \bibfield  {author} {\bibinfo {author} {\bibfnamefont {J.~T.}\ \bibnamefont
  {Ye}}, \bibinfo {author} {\bibfnamefont {Y.~J.}\ \bibnamefont {Zhang}},
  \bibinfo {author} {\bibfnamefont {R.}~\bibnamefont {Akashi}}, \bibinfo
  {author} {\bibfnamefont {M.~S.}\ \bibnamefont {Bahramy}}, \bibinfo {author}
  {\bibfnamefont {R.}~\bibnamefont {Arita}}, \ and\ \bibinfo {author}
  {\bibfnamefont {Y.}~\bibnamefont {Iwasa}},\ }\href {\doibase
  10.1126/science.1228006} {\bibfield  {journal} {\bibinfo  {journal}
  {Science}\ }\textbf {\bibinfo {volume} {338}},\ \bibinfo {pages} {1193}
  (\bibinfo {year} {2012})},\ \Eprint
  {http://arxiv.org/abs/https://www.science.org/doi/pdf/10.1126/science.1228006}
  {https://www.science.org/doi/pdf/10.1126/science.1228006} \BibitemShut
  {NoStop}%
\bibitem [{\citenamefont {Reyren}\ \emph {et~al.}(2007)\citenamefont {Reyren},
  \citenamefont {Thiel}, \citenamefont {Caviglia}, \citenamefont {Kourkoutis},
  \citenamefont {Hammerl}, \citenamefont {Richter}, \citenamefont {Schneider},
  \citenamefont {Kopp}, \citenamefont {Rüetschi}, \citenamefont {Jaccard},
  \citenamefont {Gabay}, \citenamefont {Muller}, \citenamefont {Triscone},\
  and\ \citenamefont {Mannhart}}]{Reyren2007}%
  \BibitemOpen
  \bibfield  {author} {\bibinfo {author} {\bibfnamefont {N.}~\bibnamefont
  {Reyren}}, \bibinfo {author} {\bibfnamefont {S.}~\bibnamefont {Thiel}},
  \bibinfo {author} {\bibfnamefont {A.~D.}\ \bibnamefont {Caviglia}}, \bibinfo
  {author} {\bibfnamefont {L.~F.}\ \bibnamefont {Kourkoutis}}, \bibinfo
  {author} {\bibfnamefont {G.}~\bibnamefont {Hammerl}}, \bibinfo {author}
  {\bibfnamefont {C.}~\bibnamefont {Richter}}, \bibinfo {author} {\bibfnamefont
  {C.~W.}\ \bibnamefont {Schneider}}, \bibinfo {author} {\bibfnamefont
  {T.}~\bibnamefont {Kopp}}, \bibinfo {author} {\bibfnamefont {A.-S.}\
  \bibnamefont {Rüetschi}}, \bibinfo {author} {\bibfnamefont {D.}~\bibnamefont
  {Jaccard}}, \bibinfo {author} {\bibfnamefont {M.}~\bibnamefont {Gabay}},
  \bibinfo {author} {\bibfnamefont {D.~A.}\ \bibnamefont {Muller}}, \bibinfo
  {author} {\bibfnamefont {J.-M.}\ \bibnamefont {Triscone}}, \ and\ \bibinfo
  {author} {\bibfnamefont {J.}~\bibnamefont {Mannhart}},\ }\href {\doibase
  10.1126/science.1146006} {\bibfield  {journal} {\bibinfo  {journal}
  {Science}\ }\textbf {\bibinfo {volume} {317}},\ \bibinfo {pages} {1196}
  (\bibinfo {year} {2007})},\ \Eprint
  {http://arxiv.org/abs/https://www.science.org/doi/pdf/10.1126/science.1146006}
  {https://www.science.org/doi/pdf/10.1126/science.1146006} \BibitemShut
  {NoStop}%
\bibitem [{\citenamefont {Ueno}\ \emph {et~al.}(2008)\citenamefont {Ueno},
  \citenamefont {Nakamura}, \citenamefont {Shimotani}, \citenamefont {Ohtomo},
  \citenamefont {Kimura}, \citenamefont {Nojima}, \citenamefont {Aoki},
  \citenamefont {Iwasa},\ and\ \citenamefont {Kawasaki}}]{Ueno2008}%
  \BibitemOpen
  \bibfield  {author} {\bibinfo {author} {\bibfnamefont {K.}~\bibnamefont
  {Ueno}}, \bibinfo {author} {\bibfnamefont {S.}~\bibnamefont {Nakamura}},
  \bibinfo {author} {\bibfnamefont {H.}~\bibnamefont {Shimotani}}, \bibinfo
  {author} {\bibfnamefont {A.}~\bibnamefont {Ohtomo}}, \bibinfo {author}
  {\bibfnamefont {N.}~\bibnamefont {Kimura}}, \bibinfo {author} {\bibfnamefont
  {T.}~\bibnamefont {Nojima}}, \bibinfo {author} {\bibfnamefont
  {H.}~\bibnamefont {Aoki}}, \bibinfo {author} {\bibfnamefont {Y.}~\bibnamefont
  {Iwasa}}, \ and\ \bibinfo {author} {\bibfnamefont {M.}~\bibnamefont
  {Kawasaki}},\ }\href {\doibase 10.1038/nmat2298} {\bibfield  {journal}
  {\bibinfo  {journal} {Nature Materials}\ }\textbf {\bibinfo {volume} {7}},\
  \bibinfo {pages} {855} (\bibinfo {year} {2008})}\BibitemShut {NoStop}%
\bibitem [{\citenamefont {Sekihara}\ \emph {et~al.}(2013)\citenamefont
  {Sekihara}, \citenamefont {Masutomi},\ and\ \citenamefont
  {Okamoto}}]{Sekihara2013}%
  \BibitemOpen
  \bibfield  {author} {\bibinfo {author} {\bibfnamefont {T.}~\bibnamefont
  {Sekihara}}, \bibinfo {author} {\bibfnamefont {R.}~\bibnamefont {Masutomi}},
  \ and\ \bibinfo {author} {\bibfnamefont {T.}~\bibnamefont {Okamoto}},\ }\href
  {\doibase 10.1103/PhysRevLett.111.057005} {\bibfield  {journal} {\bibinfo
  {journal} {Phys. Rev. Lett.}\ }\textbf {\bibinfo {volume} {111}},\ \bibinfo
  {pages} {057005} (\bibinfo {year} {2013})}\BibitemShut {NoStop}%
\bibitem [{\citenamefont {Lyanda-Geller}\ \emph {et~al.}(2015)\citenamefont
  {Lyanda-Geller}, \citenamefont {Li},\ and\ \citenamefont
  {Andreev}}]{Lyanda-Geller2015}%
  \BibitemOpen
  \bibfield  {author} {\bibinfo {author} {\bibfnamefont {Y.~B.}\ \bibnamefont
  {Lyanda-Geller}}, \bibinfo {author} {\bibfnamefont {S.}~\bibnamefont {Li}}, \
  and\ \bibinfo {author} {\bibfnamefont {A.~V.}\ \bibnamefont {Andreev}},\
  }\href {\doibase 10.1103/PhysRevB.92.241406} {\bibfield  {journal} {\bibinfo
  {journal} {Phys. Rev. B}\ }\textbf {\bibinfo {volume} {92}},\ \bibinfo
  {pages} {241406} (\bibinfo {year} {2015})}\BibitemShut {NoStop}%
\bibitem [{\citenamefont {de~Juan}\ \emph {et~al.}(2017)\citenamefont
  {de~Juan}, \citenamefont {Grushin}, \citenamefont {Morimoto},\ and\
  \citenamefont {Moore}}]{deJuan2017}%
  \BibitemOpen
  \bibfield  {author} {\bibinfo {author} {\bibfnamefont {F.}~\bibnamefont
  {de~Juan}}, \bibinfo {author} {\bibfnamefont {A.~G.}\ \bibnamefont
  {Grushin}}, \bibinfo {author} {\bibfnamefont {T.}~\bibnamefont {Morimoto}}, \
  and\ \bibinfo {author} {\bibfnamefont {J.~E.}\ \bibnamefont {Moore}},\ }\href
  {\doibase 10.1038/ncomms15995} {\bibfield  {journal} {\bibinfo  {journal}
  {Nature Communications}\ }\textbf {\bibinfo {volume} {8}},\ \bibinfo {pages}
  {15995} (\bibinfo {year} {2017})}\BibitemShut {NoStop}%
\bibitem [{\citenamefont {Rees}\ \emph {et~al.}(2020)\citenamefont {Rees},
  \citenamefont {Manna}, \citenamefont {Lu}, \citenamefont {Morimoto},
  \citenamefont {Borrmann}, \citenamefont {Felser}, \citenamefont {Moore},
  \citenamefont {Torchinsky},\ and\ \citenamefont {Orenstein}}]{Rees2020}%
  \BibitemOpen
  \bibfield  {author} {\bibinfo {author} {\bibfnamefont {D.}~\bibnamefont
  {Rees}}, \bibinfo {author} {\bibfnamefont {K.}~\bibnamefont {Manna}},
  \bibinfo {author} {\bibfnamefont {B.}~\bibnamefont {Lu}}, \bibinfo {author}
  {\bibfnamefont {T.}~\bibnamefont {Morimoto}}, \bibinfo {author}
  {\bibfnamefont {H.}~\bibnamefont {Borrmann}}, \bibinfo {author}
  {\bibfnamefont {C.}~\bibnamefont {Felser}}, \bibinfo {author} {\bibfnamefont
  {J.~E.}\ \bibnamefont {Moore}}, \bibinfo {author} {\bibfnamefont {D.~H.}\
  \bibnamefont {Torchinsky}}, \ and\ \bibinfo {author} {\bibfnamefont
  {J.}~\bibnamefont {Orenstein}},\ }\href {\doibase 10.1126/sciadv.aba0509}
  {\bibfield  {journal} {\bibinfo  {journal} {Science Advances}\ }\textbf
  {\bibinfo {volume} {6}},\ \bibinfo {pages} {eaba0509} (\bibinfo {year}
  {2020})},\ \Eprint
  {http://arxiv.org/abs/https://www.science.org/doi/pdf/10.1126/sciadv.aba0509}
  {https://www.science.org/doi/pdf/10.1126/sciadv.aba0509} \BibitemShut
  {NoStop}%
\bibitem [{\citenamefont {Ma}\ \emph {et~al.}(2017)\citenamefont {Ma},
  \citenamefont {Xu}, \citenamefont {Chan}, \citenamefont {Zhang},
  \citenamefont {Chang}, \citenamefont {Lin}, \citenamefont {Xie},
  \citenamefont {Palacios}, \citenamefont {Lin}, \citenamefont {Jia},
  \citenamefont {Lee}, \citenamefont {Jarillo-Herrero},\ and\ \citenamefont
  {Gedik}}]{Ma2017}%
  \BibitemOpen
  \bibfield  {author} {\bibinfo {author} {\bibfnamefont {Q.}~\bibnamefont
  {Ma}}, \bibinfo {author} {\bibfnamefont {S.-Y.}\ \bibnamefont {Xu}}, \bibinfo
  {author} {\bibfnamefont {C.-K.}\ \bibnamefont {Chan}}, \bibinfo {author}
  {\bibfnamefont {C.-L.}\ \bibnamefont {Zhang}}, \bibinfo {author}
  {\bibfnamefont {G.}~\bibnamefont {Chang}}, \bibinfo {author} {\bibfnamefont
  {Y.}~\bibnamefont {Lin}}, \bibinfo {author} {\bibfnamefont {W.}~\bibnamefont
  {Xie}}, \bibinfo {author} {\bibfnamefont {T.}~\bibnamefont {Palacios}},
  \bibinfo {author} {\bibfnamefont {H.}~\bibnamefont {Lin}}, \bibinfo {author}
  {\bibfnamefont {S.}~\bibnamefont {Jia}}, \bibinfo {author} {\bibfnamefont
  {P.~A.}\ \bibnamefont {Lee}}, \bibinfo {author} {\bibfnamefont
  {P.}~\bibnamefont {Jarillo-Herrero}}, \ and\ \bibinfo {author} {\bibfnamefont
  {N.}~\bibnamefont {Gedik}},\ }\href {\doibase 10.1038/nphys4146} {\bibfield
  {journal} {\bibinfo  {journal} {Nature Physics}\ }\textbf {\bibinfo {volume}
  {13}},\ \bibinfo {pages} {842} (\bibinfo {year} {2017})}\BibitemShut
  {NoStop}%
\bibitem [{\citenamefont {Huang}\ and\ \citenamefont
  {Hoffman}(2017)}]{Huang2017}%
  \BibitemOpen
  \bibfield  {author} {\bibinfo {author} {\bibfnamefont {D.}~\bibnamefont
  {Huang}}\ and\ \bibinfo {author} {\bibfnamefont {J.~E.}\ \bibnamefont
  {Hoffman}},\ }\href {\doibase 10.1146/annurev-conmatphys-031016-025242}
  {\bibfield  {journal} {\bibinfo  {journal} {Annual Review of Condensed Matter
  Physics}\ }\textbf {\bibinfo {volume} {8}},\ \bibinfo {pages} {311} (\bibinfo
  {year} {2017})},\ \Eprint
  {http://arxiv.org/abs/https://doi.org/10.1146/annurev-conmatphys-031016-025242}
  {https://doi.org/10.1146/annurev-conmatphys-031016-025242} \BibitemShut
  {NoStop}%
\bibitem [{\citenamefont {He}\ \emph {et~al.}(2013)\citenamefont {He},
  \citenamefont {He}, \citenamefont {Zhang}, \citenamefont {Zhao},
  \citenamefont {Liu}, \citenamefont {Liu}, \citenamefont {Mou}, \citenamefont
  {Ou}, \citenamefont {Wang}, \citenamefont {Li}, \citenamefont {Wang},
  \citenamefont {Peng}, \citenamefont {Liu}, \citenamefont {Chen},
  \citenamefont {Yu}, \citenamefont {Liu}, \citenamefont {Dong}, \citenamefont
  {Zhang}, \citenamefont {Chen}, \citenamefont {Xu}, \citenamefont {Chen},
  \citenamefont {Ma}, \citenamefont {Xue},\ and\ \citenamefont
  {Zhou}}]{He2013}%
  \BibitemOpen
  \bibfield  {author} {\bibinfo {author} {\bibfnamefont {S.}~\bibnamefont
  {He}}, \bibinfo {author} {\bibfnamefont {J.}~\bibnamefont {He}}, \bibinfo
  {author} {\bibfnamefont {W.}~\bibnamefont {Zhang}}, \bibinfo {author}
  {\bibfnamefont {L.}~\bibnamefont {Zhao}}, \bibinfo {author} {\bibfnamefont
  {D.}~\bibnamefont {Liu}}, \bibinfo {author} {\bibfnamefont {X.}~\bibnamefont
  {Liu}}, \bibinfo {author} {\bibfnamefont {D.}~\bibnamefont {Mou}}, \bibinfo
  {author} {\bibfnamefont {Y.-B.}\ \bibnamefont {Ou}}, \bibinfo {author}
  {\bibfnamefont {Q.-Y.}\ \bibnamefont {Wang}}, \bibinfo {author}
  {\bibfnamefont {Z.}~\bibnamefont {Li}}, \bibinfo {author} {\bibfnamefont
  {L.}~\bibnamefont {Wang}}, \bibinfo {author} {\bibfnamefont {Y.}~\bibnamefont
  {Peng}}, \bibinfo {author} {\bibfnamefont {Y.}~\bibnamefont {Liu}}, \bibinfo
  {author} {\bibfnamefont {C.}~\bibnamefont {Chen}}, \bibinfo {author}
  {\bibfnamefont {L.}~\bibnamefont {Yu}}, \bibinfo {author} {\bibfnamefont
  {G.}~\bibnamefont {Liu}}, \bibinfo {author} {\bibfnamefont {X.}~\bibnamefont
  {Dong}}, \bibinfo {author} {\bibfnamefont {J.}~\bibnamefont {Zhang}},
  \bibinfo {author} {\bibfnamefont {C.}~\bibnamefont {Chen}}, \bibinfo {author}
  {\bibfnamefont {Z.}~\bibnamefont {Xu}}, \bibinfo {author} {\bibfnamefont
  {X.}~\bibnamefont {Chen}}, \bibinfo {author} {\bibfnamefont {X.}~\bibnamefont
  {Ma}}, \bibinfo {author} {\bibfnamefont {Q.}~\bibnamefont {Xue}}, \ and\
  \bibinfo {author} {\bibfnamefont {X.~J.}\ \bibnamefont {Zhou}},\ }\href
  {\doibase 10.1038/nmat3648} {\bibfield  {journal} {\bibinfo  {journal}
  {Nature Materials}\ }\textbf {\bibinfo {volume} {12}},\ \bibinfo {pages}
  {605} (\bibinfo {year} {2013})}\BibitemShut {NoStop}%
\bibitem [{\citenamefont {Tan}\ \emph {et~al.}(2013)\citenamefont {Tan},
  \citenamefont {Zhang}, \citenamefont {Xia}, \citenamefont {Ye}, \citenamefont
  {Chen}, \citenamefont {Xie}, \citenamefont {Peng}, \citenamefont {Xu},
  \citenamefont {Fan}, \citenamefont {Xu}, \citenamefont {Jiang}, \citenamefont
  {Zhang}, \citenamefont {Lai}, \citenamefont {Xiang}, \citenamefont {Hu},
  \citenamefont {Xie},\ and\ \citenamefont {Feng}}]{Tan2013}%
  \BibitemOpen
  \bibfield  {author} {\bibinfo {author} {\bibfnamefont {S.}~\bibnamefont
  {Tan}}, \bibinfo {author} {\bibfnamefont {Y.}~\bibnamefont {Zhang}}, \bibinfo
  {author} {\bibfnamefont {M.}~\bibnamefont {Xia}}, \bibinfo {author}
  {\bibfnamefont {Z.}~\bibnamefont {Ye}}, \bibinfo {author} {\bibfnamefont
  {F.}~\bibnamefont {Chen}}, \bibinfo {author} {\bibfnamefont {X.}~\bibnamefont
  {Xie}}, \bibinfo {author} {\bibfnamefont {R.}~\bibnamefont {Peng}}, \bibinfo
  {author} {\bibfnamefont {D.}~\bibnamefont {Xu}}, \bibinfo {author}
  {\bibfnamefont {Q.}~\bibnamefont {Fan}}, \bibinfo {author} {\bibfnamefont
  {H.}~\bibnamefont {Xu}}, \bibinfo {author} {\bibfnamefont {J.}~\bibnamefont
  {Jiang}}, \bibinfo {author} {\bibfnamefont {T.}~\bibnamefont {Zhang}},
  \bibinfo {author} {\bibfnamefont {X.}~\bibnamefont {Lai}}, \bibinfo {author}
  {\bibfnamefont {T.}~\bibnamefont {Xiang}}, \bibinfo {author} {\bibfnamefont
  {J.}~\bibnamefont {Hu}}, \bibinfo {author} {\bibfnamefont {B.}~\bibnamefont
  {Xie}}, \ and\ \bibinfo {author} {\bibfnamefont {D.}~\bibnamefont {Feng}},\
  }\href {\doibase 10.1038/nmat3654} {\bibfield  {journal} {\bibinfo  {journal}
  {Nature Materials}\ }\textbf {\bibinfo {volume} {12}},\ \bibinfo {pages}
  {634} (\bibinfo {year} {2013})}\BibitemShut {NoStop}%
\bibitem [{\citenamefont {Shibauchi}\ \emph {et~al.}(2020)\citenamefont
  {Shibauchi}, \citenamefont {Hanaguri},\ and\ \citenamefont
  {Matsuda}}]{Shibauchi_FeSe_review}%
  \BibitemOpen
  \bibfield  {author} {\bibinfo {author} {\bibfnamefont {T.}~\bibnamefont
  {Shibauchi}}, \bibinfo {author} {\bibfnamefont {T.}~\bibnamefont {Hanaguri}},
  \ and\ \bibinfo {author} {\bibfnamefont {Y.}~\bibnamefont {Matsuda}},\ }\href
  {\doibase 10.7566/JPSJ.89.102002} {\bibfield  {journal} {\bibinfo  {journal}
  {Journal of the Physical Society of Japan}\ }\textbf {\bibinfo {volume}
  {89}},\ \bibinfo {pages} {102002} (\bibinfo {year} {2020})},\ \Eprint
  {http://arxiv.org/abs/https://doi.org/10.7566/JPSJ.89.102002}
  {https://doi.org/10.7566/JPSJ.89.102002} \BibitemShut {NoStop}%
\bibitem [{\citenamefont {Zakeri}\ \emph {et~al.}(2023)\citenamefont {Zakeri},
  \citenamefont {Rau}, \citenamefont {Jandke}, \citenamefont {Yang},
  \citenamefont {Wulfhekel},\ and\ \citenamefont {Berthod}}]{Zakeri2023}%
  \BibitemOpen
  \bibfield  {author} {\bibinfo {author} {\bibfnamefont {K.}~\bibnamefont
  {Zakeri}}, \bibinfo {author} {\bibfnamefont {D.}~\bibnamefont {Rau}},
  \bibinfo {author} {\bibfnamefont {J.}~\bibnamefont {Jandke}}, \bibinfo
  {author} {\bibfnamefont {F.}~\bibnamefont {Yang}}, \bibinfo {author}
  {\bibfnamefont {W.}~\bibnamefont {Wulfhekel}}, \ and\ \bibinfo {author}
  {\bibfnamefont {C.}~\bibnamefont {Berthod}},\ }\href {\doibase
  10.1021/acsnano.3c02876} {\bibfield  {journal} {\bibinfo  {journal} {ACS
  Nano}\ }\textbf {\bibinfo {volume} {17}},\ \bibinfo {pages} {9575} (\bibinfo
  {year} {2023})}\BibitemShut {NoStop}%
\bibitem [{\citenamefont {Kitamura}\ \emph {et~al.}(2021)\citenamefont
  {Kitamura}, \citenamefont {Ishizuka}, \citenamefont {Daido},\ and\
  \citenamefont {Yanase}}]{Kitamura2021}%
  \BibitemOpen
  \bibfield  {author} {\bibinfo {author} {\bibfnamefont {T.}~\bibnamefont
  {Kitamura}}, \bibinfo {author} {\bibfnamefont {J.}~\bibnamefont {Ishizuka}},
  \bibinfo {author} {\bibfnamefont {A.}~\bibnamefont {Daido}}, \ and\ \bibinfo
  {author} {\bibfnamefont {Y.}~\bibnamefont {Yanase}},\ }\href {\doibase
  10.1103/PhysRevB.103.245114} {\bibfield  {journal} {\bibinfo  {journal}
  {Phys. Rev. B}\ }\textbf {\bibinfo {volume} {103}},\ \bibinfo {pages}
  {245114} (\bibinfo {year} {2021})}\BibitemShut {NoStop}%
\bibitem [{\citenamefont {Kitamura}\ \emph
  {et~al.}(2022{\natexlab{a}})\citenamefont {Kitamura}, \citenamefont
  {Yamashita}, \citenamefont {Ishizuka}, \citenamefont {Daido},\ and\
  \citenamefont {Yanase}}]{Kitamura2022a}%
  \BibitemOpen
  \bibfield  {author} {\bibinfo {author} {\bibfnamefont {T.}~\bibnamefont
  {Kitamura}}, \bibinfo {author} {\bibfnamefont {T.}~\bibnamefont {Yamashita}},
  \bibinfo {author} {\bibfnamefont {J.}~\bibnamefont {Ishizuka}}, \bibinfo
  {author} {\bibfnamefont {A.}~\bibnamefont {Daido}}, \ and\ \bibinfo {author}
  {\bibfnamefont {Y.}~\bibnamefont {Yanase}},\ }\href {\doibase
  10.1103/PhysRevResearch.4.023232} {\bibfield  {journal} {\bibinfo  {journal}
  {Phys. Rev. Res.}\ }\textbf {\bibinfo {volume} {4}},\ \bibinfo {pages}
  {023232} (\bibinfo {year} {2022}{\natexlab{a}})}\BibitemShut {NoStop}%
\bibitem [{\citenamefont {Kitamura}\ \emph
  {et~al.}(2022{\natexlab{b}})\citenamefont {Kitamura}, \citenamefont {Daido},\
  and\ \citenamefont {Yanase}}]{Kitamura2022b}%
  \BibitemOpen
  \bibfield  {author} {\bibinfo {author} {\bibfnamefont {T.}~\bibnamefont
  {Kitamura}}, \bibinfo {author} {\bibfnamefont {A.}~\bibnamefont {Daido}}, \
  and\ \bibinfo {author} {\bibfnamefont {Y.}~\bibnamefont {Yanase}},\ }\href
  {\doibase 10.1103/PhysRevB.106.184507} {\bibfield  {journal} {\bibinfo
  {journal} {Phys. Rev. B}\ }\textbf {\bibinfo {volume} {106}},\ \bibinfo
  {pages} {184507} (\bibinfo {year} {2022}{\natexlab{b}})}\BibitemShut
  {NoStop}%
\bibitem [{\citenamefont {Meng}\ and\ \citenamefont
  {Balents}(2012)}]{Tobias2012}%
  \BibitemOpen
  \bibfield  {author} {\bibinfo {author} {\bibfnamefont {T.}~\bibnamefont
  {Meng}}\ and\ \bibinfo {author} {\bibfnamefont {L.}~\bibnamefont {Balents}},\
  }\href {\doibase 10.1103/PhysRevB.86.054504} {\bibfield  {journal} {\bibinfo
  {journal} {Phys. Rev. B}\ }\textbf {\bibinfo {volume} {86}},\ \bibinfo
  {pages} {054504} (\bibinfo {year} {2012})}\BibitemShut {NoStop}%
\bibitem [{\citenamefont {Meng}\ and\ \citenamefont
  {Balents}(2017)}]{Erratum_Tobias2012}%
  \BibitemOpen
  \bibfield  {author} {\bibinfo {author} {\bibfnamefont {T.}~\bibnamefont
  {Meng}}\ and\ \bibinfo {author} {\bibfnamefont {L.}~\bibnamefont {Balents}},\
  }\href {\doibase 10.1103/PhysRevB.96.019901} {\bibfield  {journal} {\bibinfo
  {journal} {Phys. Rev. B}\ }\textbf {\bibinfo {volume} {96}},\ \bibinfo
  {pages} {019901} (\bibinfo {year} {2017})}\BibitemShut {NoStop}%
\bibitem [{\citenamefont {Fatemi}\ \emph {et~al.}(2018)\citenamefont {Fatemi},
  \citenamefont {Wu}, \citenamefont {Cao}, \citenamefont {Bretheau},
  \citenamefont {Gibson}, \citenamefont {Watanabe}, \citenamefont {Taniguchi},
  \citenamefont {Cava},\ and\ \citenamefont {Jarillo-Herrero}}]{Fatemi2018}%
  \BibitemOpen
  \bibfield  {author} {\bibinfo {author} {\bibfnamefont {V.}~\bibnamefont
  {Fatemi}}, \bibinfo {author} {\bibfnamefont {S.}~\bibnamefont {Wu}}, \bibinfo
  {author} {\bibfnamefont {Y.}~\bibnamefont {Cao}}, \bibinfo {author}
  {\bibfnamefont {L.}~\bibnamefont {Bretheau}}, \bibinfo {author}
  {\bibfnamefont {Q.~D.}\ \bibnamefont {Gibson}}, \bibinfo {author}
  {\bibfnamefont {K.}~\bibnamefont {Watanabe}}, \bibinfo {author}
  {\bibfnamefont {T.}~\bibnamefont {Taniguchi}}, \bibinfo {author}
  {\bibfnamefont {R.~J.}\ \bibnamefont {Cava}}, \ and\ \bibinfo {author}
  {\bibfnamefont {P.}~\bibnamefont {Jarillo-Herrero}},\ }\href {\doibase
  10.1126/science.aar4642} {\bibfield  {journal} {\bibinfo  {journal}
  {Science}\ }\textbf {\bibinfo {volume} {362}},\ \bibinfo {pages} {926}
  (\bibinfo {year} {2018})},\ \Eprint
  {http://arxiv.org/abs/https://www.science.org/doi/pdf/10.1126/science.aar4642}
  {https://www.science.org/doi/pdf/10.1126/science.aar4642} \BibitemShut
  {NoStop}%
\bibitem [{\citenamefont {Sajadi}\ \emph {et~al.}(2018)\citenamefont {Sajadi},
  \citenamefont {Palomaki}, \citenamefont {Fei}, \citenamefont {Zhao},
  \citenamefont {Bement}, \citenamefont {Olsen}, \citenamefont {Luescher},
  \citenamefont {Xu}, \citenamefont {Folk},\ and\ \citenamefont
  {Cobden}}]{Sajadi2018}%
  \BibitemOpen
  \bibfield  {author} {\bibinfo {author} {\bibfnamefont {E.}~\bibnamefont
  {Sajadi}}, \bibinfo {author} {\bibfnamefont {T.}~\bibnamefont {Palomaki}},
  \bibinfo {author} {\bibfnamefont {Z.}~\bibnamefont {Fei}}, \bibinfo {author}
  {\bibfnamefont {W.}~\bibnamefont {Zhao}}, \bibinfo {author} {\bibfnamefont
  {P.}~\bibnamefont {Bement}}, \bibinfo {author} {\bibfnamefont
  {C.}~\bibnamefont {Olsen}}, \bibinfo {author} {\bibfnamefont
  {S.}~\bibnamefont {Luescher}}, \bibinfo {author} {\bibfnamefont
  {X.}~\bibnamefont {Xu}}, \bibinfo {author} {\bibfnamefont {J.~A.}\
  \bibnamefont {Folk}}, \ and\ \bibinfo {author} {\bibfnamefont {D.~H.}\
  \bibnamefont {Cobden}},\ }\href {\doibase 10.1126/science.aar4426} {\bibfield
   {journal} {\bibinfo  {journal} {Science}\ }\textbf {\bibinfo {volume}
  {362}},\ \bibinfo {pages} {922} (\bibinfo {year} {2018})},\ \Eprint
  {http://arxiv.org/abs/https://www.science.org/doi/pdf/10.1126/science.aar4426}
  {https://www.science.org/doi/pdf/10.1126/science.aar4426} \BibitemShut
  {NoStop}%
\bibitem [{\citenamefont {Wu}\ \emph {et~al.}(2018)\citenamefont {Wu},
  \citenamefont {Fatemi}, \citenamefont {Gibson}, \citenamefont {Watanabe},
  \citenamefont {Taniguchi}, \citenamefont {Cava},\ and\ \citenamefont
  {Jarillo-Herrero}}]{Wu2018}%
  \BibitemOpen
  \bibfield  {author} {\bibinfo {author} {\bibfnamefont {S.}~\bibnamefont
  {Wu}}, \bibinfo {author} {\bibfnamefont {V.}~\bibnamefont {Fatemi}}, \bibinfo
  {author} {\bibfnamefont {Q.~D.}\ \bibnamefont {Gibson}}, \bibinfo {author}
  {\bibfnamefont {K.}~\bibnamefont {Watanabe}}, \bibinfo {author}
  {\bibfnamefont {T.}~\bibnamefont {Taniguchi}}, \bibinfo {author}
  {\bibfnamefont {R.~J.}\ \bibnamefont {Cava}}, \ and\ \bibinfo {author}
  {\bibfnamefont {P.}~\bibnamefont {Jarillo-Herrero}},\ }\href {\doibase
  10.1126/science.aan6003} {\bibfield  {journal} {\bibinfo  {journal}
  {Science}\ }\textbf {\bibinfo {volume} {359}},\ \bibinfo {pages} {76}
  (\bibinfo {year} {2018})},\ \Eprint
  {http://arxiv.org/abs/https://www.science.org/doi/pdf/10.1126/science.aan6003}
  {https://www.science.org/doi/pdf/10.1126/science.aan6003} \BibitemShut
  {NoStop}%
\bibitem [{\citenamefont {Fei}\ \emph {et~al.}(2017)\citenamefont {Fei},
  \citenamefont {Palomaki}, \citenamefont {Wu}, \citenamefont {Zhao},
  \citenamefont {Cai}, \citenamefont {Sun}, \citenamefont {Nguyen},
  \citenamefont {Finney}, \citenamefont {Xu},\ and\ \citenamefont
  {Cobden}}]{Fei2017}%
  \BibitemOpen
  \bibfield  {author} {\bibinfo {author} {\bibfnamefont {Z.}~\bibnamefont
  {Fei}}, \bibinfo {author} {\bibfnamefont {T.}~\bibnamefont {Palomaki}},
  \bibinfo {author} {\bibfnamefont {S.}~\bibnamefont {Wu}}, \bibinfo {author}
  {\bibfnamefont {W.}~\bibnamefont {Zhao}}, \bibinfo {author} {\bibfnamefont
  {X.}~\bibnamefont {Cai}}, \bibinfo {author} {\bibfnamefont {B.}~\bibnamefont
  {Sun}}, \bibinfo {author} {\bibfnamefont {P.}~\bibnamefont {Nguyen}},
  \bibinfo {author} {\bibfnamefont {J.}~\bibnamefont {Finney}}, \bibinfo
  {author} {\bibfnamefont {X.}~\bibnamefont {Xu}}, \ and\ \bibinfo {author}
  {\bibfnamefont {D.~H.}\ \bibnamefont {Cobden}},\ }\href {\doibase
  10.1038/nphys4091} {\bibfield  {journal} {\bibinfo  {journal} {Nature
  Physics}\ }\textbf {\bibinfo {volume} {13}},\ \bibinfo {pages} {677}
  (\bibinfo {year} {2017})}\BibitemShut {NoStop}%
\bibitem [{\citenamefont {Xu}\ \emph {et~al.}(2018)\citenamefont {Xu},
  \citenamefont {Ma}, \citenamefont {Shen}, \citenamefont {Fatemi},
  \citenamefont {Wu}, \citenamefont {Chang}, \citenamefont {Chang},
  \citenamefont {Valdivia}, \citenamefont {Chan}, \citenamefont {Gibson},
  \citenamefont {Zhou}, \citenamefont {Liu}, \citenamefont {Watanabe},
  \citenamefont {Taniguchi}, \citenamefont {Lin}, \citenamefont {Cava},
  \citenamefont {Fu}, \citenamefont {Gedik},\ and\ \citenamefont
  {Jarillo-Herrero}}]{Xu2018}%
  \BibitemOpen
  \bibfield  {author} {\bibinfo {author} {\bibfnamefont {S.-Y.}\ \bibnamefont
  {Xu}}, \bibinfo {author} {\bibfnamefont {Q.}~\bibnamefont {Ma}}, \bibinfo
  {author} {\bibfnamefont {H.}~\bibnamefont {Shen}}, \bibinfo {author}
  {\bibfnamefont {V.}~\bibnamefont {Fatemi}}, \bibinfo {author} {\bibfnamefont
  {S.}~\bibnamefont {Wu}}, \bibinfo {author} {\bibfnamefont {T.-R.}\
  \bibnamefont {Chang}}, \bibinfo {author} {\bibfnamefont {G.}~\bibnamefont
  {Chang}}, \bibinfo {author} {\bibfnamefont {A.~M.~M.}\ \bibnamefont
  {Valdivia}}, \bibinfo {author} {\bibfnamefont {C.-K.}\ \bibnamefont {Chan}},
  \bibinfo {author} {\bibfnamefont {Q.~D.}\ \bibnamefont {Gibson}}, \bibinfo
  {author} {\bibfnamefont {J.}~\bibnamefont {Zhou}}, \bibinfo {author}
  {\bibfnamefont {Z.}~\bibnamefont {Liu}}, \bibinfo {author} {\bibfnamefont
  {K.}~\bibnamefont {Watanabe}}, \bibinfo {author} {\bibfnamefont
  {T.}~\bibnamefont {Taniguchi}}, \bibinfo {author} {\bibfnamefont
  {H.}~\bibnamefont {Lin}}, \bibinfo {author} {\bibfnamefont {R.~J.}\
  \bibnamefont {Cava}}, \bibinfo {author} {\bibfnamefont {L.}~\bibnamefont
  {Fu}}, \bibinfo {author} {\bibfnamefont {N.}~\bibnamefont {Gedik}}, \ and\
  \bibinfo {author} {\bibfnamefont {P.}~\bibnamefont {Jarillo-Herrero}},\
  }\href {\doibase 10.1038/s41567-018-0189-6} {\bibfield  {journal} {\bibinfo
  {journal} {Nature Physics}\ }\textbf {\bibinfo {volume} {14}},\ \bibinfo
  {pages} {900} (\bibinfo {year} {2018})}\BibitemShut {NoStop}%
\bibitem [{\citenamefont {Papaj}\ and\ \citenamefont
  {Moore}(2022)}]{Papaj2022}%
  \BibitemOpen
  \bibfield  {author} {\bibinfo {author} {\bibfnamefont {M.}~\bibnamefont
  {Papaj}}\ and\ \bibinfo {author} {\bibfnamefont {J.~E.}\ \bibnamefont
  {Moore}},\ }\href {\doibase 10.1103/PhysRevB.106.L220504} {\bibfield
  {journal} {\bibinfo  {journal} {Phys. Rev. B}\ }\textbf {\bibinfo {volume}
  {106}},\ \bibinfo {pages} {L220504} (\bibinfo {year} {2022})}\BibitemShut
  {NoStop}%
\bibitem [{\citenamefont {Dai}\ and\ \citenamefont {Lee}(2017)}]{Dai2017}%
  \BibitemOpen
  \bibfield  {author} {\bibinfo {author} {\bibfnamefont {Z.}~\bibnamefont
  {Dai}}\ and\ \bibinfo {author} {\bibfnamefont {P.~A.}\ \bibnamefont {Lee}},\
  }\href {\doibase 10.1103/PhysRevB.95.014506} {\bibfield  {journal} {\bibinfo
  {journal} {Phys. Rev. B}\ }\textbf {\bibinfo {volume} {95}},\ \bibinfo
  {pages} {014506} (\bibinfo {year} {2017})}\BibitemShut {NoStop}%
\bibitem [{\citenamefont {Nambu}(1960)}]{Nambu1960}%
  \BibitemOpen
  \bibfield  {author} {\bibinfo {author} {\bibfnamefont {Y.}~\bibnamefont
  {Nambu}},\ }\href {\doibase 10.1103/PhysRev.117.648} {\bibfield  {journal}
  {\bibinfo  {journal} {Phys. Rev.}\ }\textbf {\bibinfo {volume} {117}},\
  \bibinfo {pages} {648} (\bibinfo {year} {1960})}\BibitemShut {NoStop}%
\end{thebibliography}%

\end{document}